\documentclass[preprint,12pt]{elsarticle}

\usepackage{amssymb}
\usepackage{lineno}
\usepackage{amsmath}
\usepackage{hyperref}
\usepackage{booktabs}
\usepackage{tabularx}
\usepackage{placeins}
\usepackage{subcaption}
\usepackage{soul}
\usepackage{xcolor}
\sethlcolor{yellow}
\usepackage{comment}
\usepackage{bm}
\usepackage{bookmark}
\usepackage{makecell}
\usepackage{tikz}
\usepackage{lipsum}

\ExplSyntaxOn
\clist_const:Nn \l_soul_protect_clist { cite , citep , ref , eqref }

\NewDocumentCommand{\robusthln}{m}{
  \tl_set:Nn \l_tmpa_tl { #1 }
  \clist_map_inline:Nn \l_soul_protect_clist {
    \regex_replace_all:nnN { (\c{##1}[^{}]*\{[^{}]*\}) } %
                           { \c{mbox}\{\1\} } \l_tmpa_tl
  }
  \hl\l_tmpa_tl
}
\ExplSyntaxOff
\newcommand{\mathcolorbox}[2]{\ensuremath{#2}}

\journal{Developments in the Built Environment}

\makeatletter
\g@addto@macro\ps@pprintTitle{%
  \def\@oddhead{\hfil Accepted by \textit{Developments in the Built Environment}\hfil}%
  \let\@evenhead\@oddhead
}
\makeatother

\begin{document}

\begin{frontmatter}

\title{Near-real-time, meter-scale 3D urban wind modeling for low-altitude micrometeorology: numerical verification of a GPU-accelerated lattice Boltzmann framework}

\author[label1]{Shuai Han}
\author[label3,label4]{Huanxia Wei\corref{cor1}}
\ead{huanxia.wei@manchester.ac.uk}
\cortext[cor1]{Corresponding author.}
\author[label2,label4]{Yue Cao}
\author[label5]{Dalin Liu}
\author[label7]{Lin Wen}
\author[label4]{Chao Xia}
\author[label8]{Shuolin Xiao}
\author[label10]{Yingying Xing}
\author[label4]{Qing Jia}
\author[label6]{Wenguang Liang}
\author[label4,label9]{Zhigang Yang}

\affiliation[label1]{
            organization={National Meteorological Information Centre},
            addressline={China Meteorological Administration}, 
            city={Beijing}, 
            postcode={100081}, 
            country={China}}

\affiliation[label2]{
            organization={Department of Civil and Environmental Engineering},
            addressline={University of California Berkeley}, 
            city={Berkeley}, 
            postcode={94704}, 
            state={CA}, 
            country={United States}}
         
\affiliation[label3]{
            organization={Department of Mechanical and Aerospace Engineering},
            addressline={The University of Manchester}, 
            city={Manchester}, 
            postcode={M13 9PL}, 
            country={UK}}
            
\affiliation[label5]{
            organization={Department of Building Environment},
            addressline={National University of Singapore}, 
            postcode={119077}, 
            country={Singapore}}

\affiliation[label4]{
            organization={College of Automotive and Energy Engineering},
            addressline={Tongji University}, 
            city={Shanghai}, 
            postcode={201804}, 
            country={China}}
            
\affiliation[label6]{
            organization={Beijing Fengyun Meteorological Science and Technology Development Co., Ltd.},
            city={Beijing}, 
            postcode={100081}, 
            country={China}}

\affiliation[label7]{
            organization={School of Artificial Intelligence},
            addressline={Sun Yat-sen University},
            city={Zhuhai},
            postcode={519000}, 
            state={Guangdong Province},
            country={China}}  
            
\affiliation[label8]{
            organization={Ralph O'Connor Sustainable Energy Institute},
            addressline={Johns Hopkins University},
            city={Baltimore},
            postcode={21218}, 
            state={MD},
            country={United States}}

\affiliation[label9]{
                      organization={Beijing Aeronautical Science and Technology Research Institute},
            addressline={The Commercial Aircraft Corporation of China},
            city={Beijing},
            postcode={102211}, 
            country={China}}  
\affiliation[label10]{
  organization={The Key Laboratory of Road and Traffic Engineering of Ministry of Education, Tongji University},
  city={Shanghai},
  postcode={201804},
  country={China}}
            
\begin{abstract}
This study presents a near-real-time, meter-scale three-dimensional urban wind simulation framework for low-altitude flight events in complex urban meteorological environments. It reconstructs high-resolution wind fields by combining sparse observations with efficient microscale flow modeling. The framework integrates lattice Boltzmann method large-eddy simulation (LBM-LES), high-fidelity urban morphology reconstruction that explicitly resolves real building details, and observation-driven boundary assimilation into a rapid end-to-end pipeline for realistic urban domains. Multi-site Doppler lidar measurements from dense urban Guangzhou, China, are used for evaluation. {The system reconstructs three-dimensional wind fields at 5 m resolution over kilometer-scale domains in approximately 5 min on a single NVIDIA A800 80G GPU.} Robustness and accuracy are tested through controlled observation reduction, independent validation against withheld lidar stations, and sensitivity analyses of grid resolution and precursor domain extent. Results show stable reproduction of vertical wind structures and key local flow features under complex morphology and limited observations, providing a scalable pathway for near-real-time urban wind reconstruction.
\end{abstract}






\end{frontmatter}


\section{Introduction\label{sec:1_intro}}
{The rapid rise of the low-altitude economy and urban air mobility (UAM) is reshaping how cities move people and deliver services \cite{garrow2021uamreview,bauranov2021airspace,TenQuestion}.} Low-altitude operations such as drone logistics, urban air mobility, emergency response, inspection, and surveying typically fly close to the ground surface, demand strong timeliness, and unfold in highly complex environments \cite{steiner2019uam,reiche2021weatherbarriers,garrow2021uamreview}. As a result, meteorological conditions have evolved from an important constraint in conventional aviation support into a primary basis for developing and exploiting the spatiotemporal resources of low-altitude airspace \cite{UC-Drones,reiche2021weatherbarriers}. In particular, the three-dimensional wind field and its rapid variability affect vehicle stability, control margins, trajectory tracking, energy consumption, and mission success. They also define safe operating envelopes over complex urban surfaces \cite{steiner2019uam,reiche2021weatherbarriers}. In megacities, the combined effects of building clusters, street canyons, green space, and water bodies modify near surface momentum and heat exchange, generating strong spatial heterogeneity and multiscale coupling \cite{grimmondoke1999aerodynamic,masson2000teb,martilli2002urbanexchange,bowler2010greening}. {Street-canyon geometry has long been recognized as a primary control on canyon-scale flow channeling and ventilation pathways \cite{oke1988streetdesign}. Reviews of street-canyon flow and dispersion further document recirculation zones, flow separation, and downstream reattachment generated by urban canyon vortices \cite{vardoulakis2003streetcanyon}. At the building and district scales, computational wind engineering studies report local wind acceleration near corners and roof edges together with separated wakes around bluff bodies \cite{blocken2014cwe}, while the Architectural Institute of Japan (AIJ) practical guidelines for computational fluid dynamics (CFD) identify separation and reattachment around buildings as core validation targets for urban flow simulations \cite{tominaga2008aijguidelines}. Recent evaluation of a multilayer urban canopy model across contrasting urban contexts likewise confirms that representing urban morphological effects remains essential for reproducing these microscale wind responses \cite{EDT-MLUCM}.}
{Aircraft participating in the low-altitude economy and UAM often operate within the low-altitude airspace of 0--1000 m \cite{onek1,onek2,onek3}, which coincides with the region characterized by these strong turbulence and unsteady flow phenomenon \cite{facingunsteady}, further underscoring the importance of acquiring small-scale flow information for general aviation \cite{Yew2025Review}.}
High-quality three-dimensional wind simulation and refined wind field analysis have consequently become central challenges for urban meteorological support and the development of low-altitude meteorological service capability \cite{blocken2015urbanphysics,schar2019kmscale}.

Operational monitoring and warning for strong winds rely mainly on observation systems such as Doppler light detection and ranging (lidar) systems, meteorological towers, and automatic weather stations \cite{muller2013sensorscity}. Mature services have been established in high impact sectors including coastal ports, urban rail transit, and construction sites \cite{garrow2021uamreview,muller2013sensorscity}. These systems provide reliable time series observations that support the identification of regional strong wind events and the issuance of warnings \cite{verkaik2000gustiness}. However, conventional wind observing systems face intrinsic limits in representativeness \cite{muller2013sensorscity}. On the one hand, sensitivity, saturation effects, and instrument inertia can introduce dynamic errors when measuring intense gusts or rapidly changing flow \cite{miller2013dines}. On the other hand, point-based observations struggle to capture the sharp spatial contrasts of small-scale urban winds \cite{grimmondoke1999aerodynamic}. Given constraints on space, staffing, maintenance, and equipment cost, engineering a dense enough network to reconstruct accurate three-dimensional winds over a small urban area is not realistic \cite{muller2013sensorscity}. The resulting service granularity remains insufficient for low-altitude flight \cite{steiner2019uam,reiche2021weatherbarriers}. For example, when strong winds approach, current products often cannot provide actionable quantitative guidance on the differing small-scale risks across specific corridors or route segments, which limits professional and fine-grained low-altitude meteorological support \cite{steiner2019uam,reiche2021weatherbarriers}.

Numerical weather prediction models improve service coverage through gridded forecasts and are a key pathway to continuous depiction of regional wind fields \cite{bauer2015nwpquiet,prein2015cpmreview}. The forecast quality of fine scale wind structures depends strongly on how well processes such as deep convection, boundary layer turbulence, and interactions with urban surfaces are represented \cite{schar2019kmscale,masson2000teb,martilli2002urbanexchange}. {Many studies have shown that increasing horizontal resolution helps resolve the physical processes and detailed structures of small-scale weather systems \cite{roberts2008scaleselective,schwartz2009nextday,prein2021sensitivity}.} Yet convective storms are highly localized, abrupt, and fast evolving, and reliable prediction often requires robust subkilometre high resolution simulation \cite{prein2015cpmreview,schar2019kmscale}. In recent years, advances in supercomputing and parallelization have pushed mesoscale model resolution toward the kilometre scale, but further refinement to subkilometre and finer scales faces substantial challenges due to limits in dynamical frameworks and vertical coordinate systems \cite{bauer2015nwpquiet,schar2019kmscale}. Finer-scale simulations demand higher-fidelity terrain and land-use data and surface parameterizations that represent urban morphology (e.g., building arrays, street canyons, and vegetation) \cite{EDT-Canyon}. These elements are required to capture urban impacts on local winds at minute-level temporal resolution and meter-scale spatial resolution \cite{grimmondoke1999aerodynamic,masson2000teb,martilli2002urbanexchange}. Traditional mesoscale models often struggle to maintain both physical consistency and computational affordability in this regime \cite{bauer2015nwpquiet,schar2019kmscale}.

{For urban microscale problems, CFD provides an effective means to represent complex geometric boundaries and local turbulent structures \cite{EDT-onebd}, delivering high resolution wind and temperature fields that support risk identification for low-altitude corridors \cite{POF-UAV}, evaluation of takeoff and landing site suitability \cite{SAE}, building and pedestrian ventilation \cite{EDT-indoor}, and urban environmental management \cite{blocken2013microdispersion}.} With improved hardware performance and higher efficiency in high performance parallel computing, real-time or near-real-time simulation over complex urban districts is becoming increasingly feasible \cite{latt2021palabos,zhang2025thmusic}. Nonetheless, key trade-offs remain. Large-eddy simulation (LES) resolves large scale turbulent motions and uses subgrid scale models for smaller scales, typically achieving high turbulence fidelity, but at prohibitive computational cost for large urban domains and operational timeliness requirements \cite{meneveau2000les,blocken2014cwe}. Reynolds averaged Navier-Stokes simulations (RANS) reduce cost through flow averaging and are easier to deploy, but rely on Reynolds stress closure and therefore introduce turbulence model error \cite{blocken2014cwe,tominaga2008aijguidelines}. Meanwhile, urban roughness and surface parameterizations remain immature, leading to unstable accuracy in complex urban morphologies and limiting their ability to support the fine-grained risk quantification required for low-altitude operations \cite{grimmondoke1999aerodynamic,masson2000teb,martilli2002urbanexchange,blocken2015urbanphysics}.

The lattice Boltzmann method (LBM), which has developed rapidly in recent years, is a mesoscopic computational fluid dynamics approach with relatively simple formulations of fluid interactions, convenient treatment of complex boundaries, high parallel efficiency, and a clear implementation pathway \cite{aidun2010lbmreview,geier2006cascaded}. These features suggest potential advantages for complex geometry and large-scale parallel computation \cite{latt2021palabos,schornbaum2016massivelyparallel}. For the microscale three-dimensional wind field needs of low-altitude operations in cities, the efficiency and boundary handling properties of lattice Boltzmann methods may offer a promising technical option that balances high resolution with operational usability \cite{zhang2025thmusic,UC-GPU}. However, from the perspective of meteorological support for the low-altitude economy, a clear discontinuity remains between existing research and operational systems \cite{blocken2015urbanphysics,steiner2019uam,zhang2025thmusic}. Existing urban microscale wind studies have addressed a broad range of built-environment applications, including urban wind-resource evaluation and wind-turbine feasibility \cite{tasneem2020urbanWindResourceReview}, morphology-based classification of local ventilation performance zones \cite{li2023localVentilationPerformanceZones}, physics-based assessment of wind damage to building clusters \cite{gu2023cimWindDamageBuildingClusters}, and wind-induced interference effects around high-rise buildings \cite{wang2025windInducedInterferenceKAN}. These studies collectively highlight the importance of resolving urban morphology, local ventilation pathways, and building-scale flow interactions. However, their problem settings are still mainly oriented toward wind-environment assessment, design support, or post-event risk evaluation, rather than observation-constrained, corridor-oriented low-altitude meteorological services. Boundary forcing, the precision of urban morphology data, and standards for systematic validation against observations also remain insufficiently unified, making results difficult to translate into reusable city-scale products \cite{tominaga2008aijguidelines,blocken2013microdispersion,blocken2015urbanphysics}. Operational low-altitude meteorological services, by contrast, emphasize minute level updates, metre to hectometre scale spatial guidance, and interpretable risk factor outputs \cite{steiner2019uam,reiche2021weatherbarriers}. Yet current observation networks cannot provide sufficiently dense three-dimensional constraints, conventional mesoscale models cannot resolve urban morphological detail, and classical microscale methods have not resolved the core tension between cost and accuracy \cite{muller2013sensorscity,schar2019kmscale,blocken2014cwe,meneveau2000les}.
Zhang et al. \cite{zhang2025thmusic} developed \textsc{TH-MuSiC}, a high-efficiency platform coupling the Weather Research and Forecasting model with LBM. However, the end-to-end workflow still requires tens of minutes for domains with tens of millions of cells.
Overall, a key research gap persists: a technical route that can stably produce high resolution three-dimensional wind fields over complex urban surfaces at an affordable computational cost, while synergizing with existing meteorological observation and forecasting systems \cite{bauer2015nwpquiet,blocken2015urbanphysics}. Moreover, the numerical methods \cite{EDT-Wallfunction}, mesh strategy \cite{EDT-LBMMesh} and domain configurations are also worth noting to ensure the reliability.

In this study, we develop and verify a near-real-time, meter-scale three-dimensional urban wind reconstruction framework aimed at providing operational meteorological support for low-altitude applications over complex city environments. The framework integrates high-fidelity urban morphology reconstruction with a GPU-accelerated LBM-LES solver powered by \textsc{FluidX3D}.
Sparse multi-site Doppler lidar measurements are used to build observation-driven boundary conditions via gridded volumetric wind reconstruction and conservative boundary correction, forming an end-to-end pipeline that can deliver kilometer-scale 3D wind fields within minutes.
Using a dense lidar deployment over Haixinsha Island and its surroundings in Guangzhou, we conduct numerical verifications via independent station holdout and progressive reductions of observational inputs to assess robustness under limited constraints, and further examine the sensitivity of reconstructed winds to lattice resolution and precursor-domain extent to clarify the dominant numerical factors governing accuracy and stability. 

The remainder of this paper is organized as follows. Section~\ref{2_numerical} describes the integrated preprocessing, boundary reconstruction, and the GPU LBM-LES implementation. Section~\ref{sec:3_simulations} introduces the study area, lidar observations, and experimental design. Section~\ref{sec:4_results} presents validation results and sensitivity analyses. Section~\ref{sec:4_concl} summarizes the main findings and discusses implications for near-real-time low-altitude meteorological services.

\section{Numerical methods\label{2_numerical}}

We developed a comprehensive simulation pipeline that integrates geospatial preprocessing, conservative boundary reconstruction, voxel-based geometry handling, and \textsc{FluidX3D}, the GPU-accelerated LBM kernel. The solver incorporates LES closure and Coriolis forcing to accurately capture atmospheric flow dynamics.

\subsection{Geospatial preprocessing and boundary reconstruction}

To ensure geometric fidelity, all geospatial inputs are first transformed into a metric projected coordinate system suitable for uniform Cartesian grids. Because map projection can distort rectangular longitude-latitude extents, the computational domain is re-clipped in the projected space to recover an axis-aligned rectangular region compatible with structured meshing. Terrain elevations derived from the Digital Elevation Model (DEM) are interpolated from scattered samples to a regular raster using Inverse Distance Weighting (IDW), followed by smoothing with a discrete Gaussian kernel. This process suppresses small-scale noise while retaining large-scale topographic gradients. During subsequent processing, terrain height values are retrieved via fast bilinear interpolation on this precomputed raster.
Details of the terrain-layer construction are provided in Supplementary Material~\ref{supp:TEInt}.

Building placement is rendered terrain-consistent by aligning the building base to the local minimum terrain elevation over its footprint, thereby avoiding gaps or solid overlaps at the ground interface. Both buildings and terrain are mapped onto the Cartesian lattice through voxelization. A ray-tracing-based occupancy test efficiently classifies solid and fluid cells over large-scale domains. The final solid mask is assembled by the Boolean merging of a fixed-thickness base layer, the terrain layer, and the building layer, yielding a unified no-slip boundary representation. Since Cartesian voxelization introduces a staircase approximation, spatial resolution is treated as a governing modeling parameter for geometric fidelity and near-wall momentum exchange.

Boundary wind conditions are derived via a two-stage downscaling strategy designed to balance accuracy and throughput. First, the irregular point field in projected space is interpolated to an intermediate Cartesian grid using Delaunay triangulation, employing barycentric coordinates for linear interpolation to preserve local gradients without extraneous smoothing. For targets outside the triangulation convex hull, the procedure reverts to a nearest-neighbor assignment accelerated by a k-dimensional (k-d) tree. Second, the intermediate boundary field is mapped onto the CFD boundary nodes. This step utilizes either a nearest scheme for maximum computational speed or an improved k-Nearest Neighbor (kNN) reconstruction for smoother, more stable fields. The improved method employs a fixed neighborhood size, inverse distance squared weighting, and an adaptive stabilization term to prevent numerical sensitivity at minimal distances, while applying a consistent weight set across all velocity components to maintain vector coherence. The full two-stage velocity-interpolation procedure is shown in Supplementary Material~\ref{supp:VInt}.

To mitigate spurious compressibility and net volume imbalance inherent in imperfect atmospheric boundary data, we apply a mass flux consistency correction at each boundary. This correction eliminates mean normal flux bias by applying the minimal normal velocity adjustment required to enforce near-zero net volume flow, ensuring the interpolated boundary field remains as close to the original data as possible.

\subsection{Lattice Boltzmann flow solver and implementation}

The refined urban wind simulation employs the mesoscopic LBM based on \textsc{FluidX3D}. Compared to traditional CFD solvers based on macroscopic continuum mechanics, LBM offers distinct advantages, including algorithmic simplicity, minimal numerical dissipation, and inherent capability in handling complex boundaries. The method utilizes an explicit time-integration scheme where the non-linear collision term is local and the linear streaming step requires no interpolation. These characteristics eliminate the need for global matrix inversions and guarantee high computational efficiency and ease of parallelization.
In the LBM framework, the continuous phase space is discretized by a finite set of virtual particle velocities. This discrete velocity scheme serves two critical functions: it reduces the infinite motion modes of real fluid molecules to a computable finite set, and it discretizes continuous space by defining the distance virtual particles travel per time step. This approach replaces the physical mesh and finite-difference discretizations used in traditional CFD, thereby reducing errors associated with mesh generation and discretization schemes.

In this study, we adopt the D3Q19 discrete velocity model, where $D$ denotes the spatial dimension and $Q$ represents the number of discrete velocity directions. The D3Q19 model is widely validated for simulating three-dimensional flows across various urban scales, ranging from isolated buildings to complex street canyons and city blocks. The flow field evolution is governed by the Single Relaxation Time (SRT) collision operator on a uniform lattice. Subgrid-scale stresses are modeled using the Smagorinsky-Lilly model within an LES framework. Open boundaries are managed via equilibrium-based velocity constraints and reflection suppression to minimize artificial wave return, while ground, terrain, and buildings are imposed as no-slip walls.
A dimensionless scaling scheme adaptively matches the reference wind speed to a target LBM Mach number, ensuring numerical stability under high winds.
Earth rotation effects are incorporated through a Coriolis body force evaluated directly within the GPU kernel. The rotation vector is expressed in the local East-North-Up (ENU) frame at the domain latitude, consistently converted into lattice units, and integrated using a Guo-type forcing discretization. A complete description of the LBM formulation and the \textsc{FluidX3D}-based implementation is provided in Supplementary Material~\ref{supp:LBM}. {The physical boundary condition and numerical formulation of the immersed-boundary method are provided in Supplementary Material~\ref{supp:surface-momentum-ibm}.} The Coriolis forcing term is detailed in Supplementary Material~\ref{supp:sourceterm}.

The implementation follows a hybrid CPU-GPU architecture. The CPU handles preprocessing, constructing and caching interpolation search structures such as triangulations and k-d trees with full parallel-computing capability. Meanwhile, GPU kernels execute the computationally intensive per-step macroscopic reconstruction, source term assembly, and lattice updates. This organization minimizes synchronization overhead and data movement, enabling efficient simulation over extremely large structured grids.

{Together, Supplementary Materials~\ref{supp:Kriging}--\ref{supp:wn} document the complete workflow for boundary reconstruction, terrain and velocity processing, the LBM--LES and thermal formulations, immersed-boundary treatment, Coriolis forcing, synthetic turbulence generation, and wavenumber-spectrum analysis.}
{For clarity, the principal numerical choices and baseline settings used in the present urban wind-field calculations are summarized in Table~\ref{tab:numerical-configuration}.}

\begin{table}[!htbp]
\centering
\caption{{Numerical configuration of the baseline urban wind-field calculations.}}
\label{tab:numerical-configuration}
\footnotesize
\setlength{\tabcolsep}{5pt}
\renewcommand{\arraystretch}{1.12}
\begin{tabularx}{\linewidth}{@{}>{\raggedright\arraybackslash}p{0.25\linewidth}>{\raggedright\arraybackslash}X@{}}
\toprule
{Item} & {Configuration} \\
\midrule
{Flow lattice} & {Weakly compressible D3Q19 LBM on a uniform Cartesian lattice.} \\
{Collision and integration} & {Single-relaxation-time Bhatnagar--Gross--Krook collision with an explicit collide--stream update.} \\
{LES closure} & {Smagorinsky--Lilly subgrid model, with the effective relaxation time computed locally from nonequilibrium moments.} \\
{Baseline grid and scaling} & {5 m isotropic spacing, 116 million lattice cells, and a target lattice Mach number of 0.10.} \\
{Time step and duration} & {0.01 s and 40,000 steps, corresponding to 400 s of physical time and approximately 85 building-scale turnover times.} \\
{Open boundaries} & {Equilibrium-reconstruction velocity treatment on the four lateral faces and top, with a fixed analysis-time mean, synthetic von Karman fluctuations on active inflow cells, sponge and buffer damping, and global flux correction.} \\
{Solid boundaries} & {The lower boundary, terrain, and buildings are stationary, impermeable no-slip surfaces; regularized immersed-boundary direct forcing supplies the wall treatment and is rebuilt at each LBM step.} \\
{Rotational forcing} & {Coriolis acceleration in local ENU coordinates incorporated through Guo forcing.} \\
{Thermal coupling} & {Optional D3Q7 or D3Q19 thermal lattice with Boussinesq coupling; the present urban calculations are isothermal with buoyancy disabled.} \\
\bottomrule
\end{tabularx}
\end{table}

\subsection{\texorpdfstring{{Validation in principle}}{Validation in principle}\label{sec:validation-in-principle}}

\subsubsection{\texorpdfstring{{AIJ wind-tunnel validation}}{AIJ wind-tunnel validation}\label{sec:aij-validation}}

{The solver configuration is first evaluated against the AIJ Case A benchmark of turbulent flow around an isolated 1:1:2 square prism (Meng--Hibi experiment), a standard validation case for building-resolved urban CFD }\cite{tominaga2008aijguidelines,MengHibi1998}{. At full scale, the building has a width of }\mathcolorbox{yellow}{B=32~\mathrm{m}}{ and a height of }\mathcolorbox{yellow}{H=64~\mathrm{m}}{, and the roof-height reference speed is }\mathcolorbox{yellow}{U_H=4.49~\mathrm{m/s}}{. The calculation uses a 2~m lattice spacing, corresponding to }\mathcolorbox{yellow}{H/\Delta=32}{, and is compared with the wind-tunnel measurements on two horizontal planes, }\mathcolorbox{yellow}{z/H=0.125}{ and }\mathcolorbox{yellow}{z/H=1.25}{, and on the centreline vertical section. Table~\ref{tab:aij-case-a} reports the fraction-of-two agreement (FAC2), hit rate }\mathcolorbox{yellow}{q}{ at a tolerance of 0.25, correlation }\mathcolorbox{yellow}{r}{, and normalized root-mean-square error (NRMSE) for the normalized mean speed, together with FAC2 and }\mathcolorbox{yellow}{r}{ for the normalized turbulent kinetic energy (TKE).}

\begin{table}[!htbp]
\centering
\caption{{AIJ Case A single-prism validation. The hit rate }\mathcolorbox{yellow}{q}{ uses a tolerance of 0.25. The TKE FAC2 excludes near-zero measurement points (}\mathcolorbox{yellow}{|k/U_H^2|\leq 0.05}{), leaving 26--30 of the 60 or 66 points per plane.}}
\label{tab:aij-case-a}
\footnotesize
\setlength{\tabcolsep}{5pt}
\renewcommand{\arraystretch}{1.1}
\begin{tabular}{lcccccc}
\toprule
 & \multicolumn{4}{c}{{Mean speed }\mathcolorbox{yellow}{U/U_H}} & \multicolumn{2}{c}{{TKE }\mathcolorbox{yellow}{k/U_H^2}} \\
\cmidrule(lr){2-5}\cmidrule(lr){6-7}
{Plane} & {FAC2} & \mathcolorbox{yellow}{q} & \mathcolorbox{yellow}{r} & {NRMSE} & {FAC2} & \mathcolorbox{yellow}{r} \\
\midrule
\mathcolorbox{yellow}{z/H=0.125} & {0.74} & {0.62} & {0.78} & {0.26} & {0.31} & {0.64} \\
\mathcolorbox{yellow}{z/H=1.25}  & {0.93} & {0.97} & {0.92} & {0.13} & {0.90} & {0.88} \\
{Vertical section}    & {0.81} & {0.89} & {0.97} & {0.12} & {0.90} & {0.53} \\
\bottomrule
\end{tabular}
\end{table}

{Overall, the mean-speed comparison supports satisfactory agreement for the present solver configuration, although the performance is not uniform across all three evaluation sections. At $z/H = 1.25$, FAC2 is 0.93 and the hit rate is 0.97, while the centreline vertical-section correlation is 0.97. The TKE FAC2 is 0.90 both at $z/H = 1.25$ and in the vertical section. At the near-ground plane $z/H = 0.125$, the mean-speed metrics are comparatively lower (FAC2 = 0.74 and $q = 0.62$), while the TKE FAC2 decreases more substantially to 0.31, indicating increased difficulty in quantitatively resolving near-ground turbulence under strong wall-shear and separated-flow effects. Taken together, our results support the capability of the solver configuration for this canonical building-resolved benchmark, while identifying near-ground turbulence prediction as a remaining limitation.}

\begin{figure}[t]
    \centering
    \includegraphics[width=0.85\linewidth]{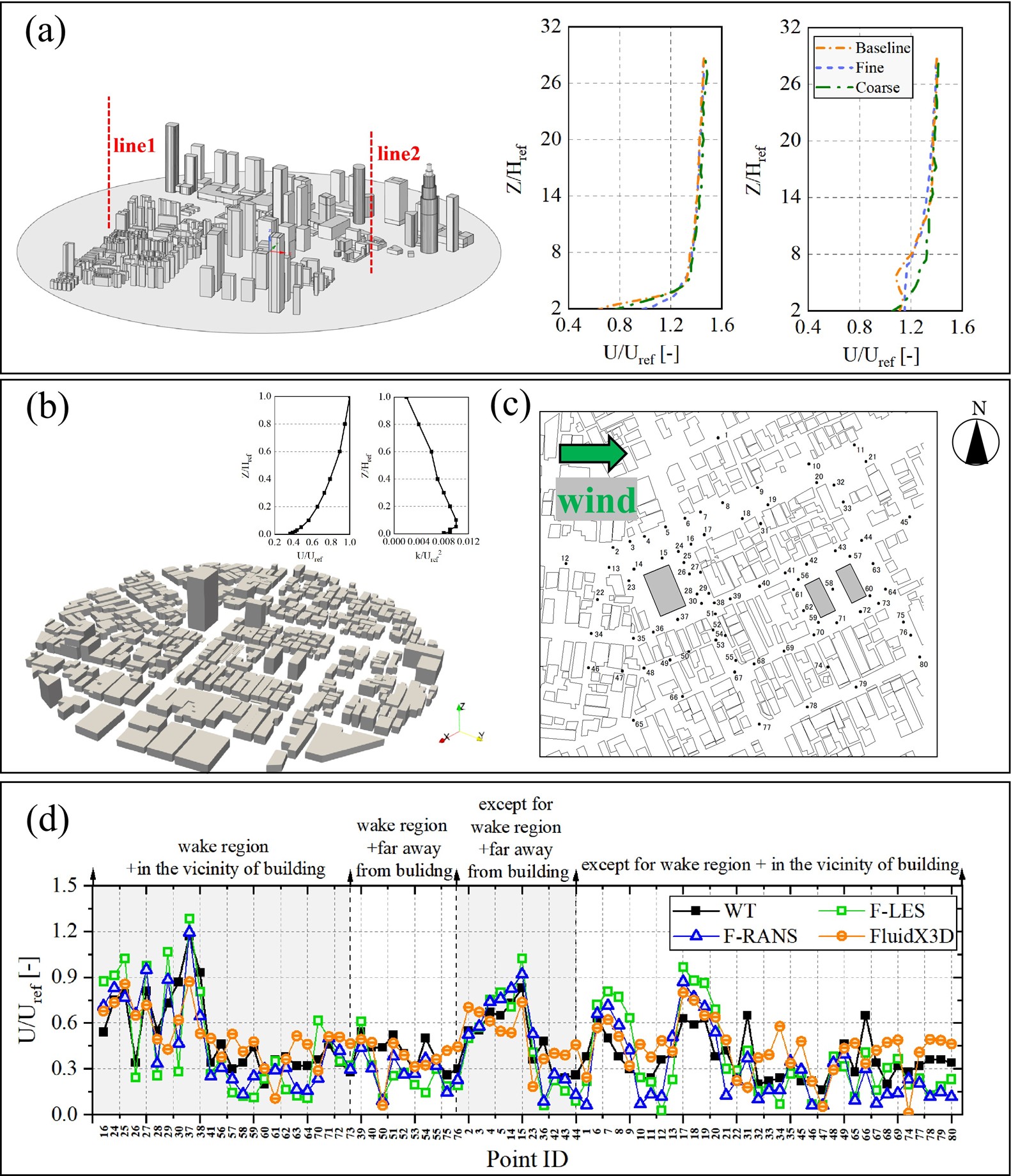}
    \caption{{AIJ Case E validation configuration and results: (a) vertical-profile grid comparison along Lines 1 and 2; (b) prescribed inflow and the geometric model; (c) the 80 wind-tunnel sampling locations; and (d) LatticeUrbanWind, Fluent RANS using the realizable k-}\mathcolorbox{yellow}{\varepsilon}{ model, and Fluent LES against the wind-tunnel measurements.}}
    \label{fig:aij-validation}
\end{figure}

{The evaluation is then extended to AIJ Case E. Figure~\ref{fig:aij-validation} presents the validation configuration and results for our platform.}
{The Case E wind-tunnel experiment represents a typical urban building district in Niigata, Japan }\cite{tominaga2008aijguidelines,tominaga2004casee}{. Wind speeds were measured at a pedestrian height of 2~m for eight wind directions; the west-wind case was selected for the comparison. A uniform 1~m cell size was used in the validation to resolve the measurement height.}

{The 80 sampling locations cover wake and non-wake regions, both near to and far from the buildings. At these locations, the normalized wind speed }$U/U_{\mathrm{ref}}${, where }$U_{\mathrm{ref}}${ is the reference wind speed, is compared among LatticeUrbanWind, Fluent RANS using the realizable }$k$-$\mathcolorbox{yellow}{\varepsilon}${ model, Fluent LES, and the wind-tunnel measurements.}

{The mean absolute error (MAE) of our method is 0.143, close to the Fluent RANS and Fluent LES values of 0.138 and 0.144, respectively, which is defined by}
\begin{equation}
\text{\mathcolorbox{yellow}{\displaystyle \mathrm{MAE}=\frac{\sum_{i=1}^{n}|p_i-o_i|}{N}}}
\end{equation}
\noindent{where }$p_i${ and }$o_i${ are the CFD and wind-tunnel values at sampling point }$i${, and }$N${ is the number of sampling points.}

\subsubsection{\texorpdfstring{{Density-current front validation}}{Density-current front validation}\label{sec:density-current-validation}}

{To assess the representation of buoyancy-driven transient flow, a two-dimensional density-current benchmark was considered }\cite{girfoglio2023validation,marras2015dynamicSGS}{. It starts from an initially quiescent, neutrally stratified atmosphere with a cold elliptical potential-temperature perturbation in a 25.6~km by 6.4~km domain. Negative buoyancy drives the cold perturbation downward and then along the lower boundary, producing an advancing front, flow head, and shear-layer roll-up. The case therefore tests coupled momentum--thermal transport and the evolution of a buoyancy-driven front.}

\begin{figure}[!htbp]
    \centering
    \includegraphics[width=\linewidth]{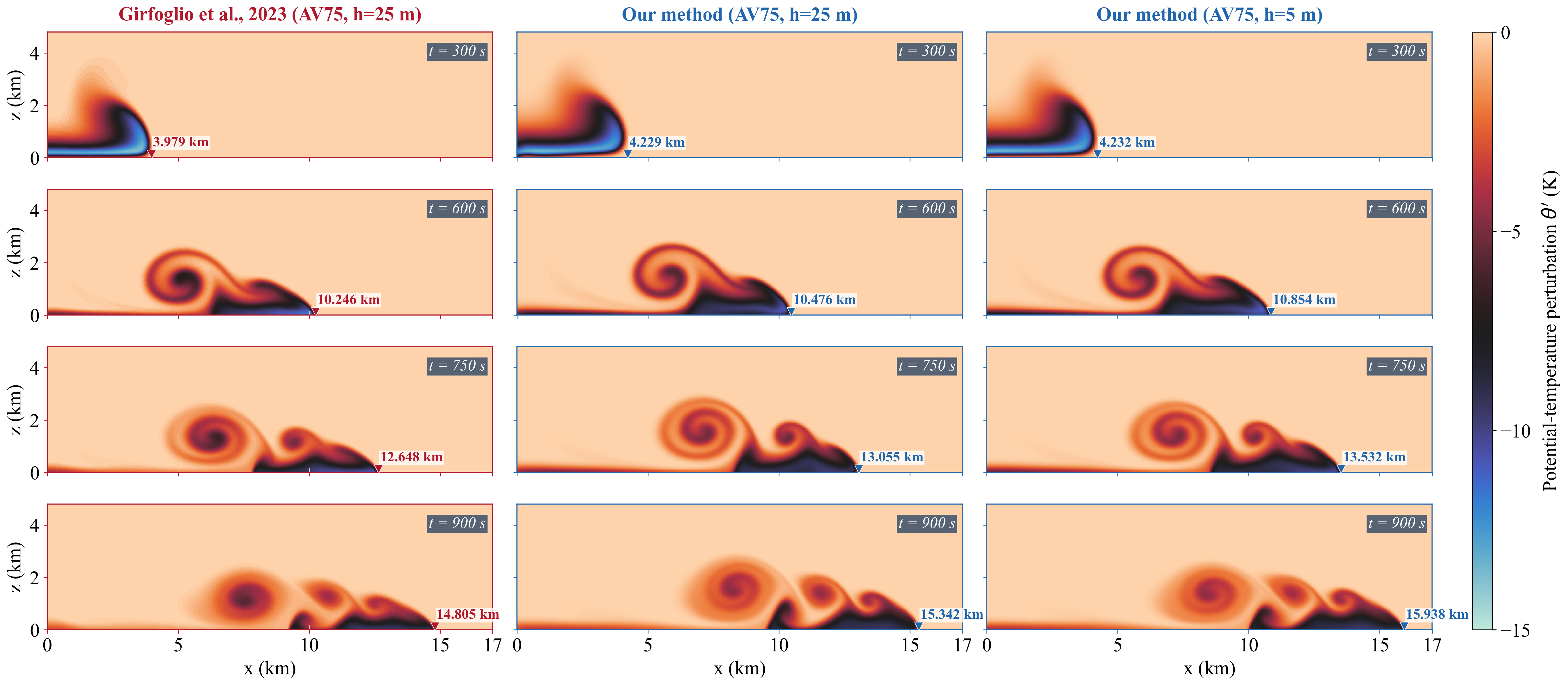}
    \caption{{Potential-temperature perturbation }\mathcolorbox{yellow}{\theta^\prime}{ for the AV75 density-current comparison at 300, 600, 750, and 900~s.}}
    \label{fig:density-current-av75}
\end{figure}

{Figure~\ref{fig:density-current-av75} compares the potential-temperature perturbation }\mathcolorbox{yellow}{\theta^\prime}{ for the published AV75 constant-dynamic-viscosity configuration (75~Pa\,s) and for our platform at 25~m and 5~m grid spacing. The comparison shows the descent of the cold perturbation, its spreading along the lower boundary, the roll-up of the head, and the downstream progression of the cold-pool front. For the 25~m comparison, the mean absolute front-position difference is 3.85\% over 300--900~s.}

{Figure~\ref{fig:density-current-les} provides a matched-effective-resolution front-position comparison with the published dynamic subgrid-scale (SGS) LES at 12.5~m grid spacing. At 600, 750, and 900~s, the reference and our-platform front positions are 10.379 and 10.346~km, 12.882 and 12.798~km, and 15.056 and 14.983~km, respectively. The corresponding absolute relative differences are 0.310\%, 0.659\%, and 0.488\%, giving a mean absolute percentage difference of 0.486\%. The paired fields show the progressive surface spreading, shear-layer roll-up, and downstream advance of the density current.}

\begin{center}
    \begin{minipage}{0.95\linewidth}
        \centering
        \includegraphics[width=0.90\linewidth]{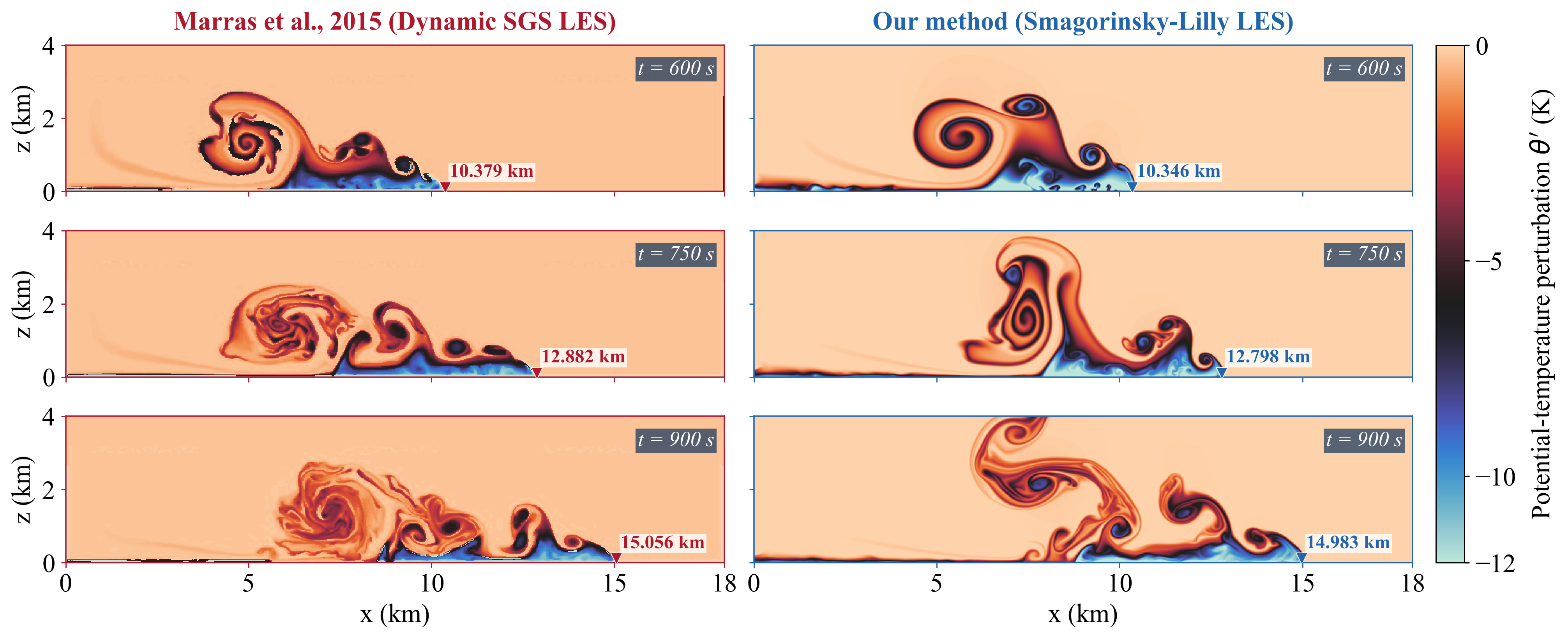}
        \captionof{figure}{{Potential-temperature perturbation }\mathcolorbox{yellow}{\theta^\prime}{ for the density-current benchmark at 600, 750, and 900~s.}}
        \label{fig:density-current-les}
    \end{minipage}
\end{center}

\clearpage
\section{Investigated region and observations\label{sec:3_simulations}}
\suppressfloats[t]

\begin{figure}[t]
        \centering
        \includegraphics[width=\linewidth]{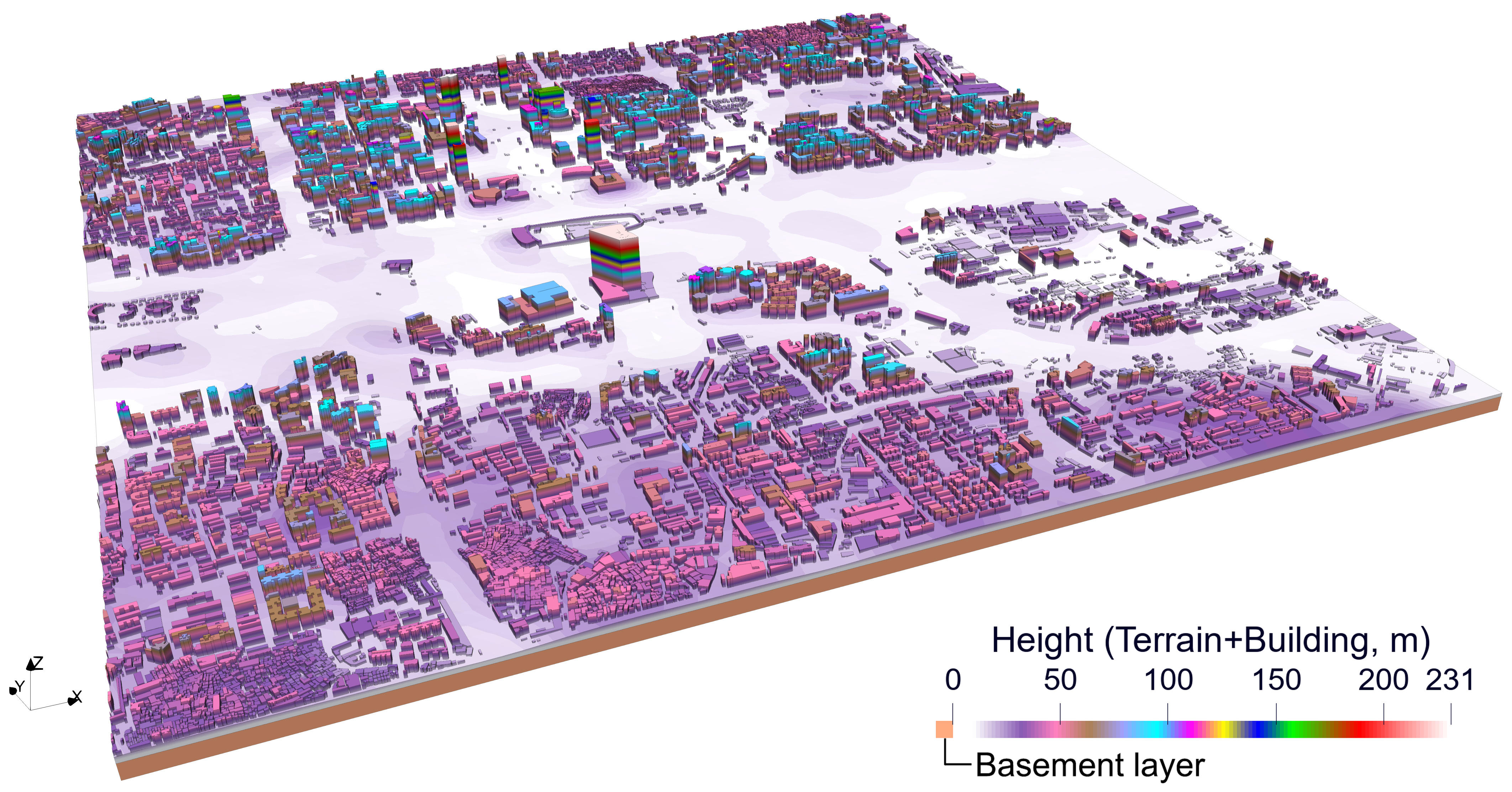}
    \caption{{Buildings in the computational region at Haixinsha Island and surrounding areas in Guangzhou, coloured by height.}}
    \label{fig:1}
\end{figure}

Geographically, the study area is situated between longitudes $113.302^\circ$E and $113.342^\circ$E, and latitudes $23.093^\circ$N and $23.133^\circ$N, encompassing Haixinsha Island and its vicinity in Guangzhou, Guangdong Province of China. {The x-axis and y-axis are first established in the east and north directions, respectively, followed by a correction rotation determined by the calculated meridian convergence angle of $-0.911^\circ$.
The simulation domain extends approximately $4025.73$ m along the x-axis and $4485.82$ m along the y-axis. }The building database used in this case is taken from the 3D-GloBFP dataset \cite{3dglobfp}, and the DEM databases are taken from the NASADEM global digital elevation model \cite{NASADEM_2021}. The buildings in the computational region are shown in Figure~\ref{fig:1}. {The model employs a uniform horizontal and vertical grid resolution of 5 m, which is later shown to be the coarsest tested resolution that still preserves the essential building geometry and near-ground profile structure in the present urban setting, while remaining computationally practical.}

\begin{table}[b]
\centering
\caption{Attributes of Doppler wind lidar stations.\label{tab:1}}
\label{tab:haixinsha_lidar_attributes}
\setlength{\tabcolsep}{7pt}
\renewcommand{\arraystretch}{1.1}
\begin{tabularx}{\columnwidth}{c c c c c c}
\toprule
No. & Station ID & \makecell{Time\\resolution} & \makecell{Vertical\\resolution (m)} & \makecell{Start\\height (m)} & \makecell{Max\\height (m)} \\
\midrule
1 & GAW102 & 1 min  & 13 & 43 & 442  \\
2 & GAW103 & 1 min  & 13 & 52 & 2455 \\
3 & GAW104 & 1 min  & 13 & 52 & 2455 \\
4 & GAW105 & 10 min & 14 & 85 & 2828 \\
5 & GAW110 & 1 min  & 53 & 42 & 1058 \\
6 & GAW111 & 1 min  & 13 & 52 & 2455 \\
\bottomrule
\end{tabularx}
\end{table}

\begin{figure}[t]
        \centering
        \includegraphics[width=\linewidth]{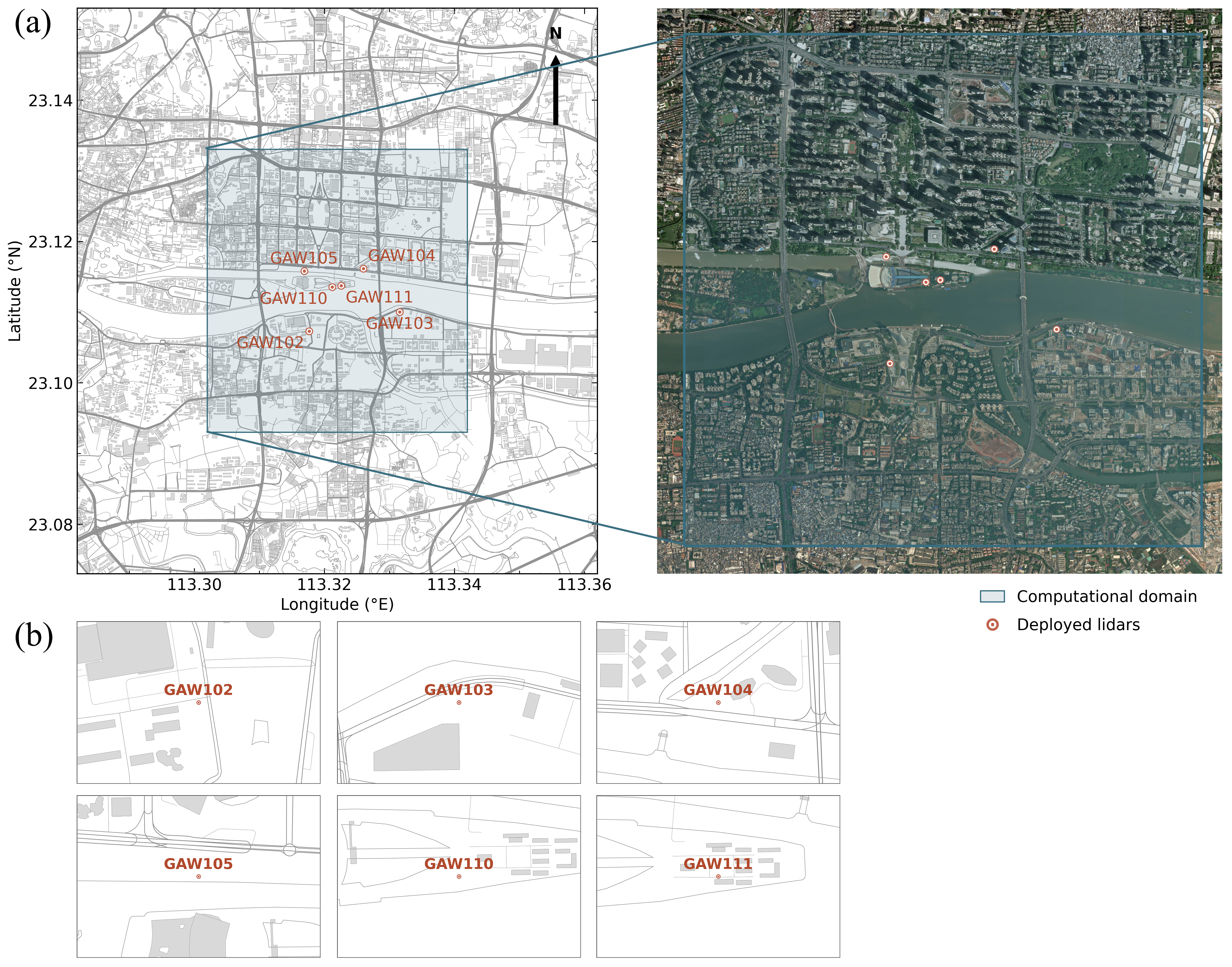}
    \caption{{Locations of deployed lidars: (a) map of the study area with aerial photo~\cite{esriWorldImagery}; and (b) surrounding buildings.}}
    \label{fig:2}
\end{figure}

Six Doppler lidar systems were deployed within this region, with the station IDs listed in Table~\ref{tab:1} and shown in Figure~\ref{fig:2}. {They are placed on open areas like lawns or parking lots to minimize the impact of surrounding buildings as much as possible.}
The equipment and observational data used in this study were provided through the Haixinsha low-altitude economy meteorological service scenario of the Guangzhou Low-Altitude Flight Meteorological Support Volunteer Observation Experiment conducted by the Guangzhou Meteorological Bureau.
These observations are employed both for boundary construction and for validation. However, within any case, they do not simultaneously serve both roles. 
{Figure~\ref{fig:2} shows the computational domain, which is dominated by built-up land, and a permanent water body traverses the computational domain.}
At each analysis time, each lidar site provides quality-controlled horizontal wind components at discrete heights. {In the present study, only these horizontal components are assimilated in the boundary reconstruction. This is primarily a methodological choice constrained by the common operational lidar product available consistently across the six sites, rather than an assumption that vertical motions are dynamically negligible. Once initialized, the LBM--LES solver still advances the full three-dimensional velocity field $(u,v,w)$ inside the CFD domain; therefore, the restriction applies only to the lidar-based boundary forcing and the direct observation-based validation. Vertical velocity can nevertheless be important in urban flows because building-induced shear layers, rooftop updrafts and downdrafts, wake recirculation, and thermally driven canopy exchange all involve non-negligible vertical transport \cite{blocken2014cwe}. In the present operational dataset, however, quality-controlled horizontal winds were the only components available with sufficient cross-site consistency for robust multi-site assimilation. The boundary is constructed using transient observation data (instantaneous wind speeds), supporting a quasi-steady time stepping strategy.} After temporal collocation across sites and filtering by quality-control flags, the irregularly distributed samples of the horizontal wind components are modeled as realizations of a spatially correlated random field and reconstructed onto a regular grid. Specifically, for each target height level, a spherical semivariogram is established in the horizontal plane and Ordinary Kriging weights are solved to obtain the minimum-variance linear estimate under the unbiasedness constraint on a regular longitude–latitude grid. The resulting gridded fields at individual height levels are then stacked to form a three-dimensional wind field at the target time. {The component-wise Kriging variances, }\mathcolorbox{yellow}{\sigma_{K,u}^2}{ and }\mathcolorbox{yellow}{\sigma_{K,v}^2}{, are evaluated as interpolation-uncertainty measures and are reported at the withheld validation locations in Section~4.4.} The use of multiple lidar sites increases spatial sampling and reduces extrapolation near the domain boundaries, thereby improving the robustness and spatial continuity of the reconstructed wind field. See Supplementary Material~\ref{supp:Kriging} for algorithm, parameters and other implement details for reproducibility.
The reconstructed wind field is then sampled on the domain boundaries to drive the LBM solver. Unless otherwise specified, we reconstruct low-altitude three-dimensional winds within the urban canopy at a spatial resolution of 5~m.

{At each analysis time, the temporally collocated mean boundary field is held fixed throughout the LBM--LES integration. On-the-fly synthetic turbulent fluctuations are superposed on this mean field; thus, the boundary does not advance a meteorological time series, but acts as a steady equilibrium-reconstruction (Dirichlet-like) driver for the unsteady flow solution, supporting a quasi-steady time-stepping strategy. The turbulence-generation formulation is detailed in Supplementary Material~\ref{supp:STG}.}

The experimental design employs a strategy of progressive data reduction by systematically decreasing the number of input lidars to validate the accuracy of low-altitude three-dimensional wind fields simulated by the LBM. Consequently, this study assesses the feasibility of reconstructing comprehensive wind fields without observational blind zones using limited and isolated data. To evaluate the simulation capabilities of the LBM model within complex urban environments, four sets of comparative experiments were established. Case 1 utilizes observations from five lidars as boundary condition inputs and reserves one for validation. The subsequent groups systematically reduce the input sources, where Cases 2, 3, and 4 employ four, three, and two lidars for input while reserving two, three, and four stations for validation, respectively. The detailed station IDs used for input and evaluation purposes are listed in Table~\ref{tab:2}. {Candidate periods were screened according to four practical criteria: absence of precipitation, high cross-site data integrity and temporal continuity across the deployed lidar network, moderate-to-strong background winds that provide a stringent verification condition for the CFD model, and relatively stable large-scale forcing without abrupt wind-direction transitions. Under these criteria, the two-hour window from 17:00 to 19:00 (UTC+08:00) on 10 September 2025 provided the best combination of data quality and suitability for independent holdout validation. After quality assurance, the specific validation instants retained for profile comparison, namely 17:10, 18:10, and 18:30, reached maximum valid observation heights of 800~m, 400~m, and 400~m, respectively. The selected case lies within the moderate-to-high wind range of the observed distribution rather than at either the calm or extreme tail, indicating that it is representative of operationally relevant windy conditions while remaining suitable for robust model verification. In addition, smooth cascading energy in wavenumber presented by Supplementary Material~\ref{supp:wn} further shows that no strong cyclonic flow patterns or large-scale extreme weather events were observed during the monitoring period.} It is noteworthy that while the raw lidar observations offer high temporal resolution ranging from minute-level to second-level frequencies, a sampling interval of 10 minutes was adopted for this simulation. This resolution was chosen to achieve an optimal balance between computational efficiency and the temporal representativeness of the data. {The 10-min interval is the cadence at which a new analysis-time boundary field is constructed; it is not a time-varying forcing applied during any individual LBM--LES integration.}

\begin{table}[t]
\centering
\caption{Experimental cases with model input stations and evaluation stations.}
\label{tab:2}
\setlength{\tabcolsep}{4pt}
\renewcommand{\arraystretch}{1.15}

\newcolumntype{C}{>{\raggedright\arraybackslash}p{0.17\columnwidth}}
\newcolumntype{L}{>{\raggedright\arraybackslash}X}

\begin{tabularx}{\columnwidth}{C L L}
\toprule
\makecell[l]{Testing case} & \makecell[l]{Model input stations} & \makecell[l]{Evaluation stations} \\
\midrule
\makecell[l]{Case 1} &
\makecell[l]{GAW102, GAW103,\\GAW104, GAW105,\\GAW110} &
\makecell[l]{GAW111} \\
\midrule
\makecell[l]{Case 2} &
\makecell[l]{GAW102, GAW103,\\GAW104, GAW105} &
\makecell[l]{GAW110, GAW111} \\
\midrule
\makecell[l]{Case 3} &
\makecell[l]{GAW102, GAW103,\\GAW104} &
\makecell[l]{GAW105, GAW110,\\GAW111} \\
\midrule
\makecell[l]{Case 4} &
\makecell[l]{GAW102, GAW103} &
\makecell[l]{GAW104, GAW105,\\GAW110, GAW111} \\
\bottomrule
\end{tabularx}
\end{table}

\begin{figure}[t]
        \centering
        \includegraphics[width=1\linewidth]{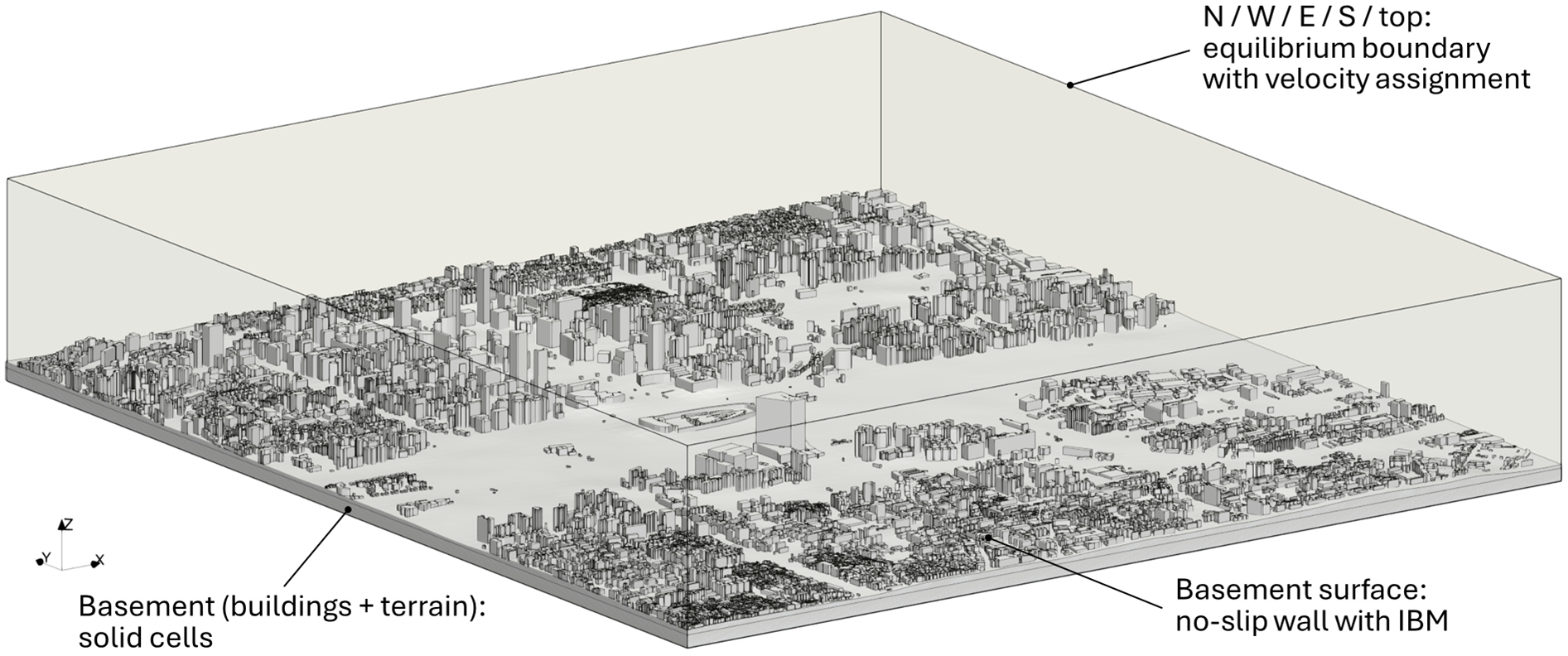}
    \caption{{LBM computational domain and boundary conditions.}}
    \label{fig:compdomain}
\end{figure}

The computational domain and boundary conditions are presented in Figure~\ref{fig:compdomain}. {The boundary conditions are configured as follows: The four lateral faces and the planar top face are prescribed as equilibrium-based open velocity boundaries, with the macroscopic boundary states obtained by face-wise mapping of the preprocessed mesoscale wind field onto the computational boundary. To reduce spurious reflection of outgoing disturbances and to improve numerical stability near the outer domain surfaces, the open-boundary treatment is further supplemented by relaxation-based damping regions adjacent to the upper and lateral boundaries, including a top sponge layer and boundary-adjacent buffer nudging toward the target boundary state; in addition, a global flux-balancing correction is applied over the full set of open boundaries. In contrast, {stationary no-slip conditions on the lower boundary, terrain, and building surfaces are imposed by the immersed-boundary method.} This configuration imposes mesoscale-informed velocity conditions on the outer domain surfaces while representing the ground and urban obstacles as impermeable no-slip walls. In total, 40000 timesteps are calculated in each case, which corresponds to approximately 85 building-scale turnover time to ensure the convergence of low-order quantities including wind speed and turbulent kinetic energy.} {The present building-scale wind-field calculations adopt an isothermal configuration: temperature is held at the reference value, so no temperature gradient or temperature-induced buoyancy force is introduced.} {Details of the immersed-boundary method are provided in Supplementary Material~\ref{supp:surface-momentum-ibm}.}

The computations were carried out on a GPU node equipped with an NVIDIA A800 80G, using a single GPU, with the complete computational workflow requiring approximately 5 minutes (domain: 4 km, resolution: 5 m, height: 800 m). In addition, for testing purposes, simulations were also performed on a cost-effective desktop GPU setup consisting of customized NVIDIA RTX 2080 Ti 22G GPU $\times$ 2, resulting in a total runtime of approximately 8 minutes. {These wall-clock times correspond specifically to the present baseline configuration. For the explicit LBM solver, the total computational cost scales primarily with the total number of lattice updates, i.e. approximately with $N_{\mathrm{cell}}\times N_{\mathrm{step}}$, where $N_{\mathrm{cell}}$ is determined by the grid-cell count for the chosen domain and $N_{\mathrm{step}}=T_{\mathrm{phys}}/\Delta t$ depends on the simulated physical duration $T_{\mathrm{phys}}$ and the adopted time-step size $\Delta t$. In the baseline case reported here, $N_{\mathrm{cell}}=1.16\times10^8$, $\Delta t_\mathrm{SI}=0.01$ s, and $N_{\mathrm{step}}=40000$, corresponding to 400 s of simulated physical time. Within this specific configuration, the reported runtimes demonstrate the computational efficiency of the proposed numerical methodology and platform.}

For the baseline case which uses 5~m grid size, key numerical parameters during the computation are listed as follows: The reference velocity in SI unit $\|\textbf{U}\|_{\text{ref,SI}}$ is 5.3205 m/s, according to an LBM mach number $Ma_{\text{LBM}}=0.10$. The time step $\Delta t_\text{SI}$ is 0.01 s. The geometry nondimensionalization factor $D^*_\text{LBM}/D_\text{SI}$ is 0.20. The lattice count is 116 million. LBM nondimensionalized volume flux before and after correction is $-2.497\times10^2$ and $-3.087\times10^{-5}$, respectively, with a mean velocity variation on the surface mesh $\overline{\Delta u^*}$ of 1.974 (positive values indicate an outflow). The maximum LBM Reynolds number is $4.05\times10^9$, and the LBM kinematic viscosity is $6.0\times 10^{-8}$.
The terrestrial convergence angle at the center of the computational region is $0.911^\circ$ based on Universal Transverse Mercator (UTM) Zone 49N.

\section{Results and discussion\label{sec:4_results}}

\subsection{{Convergence of mesh and re-voxelization}}

Mesh strategy has a significant impact on the numerical results \cite{EDT-LBMMesh,UC-Mesh}.
We re-voxelize the triangle mesh by casting axis-aligned rays through grid columns and computing ray–triangle intersections, using an odd–even parity rule along each column. To improve robustness near degeneracies, we corroborate independent passes along different ray directions for solid labeling.

Due to the adoption of a uniform, isotropic Cartesian grid, the mesh does not conform to the body geometry, which inevitably gives rise to a stair-step approximation effect. Consequently, the resulting solid grid is geometrically inconsistent with the actual building model at the boundaries. Employing a finer spatial discretization enables a more faithful reconstruction of the geometric details of the building, thereby leading to more accurate numerical predictions.
Moreover, after orthogonal projection onto the principal axes, buildings whose shorter lateral dimension is smaller than the grid resolution are neglected. As a result, the ground friction effect is underestimated \cite{nonbodyfit}. Although this deficiency can be partially compensated through boundary assimilation effect, particularly from the top boundary, the grid resolution remains the most dominant factor governing the global accuracy.
The target grid resolution required in practice is governed by the local distribution of building dimensions as well as the relative weighting between buildings and terrain. Guangzhou is characterized by predominantly flat terrain with minimal elevation variations. Therefore, we primarily focus on the stair-step approximation effects induced by buildings.

\begin{figure}[t]
    \centering

    \begin{minipage}[b]{\linewidth}
        \centering
        \begin{tikzpicture}
            \node[anchor=south west, inner sep=0] (img) at (0,0)
                {\includegraphics[width=\linewidth]{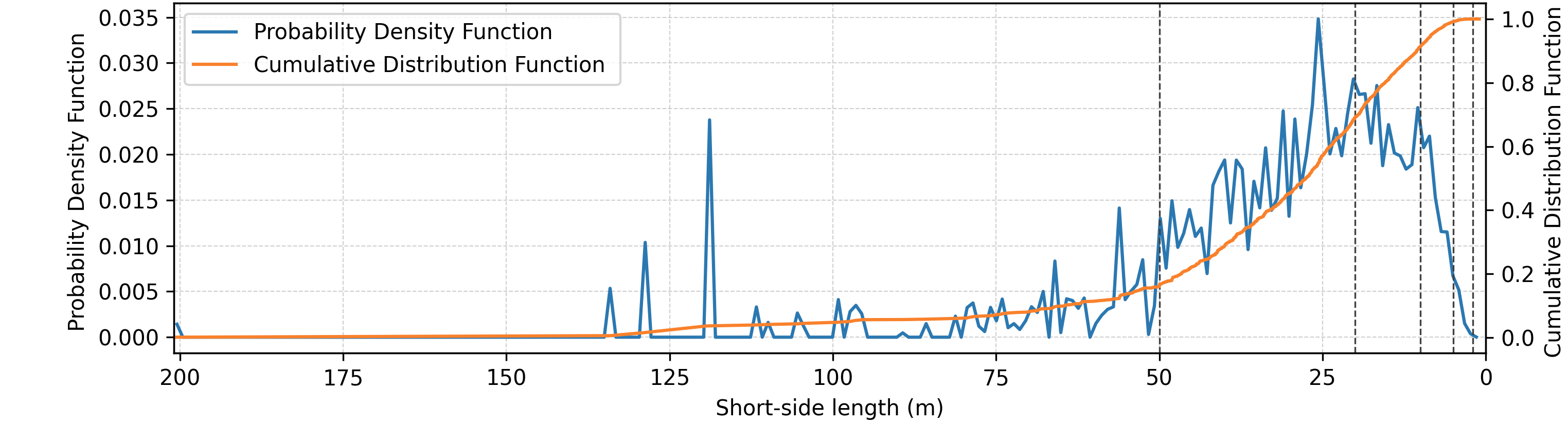}};
            \begin{scope}[x={(img.south east)}, y={(img.north west)}]
                \node[anchor=north west, inner sep=1pt, fill=white, fill opacity=0.2, text opacity=1,font=\small]
                    at (0.0,0.98) {(a)};
            \end{scope}
        \end{tikzpicture}
    \end{minipage}
    
    \begin{minipage}[b]{0.495\linewidth}
        \centering
        \begin{tikzpicture}
            \node[anchor=south west, inner sep=0] (img) at (0,0)
                {\includegraphics[width=\linewidth,height=0.5\linewidth]{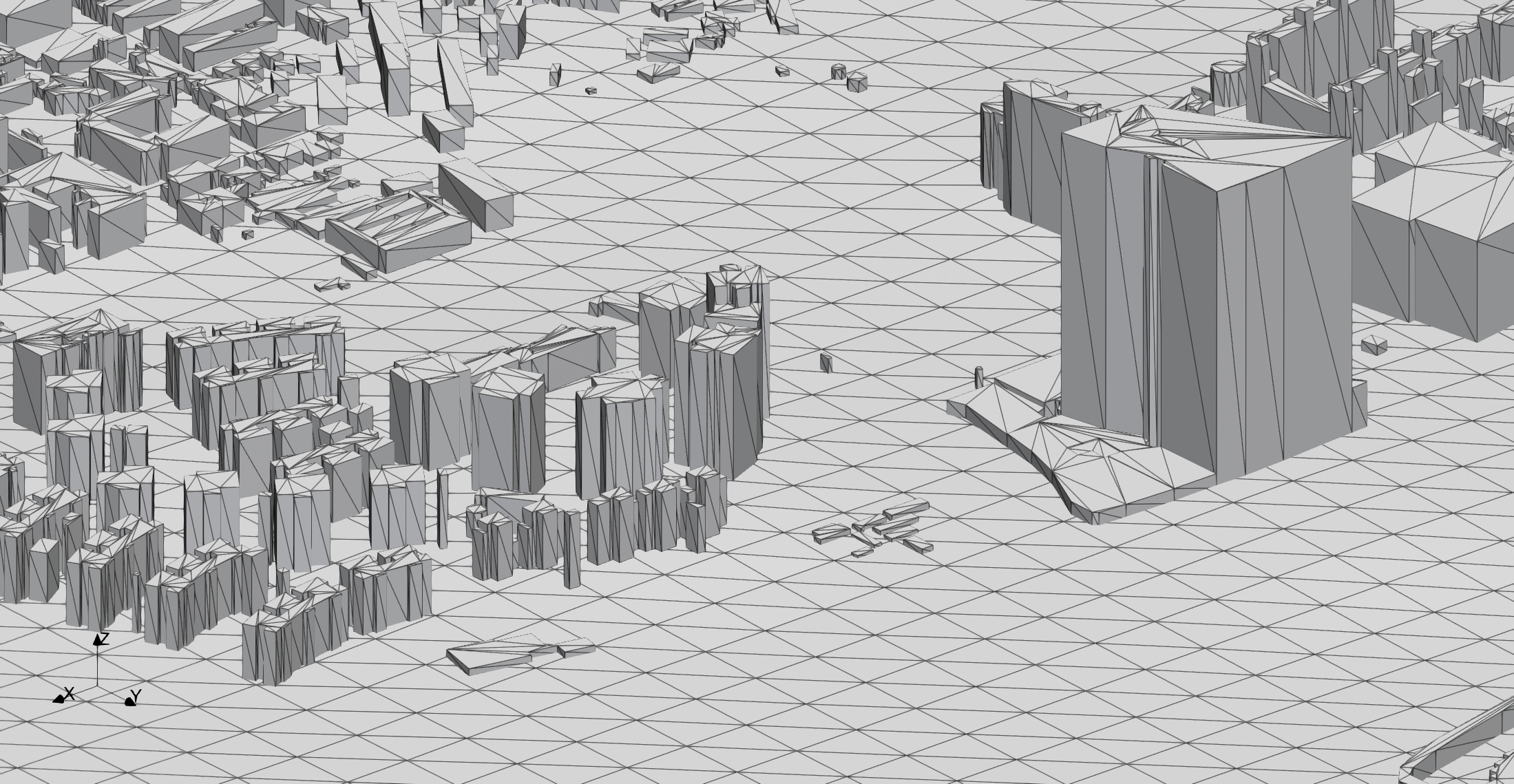}};
            \begin{scope}[x={(img.south east)}, y={(img.north west)}]
                \node[anchor=north west, inner sep=1pt, fill=white, fill opacity=0.2, text opacity=1,font=\small]
                    at (0.01,0.98) {(b)};
            \end{scope}
        \end{tikzpicture}
    \end{minipage}
    \hfill
    \begin{minipage}[b]{0.495\linewidth}
        \centering
        \begin{tikzpicture}
            \node[anchor=south west, inner sep=0] (img) at (0,0)
                {\includegraphics[width=\linewidth,height=0.5\linewidth]{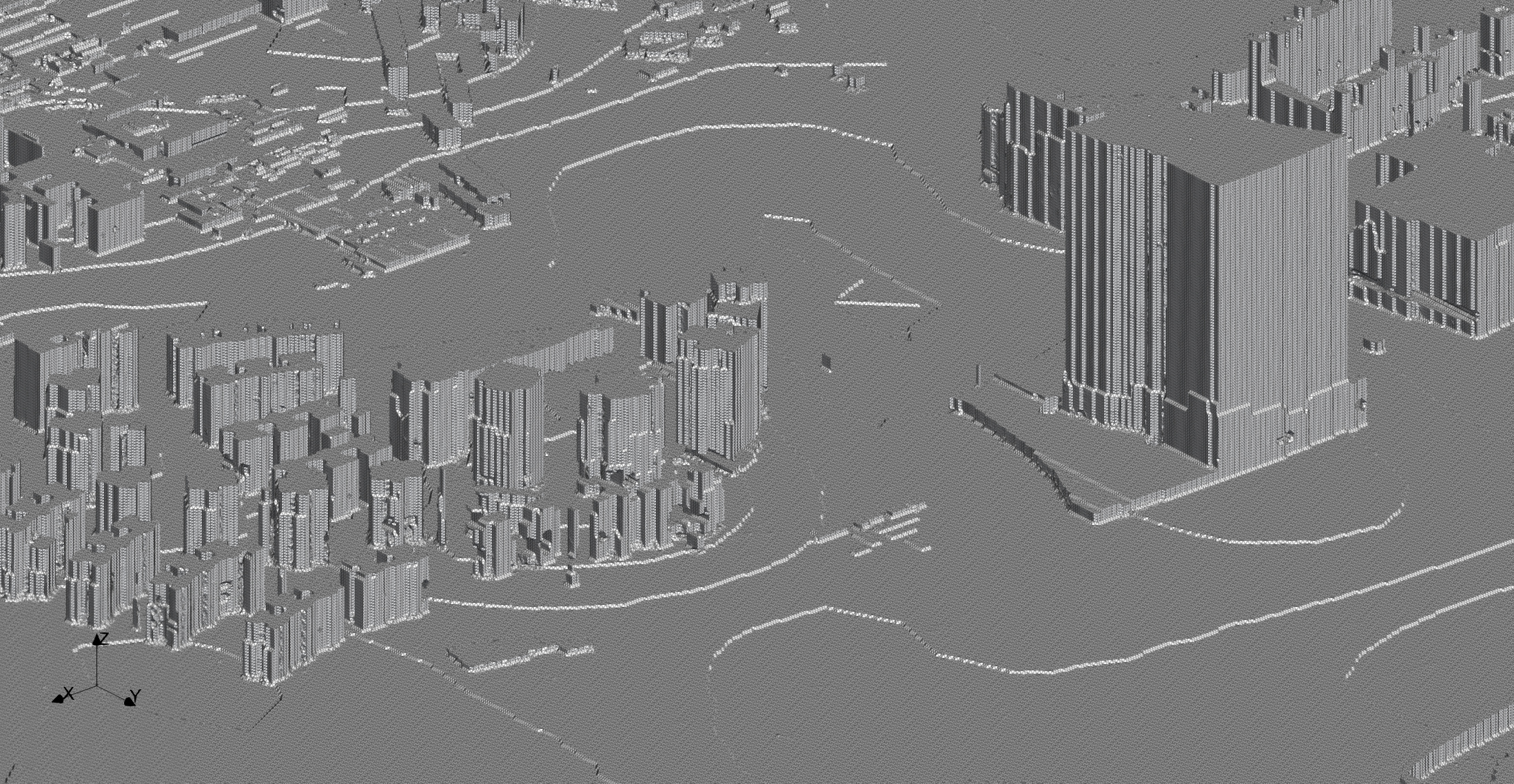}};
            \begin{scope}[x={(img.south east)}, y={(img.north west)}]
                \node[anchor=north west, inner sep=1pt, fill=white, fill opacity=0.2, text opacity=1,font=\small]
                    at (0.01,0.98) {(c)};
            \end{scope}
        \end{tikzpicture}
    \end{minipage}

    \begin{minipage}[b]{0.495\linewidth}
        \centering
        \begin{tikzpicture}
            \node[anchor=south west, inner sep=0] (img) at (0,0)
                {\includegraphics[width=\linewidth,height=0.5\linewidth]{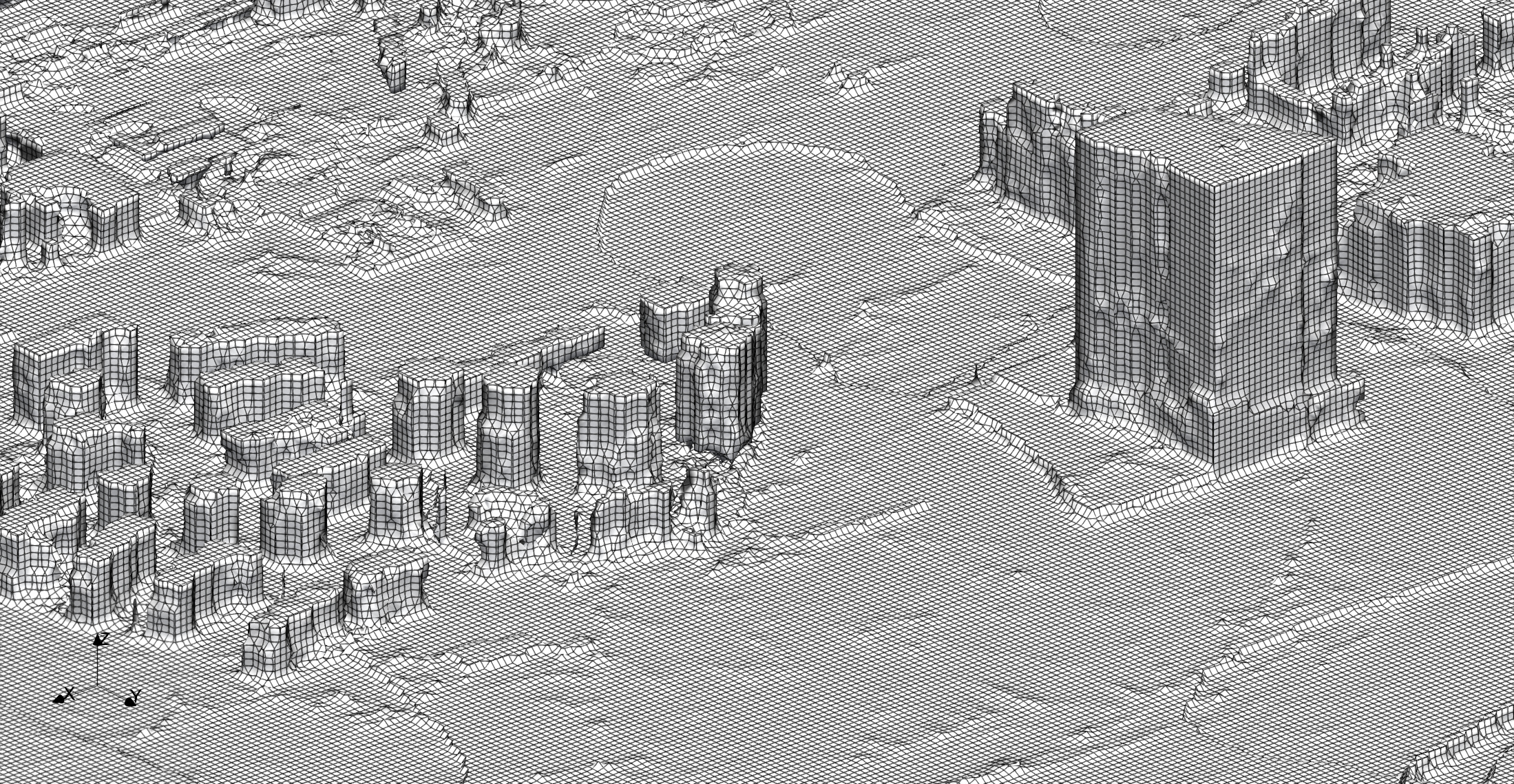}};
            \begin{scope}[x={(img.south east)}, y={(img.north west)}]
                \node[anchor=north west, inner sep=1pt, fill=white, fill opacity=0.2, text opacity=1,font=\small]
                    at (0.01,0.98) {(d)};
            \end{scope}
        \end{tikzpicture}
    \end{minipage}
    \hfill
    \begin{minipage}[b]{0.495\linewidth}
        \centering
        \begin{tikzpicture}
            \node[anchor=south west, inner sep=0] (img) at (0,0)
                {\includegraphics[width=\linewidth,height=0.5\linewidth]{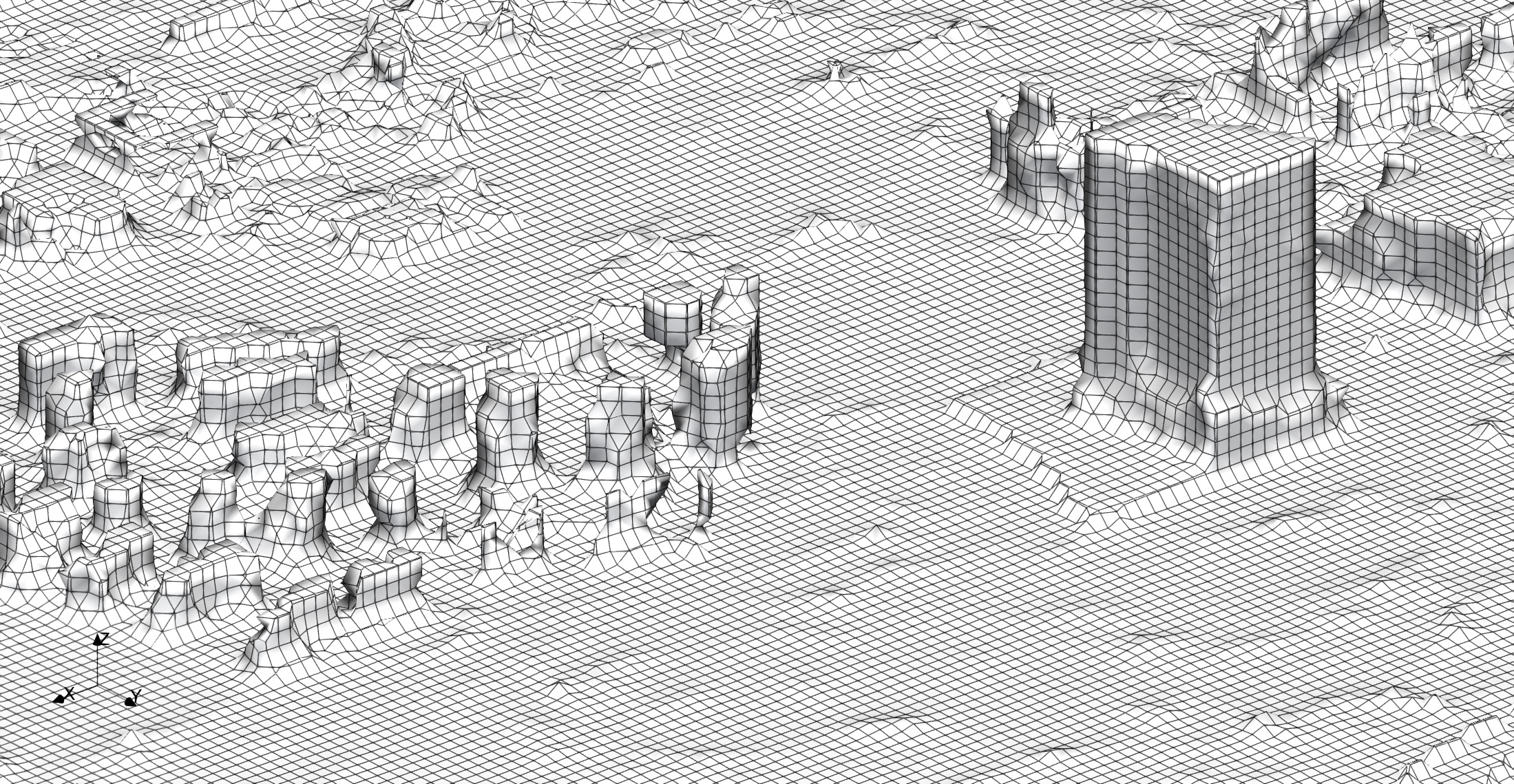}};
            \begin{scope}[x={(img.south east)}, y={(img.north west)}]
                \node[anchor=north west, inner sep=1pt, fill=white, fill opacity=0.2, text opacity=1,font=\small]
                    at (0.01,0.98) {(e)};
            \end{scope}
        \end{tikzpicture}
    \end{minipage}

    \begin{minipage}[b]{0.495\linewidth}
        \centering
        \begin{tikzpicture}
            \node[anchor=south west, inner sep=0] (img) at (0,0)
                {\includegraphics[width=\linewidth,height=0.5\linewidth]{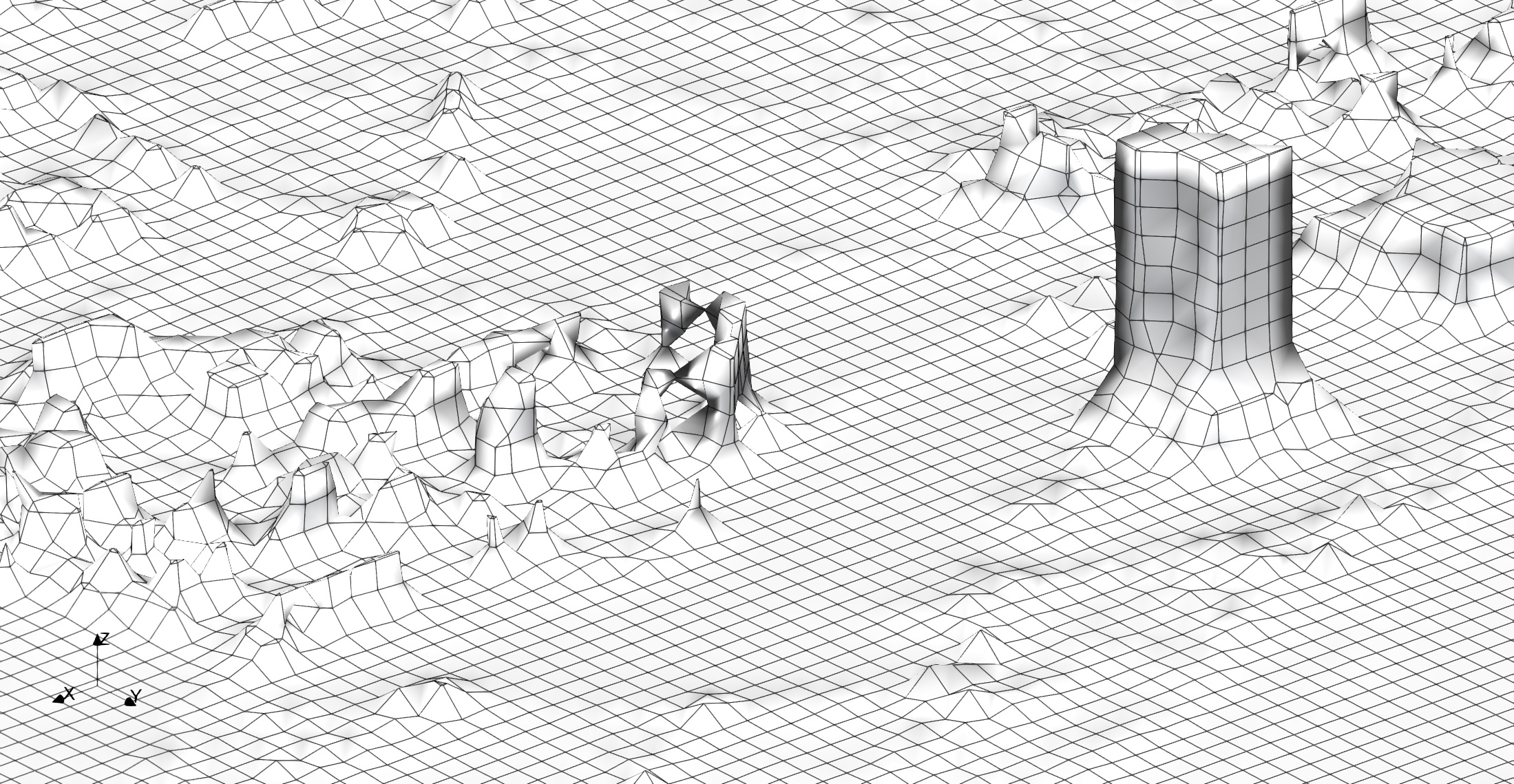}};
            \begin{scope}[x={(img.south east)}, y={(img.north west)}]
                \node[anchor=north west, inner sep=1pt, fill=white, fill opacity=0.2, text opacity=1,font=\small]
                    at (0.01,0.98) {(f)};
            \end{scope}
        \end{tikzpicture}
    \end{minipage}
    \hfill
    \begin{minipage}[b]{0.495\linewidth}
        \centering
        \begin{tikzpicture}
            \node[anchor=south west, inner sep=0] (img) at (0,0)
                {\includegraphics[width=\linewidth,height=0.5\linewidth]{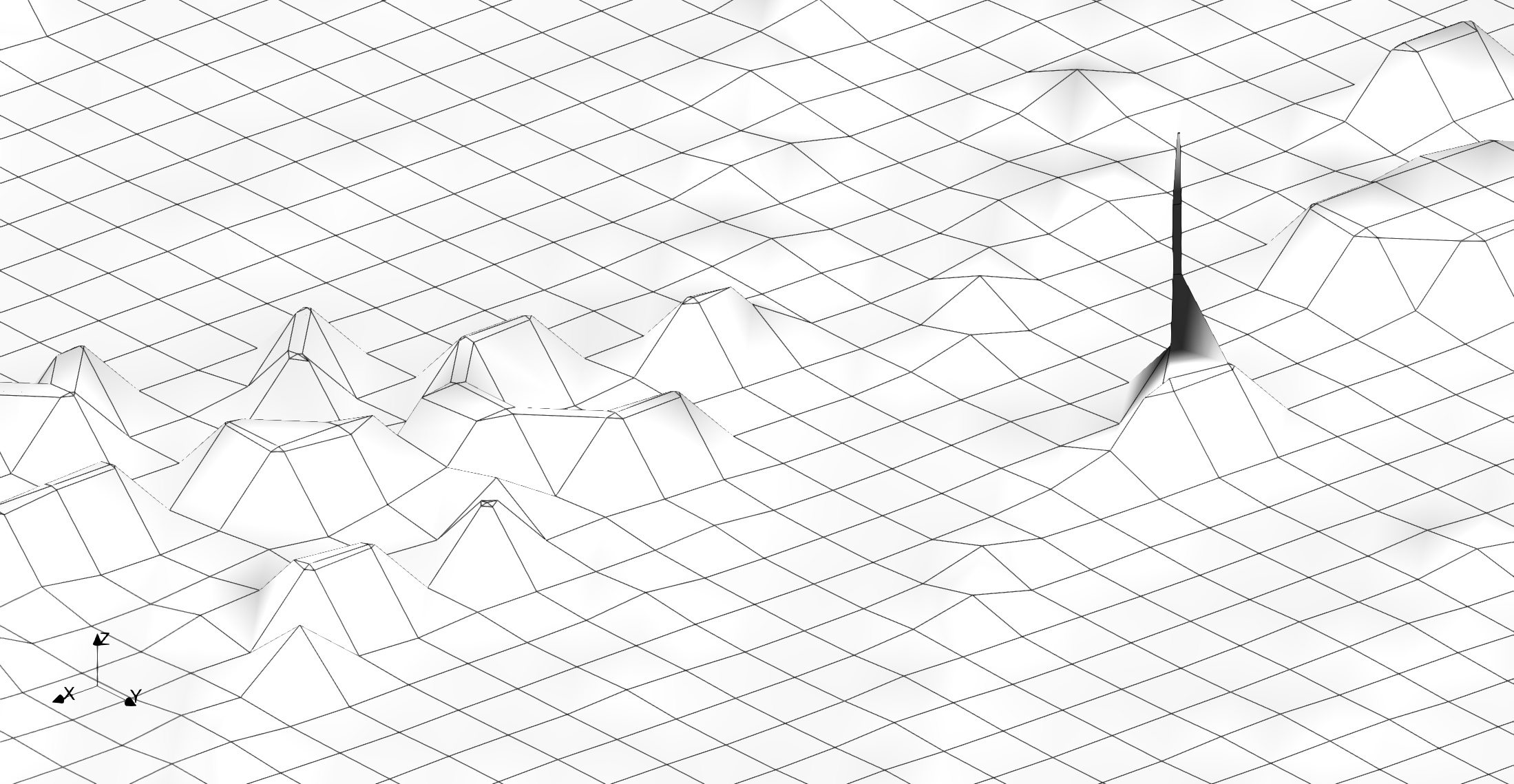}};
            \begin{scope}[x={(img.south east)}, y={(img.north west)}]
                \node[anchor=north west, inner sep=1pt, fill=white, fill opacity=0.2, text opacity=1,font=\small]
                    at (0.01,0.98) {(g)};
            \end{scope}
        \end{tikzpicture}
    \end{minipage}

\caption{Voxelized building boundaries under varying spatial resolutions at a zoomed view, comparing with original buildings. (a) Probability density function and cumulative distribution function of the short side dimensions of buildings in the computational domain (volume-weighted), with grid sizes overlapped, (b) original exact geometry of buildings, and (c-g) re-voxelized buildings characterize by $||\mathbf{U}||=10^{-7}$ m/s contour under mesh base size of (c) 2 m, (d) 5 m, (e) 10 m, (f) 20 m, and (g) 50 m.}
    \label{fig:resolution}
\end{figure}
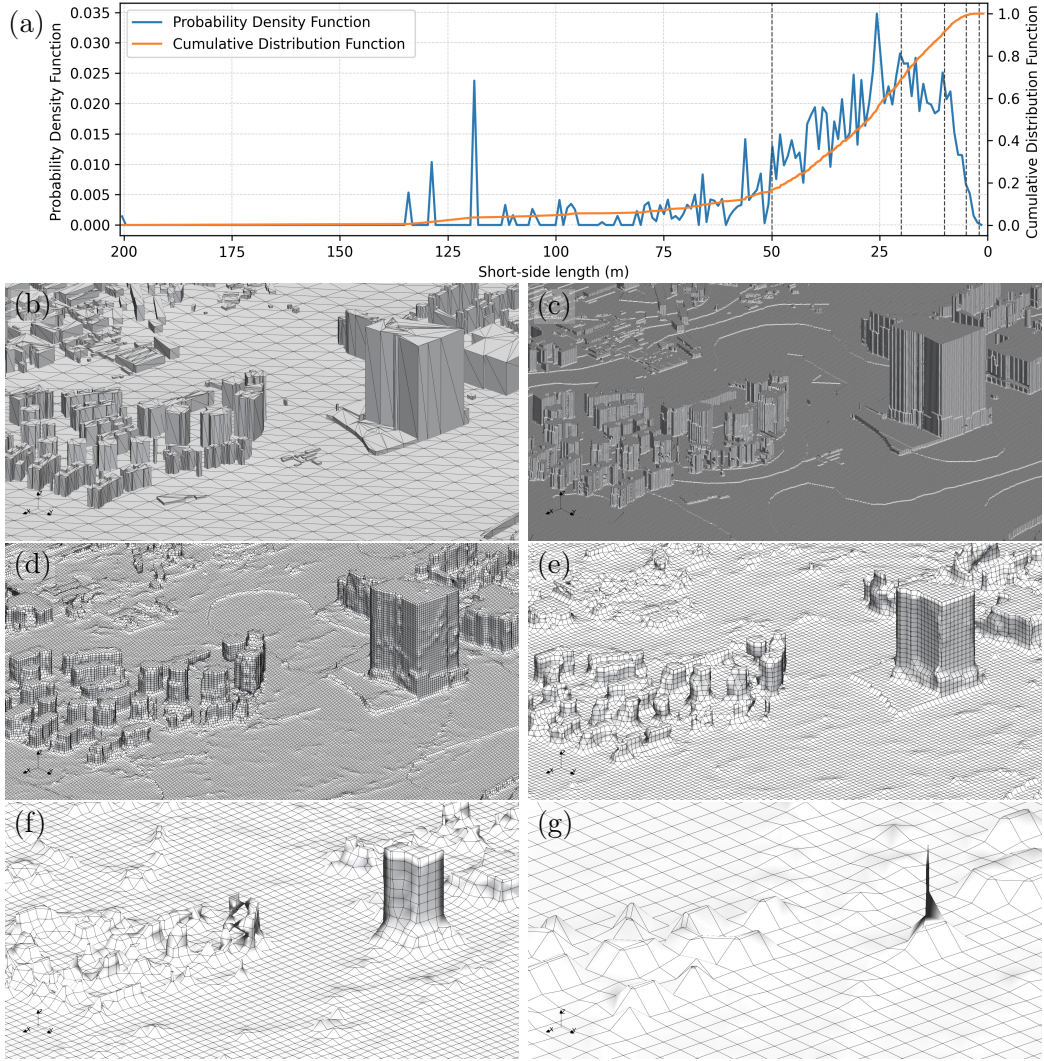

The distribution of building sizes within the computational domain is characterized using the probability density function (PDF) and the cumulative distribution function (CDF), obtained by integrating over building volume (Figure \ref{fig:resolution}(a)). It can be observed that the building population in the domain is dominated by small-scale structures: approximately 80\% (by volume) of the buildings have a shorter lateral dimension smaller than 50 m, which poses a significant challenge for accurate numerical representation. Five grid resolutions are employed, namely 2 m, 5 m, 10 m, 20 m, and 50 m, as marked in Figure~\ref{fig:resolution}(a). The original triangulated representation of the buildings is shown in Figure \ref{fig:resolution}(b). As the velocity inside solid regions is zero, isosurfaces defined by $||\mathbf{U}||=10^{-7}$ m/s is used to approximate the effective solid-envelope surface, as illustrated in Figure \ref{fig:resolution}(c)-(g).

First, when the base size is set to 2 m and 5 m, the stair-step representation exhibits a high degree of consistency with the original voxelized geometry. Both small-scale buildings and fine details of large buildings are reconstructed with good fidelity. {The building CDF indicates that in both cases more than 99\% of the solid volume is retained. However, for a fixed three-dimensional domain, the lattice count scales approximately with $\Delta x^{-3}$; therefore, reducing the grid spacing from 5 m to 2 m would increase the lattice count by approximately $(5/2)^3 \approx 15.6$, with a corresponding first-order increase in memory demand and wall-clock cost.}
When the base size increases to 10 m, noticeable discrepancies begin to emerge. While the primary buildings are still successfully reconstructed, some smaller buildings are omitted. In addition, slight volumetric deviations are observed for large buildings. At this resolution level, approximately 8\% of the target buildings have a shorter lateral dimension smaller than the grid size, leading to a risk of being unextrudable.
With a further increase of the base size to 20 m, a substantial number of small buildings lose their original height characteristics and degenerate into irregularly-shaped frictional elements. More than 30\% of the building volume falls below this resolution threshold. The volumes of major buildings are underestimated, and geometric details such as podium structures are no longer preserved.
Finally, when the base size reaches 50 m, the buildings completely lose their original geometric characteristics.

\begin{figure}[t]
    \centering

    \begin{minipage}[b]{0.49\linewidth}
        \centering
        \begin{minipage}[b]{0.48\linewidth}
            \centering
            \includegraphics[width=\linewidth]{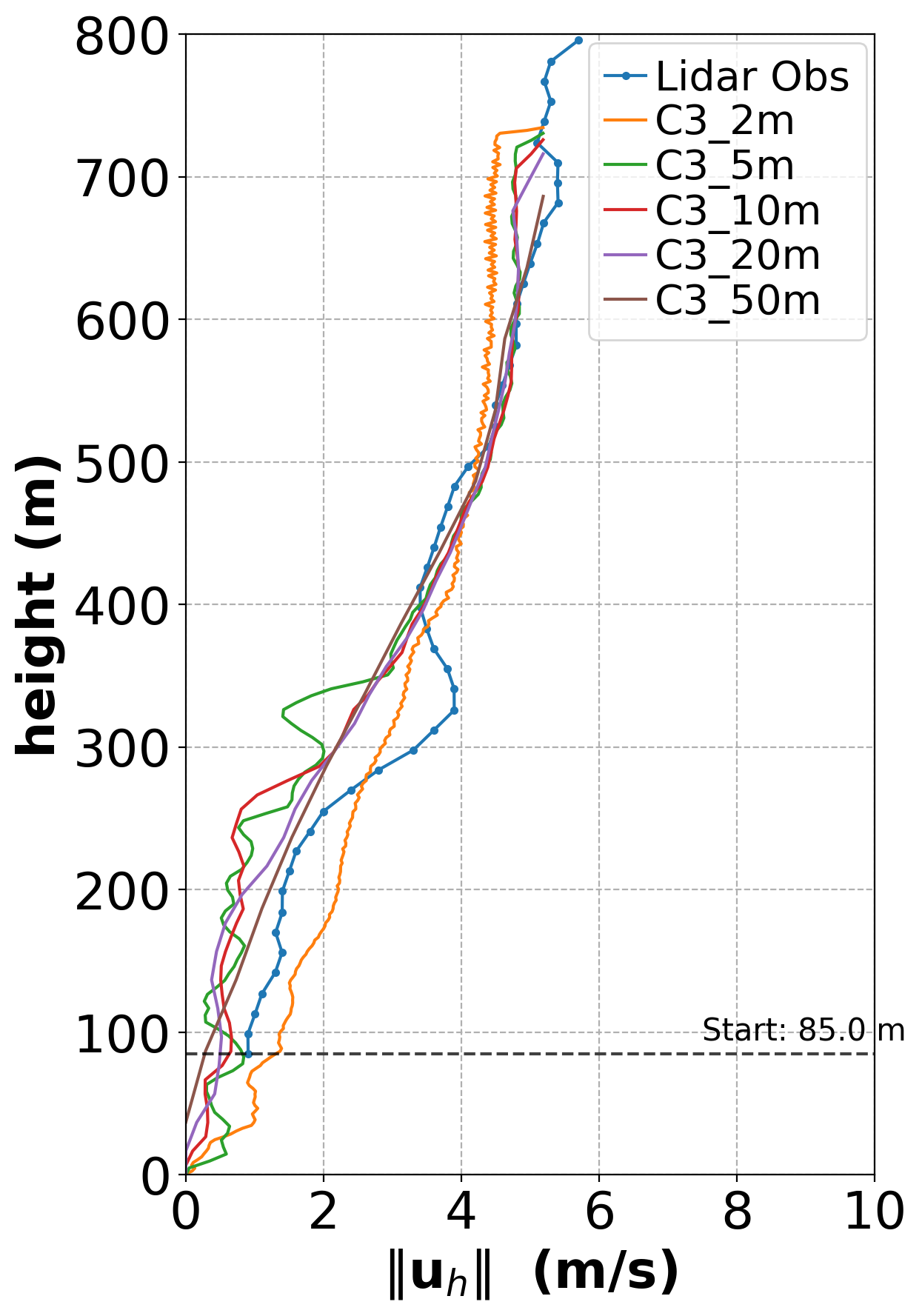}

        \par\smallskip
        \textup{(a)}
        \end{minipage}
        \hfill
        \begin{minipage}[b]{0.48\linewidth}
            \centering
            \includegraphics[width=\linewidth]{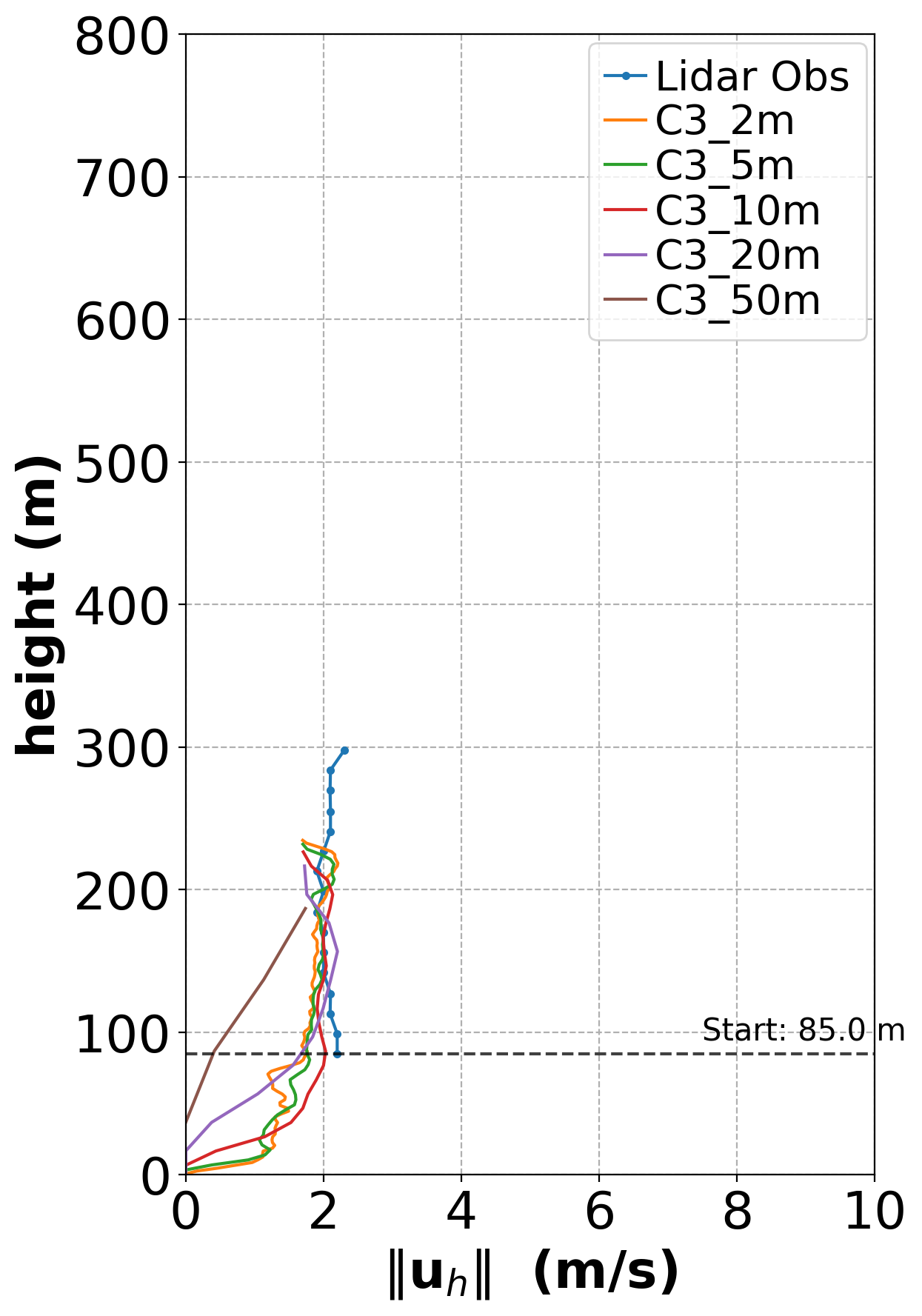}

        \par\smallskip
        \textup{(b)}
        \end{minipage}

    \end{minipage}
    \hfill
    \begin{minipage}[b]{0.49\linewidth}
        \centering
        \begin{minipage}[b]{0.48\linewidth}
            \centering
            \includegraphics[width=\linewidth]{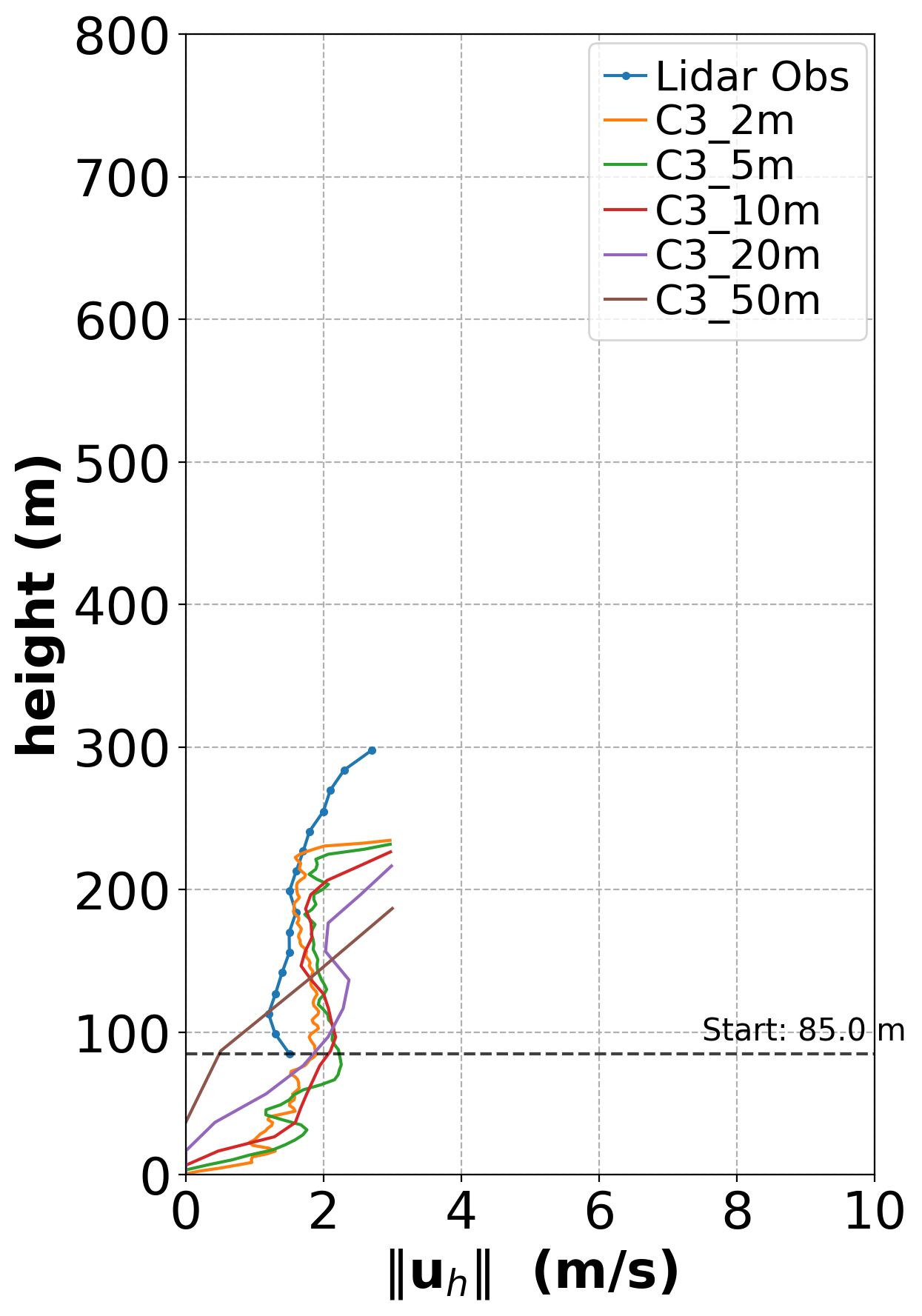}
            
        \par\smallskip
        \textup{(c)}
        \end{minipage}
        \hfill
        \begin{minipage}[b]{0.48\linewidth}
            \centering
            \includegraphics[width=\linewidth]{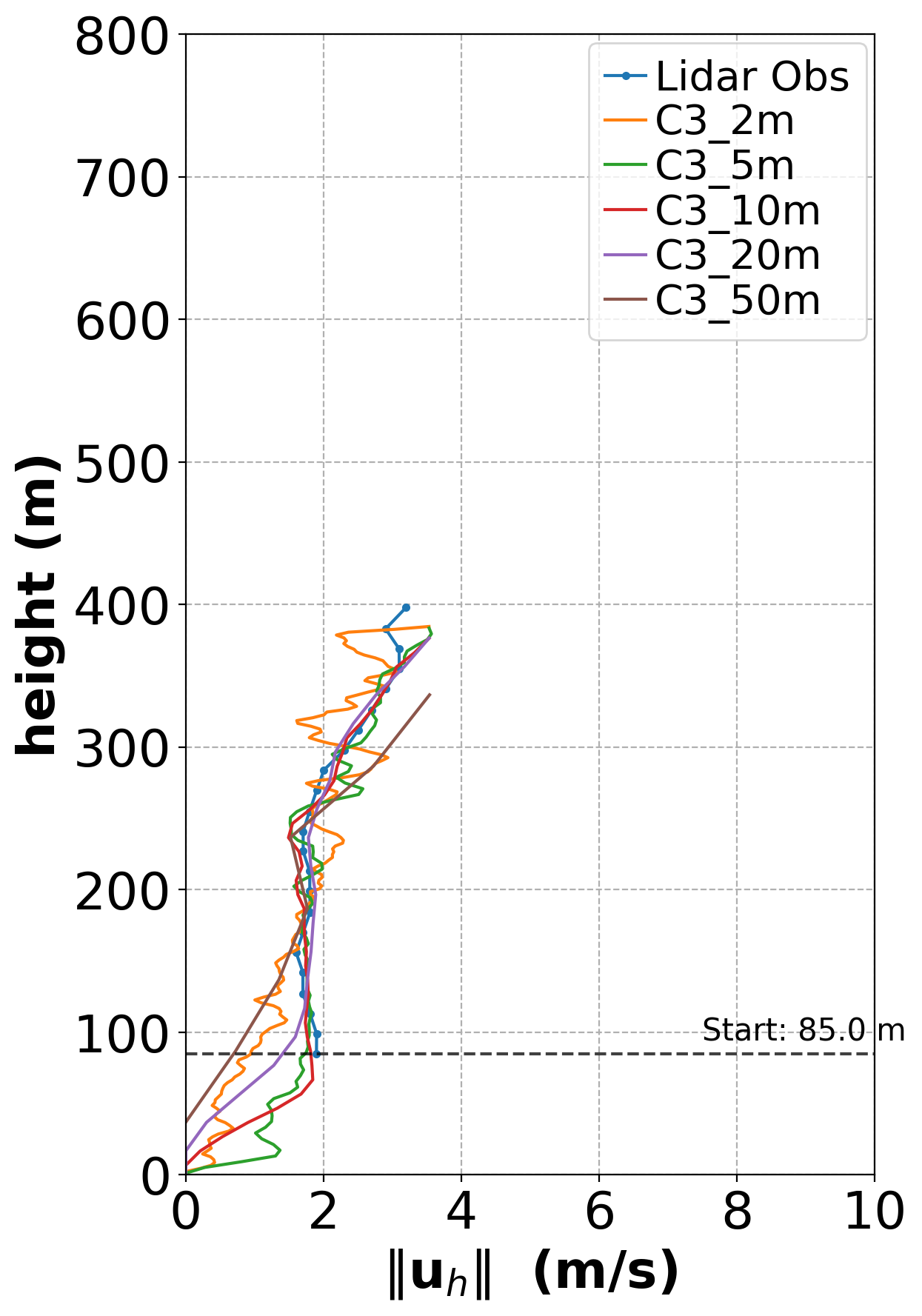}
            
        \par\smallskip
        \textup{(d)}
        \end{minipage}

    \end{minipage}
    \caption{Wind profile at GAW105 position under various mesh resolutions comparing with observations (Case 3). (a) 17:10, (b) 18:10, (c) 18:30, and (d) 18:40.}
    \label{fig:meshprofile}
\end{figure}

{This impact of lattice size is further corroborated by the horizontal wind profiles $\|\mathbf{U}_h\|=\sqrt{u^2+v^2}$ at four different time instants, shown in Figure~\ref{fig:meshprofile}.} Under conditions of strong wind speed and large velocity gradients at 09:10, the results obtained using different grid strategies are relatively similar. This is because potential flow dominates the overall flow field, and the boundary layer exhibits a larger shape factor $H_{12}$, which is negatively correlated with the fullness of the velocity profile. 
In this regime, the boundary assimilation effect overwhelms the intrinsic flow development associated with building-induced wake and separation, thereby attenuating the influence of building stair-step representation and wall shear stress errors. {By contrast, under lower-wind conditions, the differences become much more pronounced. As fewer buildings are retained in the voxelized geometry, ground friction is progressively underestimated in the 10 m, 20 m, and 50 m simulations, causing the near-ground velocity profile to become excessively underdeveloped. In comparison, the 2 m and 5 m meshes both exhibit pronounced shear effects in the ultra-low region and remain much closer to each other and to the observations. Therefore, it is shown that 5 m is the coarsest tested resolution that still preserves the essential building geometry and near-ground profile structure in this urban setting, while remaining substantially more computationally affordable than 2 m. As a result, subsequent simulations are performed using a 5 m lattice resolution.}

{To further validate the mesh strategy towards a satisfactory LES configuration, two approaches are applied. First, the mesh resolution is also evaluated
using two-point correlation in a series of continuously-distributed lattice probes, referring to Davidson \cite{Davin_validation}. The normalized two-point correlation, using the root-mean-square (RMS) velocity fluctuations for normalization, is given by}
\begin{equation}{
C_{U_i}^{\mathrm{norm}}\left(\vec{x}_0,\vec{x}\right)
=
\frac{\overline{{U_i}'\left(\vec{x}_0\right)\,{U_i}'\left(\vec{x}\right)}}
{{U_i}'_{\mathrm{RMS}}\left(\vec{x}_0\right)\,{U_i}'_{\mathrm{RMS}}\left(\vec{x}\right)},~~ \\
{U_i}'_{\mathrm{RMS}}\left(\vec{x_i}\right)
=
\overline{{U_i}'\left(\vec{x_i}\right)^2}^{1/2},
}\end{equation}
{where $\vec{x}_o$ and $\vec{x}$ are the spatial coordinates of the reference point and the current point, respectively, and ${U_i}$ is the local velocity in the $i^\text{th}$ direction. In our study, the reference point is located at (2002.5,2345.5,197.5)~m, following with other 12 continuous points spacing in single mesh size aligning with x-axis.}
\begin{figure}[b]
    \centering
    \includegraphics[width=\linewidth]{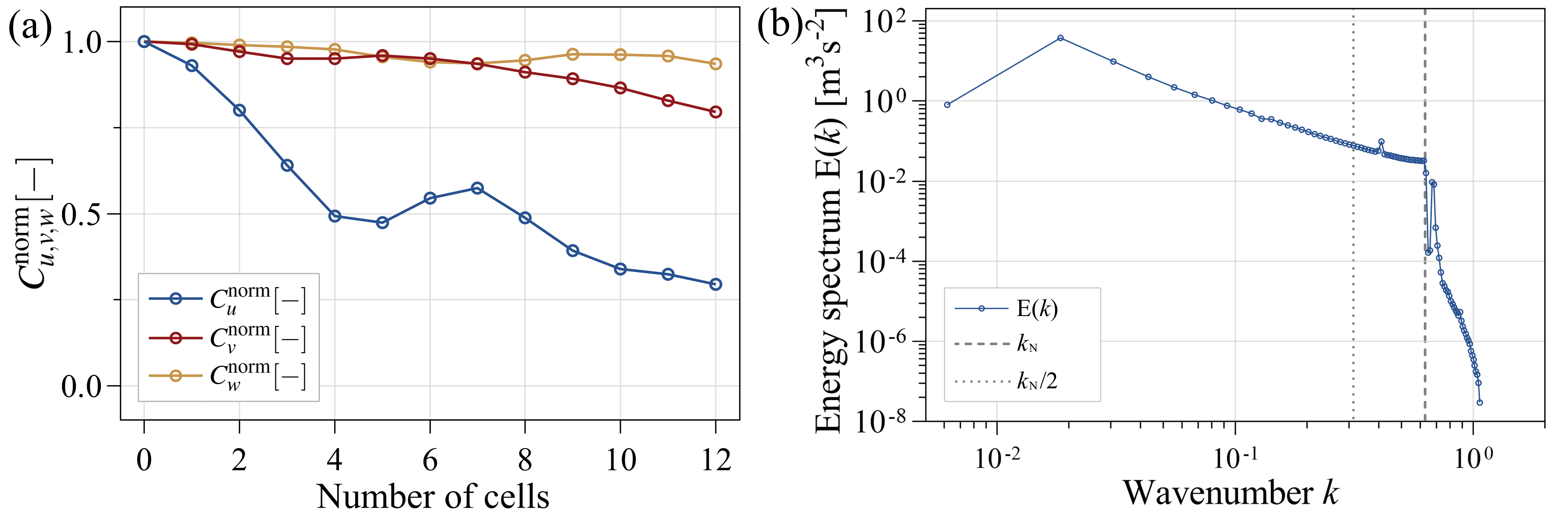}
    \caption{{Turbulence statistics verification. (a) Convergence of two-point correlation $C_{u,v,w}^{\mathrm{norm}}$ versus stride number of cells, and (b) global energy wavenumber spectrum $E(k)$.}}
    \label{fig:Mesh_kn}
\end{figure}
{It is recommended that the largest eddies should be resolved by at least
eight cells, corresponding to a positive correlation for a minimum of 8 cells. The correlation versus stride number of cells is presented in Figure~\ref{fig:Mesh_kn}(a). It is shown that, the correlation with the reference point remains high for the first 12 cells, demonstrating a reasonable resolution of the mesh in our study.}

{Furthermore, the spectral kinetic energy density at the wavenumber vector $\mathbf{k}$ is defined as}
\begin{equation}{
e(\mathbf{k})=\frac{1}{2}\left(\left|\widehat{u}(\mathbf{k})\right|^{2}+\left|\widehat{v}(\mathbf{k})\right|^{2}+\left|\widehat{w}(\mathbf{k})\right|^{2}\right),
}\end{equation}
{where $\widehat{\mathbf{u}}(\mathbf{k})=\mathcal{F}\{\mathbf{u}'(\mathbf{x})\}$ denotes the Fourier transform of the velocity-fluctuation field $\mathbf{u}'$, with $\mathbf{u}'=\mathbf{u}-\langle\mathbf{u}\rangle$. The corresponding three-dimensional isotropic (one-dimensional) energy spectrum is obtained by averaging over all wavevectors with identical magnitude $k=\|\mathbf{k}\|$, such that the total kinetic energy satisfies $\int_{0}^{\infty} E(k)\,\mathrm{d}k=\int_{\mathbb{R}^{3}} e(\mathbf{k})\,\mathrm{d}\mathbf{k}$, equivalently,}
\begin{equation}{
E(k)=\int_{\|\mathbf{k}\|=k} e(\mathbf{k})\,\mathrm{d}S_{\mathbf{k}}
= k^{2}\int_{4\pi} e\!\left(k,\Omega\right)\,\mathrm{d}\Omega,
}\end{equation}
{as shown in Figure~\ref{fig:Mesh_kn}(b).}
{To assess consistency with the Kolmogorov inertial-range scaling, we additionally marked slope $k^{-5/3}$.}
{The Nyquist wavenumber $k_N$ and the reference limit $k_N/2$ are marked. From the figure, the energy spectrum remains strictly positive and varies smoothly over the entire set of resolved wavenumber range, with no spurious oscillations and discontinuities. For wavenumbers below $k_N/2$, $E(k)$ exhibits an approximately power-law decay and the compensated spectrum forms a broad, near-horizontal plateau, consistent with Kolmogorov inertial-range -5/3 scaling. This behavior indicates that the LES resolves a physically meaningful inertial cascade over a substantial portion of the spectrum.
At high wavenumber end, there is no anomalous upturn, demonstrating no spectral pile-up (spectral blocking) or aliasing-driven energy accumulation. Similarly, 2D wavenumber spectra (Supplementary Material~\ref{supp:wn}) also prove the appropriate resolution of LES.
Overall, it is supported that the resolved range is physically consistent and that no grid-scale instabilities are present. }

\subsection{{Boundary effect and relaxation zone}}

Low-order boundary conditions (e.g., velocity Dirichlet) tend to induce the earliest departures in time-dependent high-order quantities in regions adjacent to the boundaries of urban wind-field domain, such as turbulent kinetic energy and related turbulence statistics. These errors can propagate and eventually contaminate lower-order fields such as the mean velocity profile and shear-layer structure. 
Such boundary-induced distortions have been widely reported \cite{bc-gmd,bc-Heinze,bc-Mazzaro,bc-Moeng}.
A characteristic manifestation is an unphysical over-stabilization of the near-boundary flow, which may even suppress or delay transition. One conventional strategy is the use of periodic boundary conditions. However, their applicability presupposes homogeneous inflow, which constrains their use in large-scale heterogeneous winds. Another approach is to explicitly prescribe turbulent boundary information, yet assigning turbulence variables in a manner consistent with the underlying physics is nontrivial and may readily compromise numerical stability. In contrast, employing an enlarged precursor domain (an upstream development fetch region of the computational domain) allows the influence of boundary assimilation to attenuate progressively through cascaded evolution and provides sufficient fetch for flow development (also known as relaxation zone or boundary buffer), thereby enabling reliable reproduction of both low- and high-order quantities within the region of interest. In this study, we introduce a precursor domain and systematically examine how its size affects numerical accuracy and reliability.

\begin{figure}[t]
        \centering
        \includegraphics[width=1\linewidth]{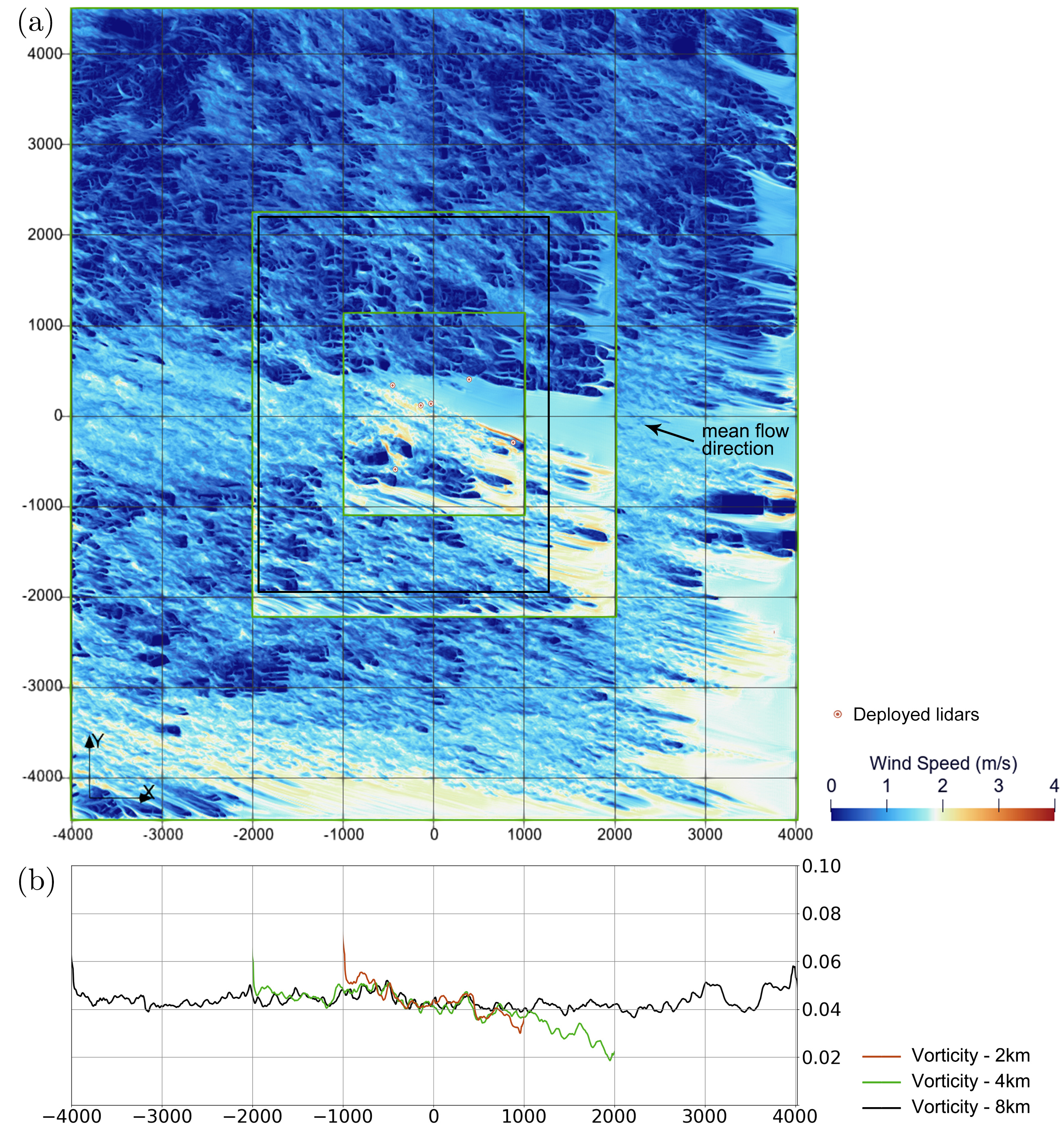}
    \caption{Computational regions and wind speed on $z=50~m$ with various precursor domain strategies at 18:10. (a) Wind speed contour with main flow direction overlapped; {the black line outlines the fully developed region, and red circles mark the lidar stations}; and (b) the distribution along x-axis of normalized y-integrated vorticity magnitude $\langle\|\omega\|\rangle_y$.}
    \label{fig:hyperdomain}
\end{figure}

{We tested three precursor domain strategies, including approximately 2008~m×2243~m (labelled 2~km), 4026~m×4486~m (labelled 4~km, baseline) and 8046~m×8987~m (labelled 8~km).} Lattice size is kept at 5~m resolution for all of them. The computational domain with sectional wind speed distribution at 50~m height is shown in Figure~\ref{fig:hyperdomain}.
To investigate the effective extent of the practically valid data region, this study does not prescribe fixed boundary buffers or assign a focus region. Instead, total computational regions are included in the analysis.

First, the downstream wind fields obtained under various precursor domain strategies exhibit substantial overlap, indicating that the cascading effects of the flow can be accurately corrected and that structural wind effects are reliably reconstructed. Consequently, within the fully developed region, the flow-field structure demonstrates a high level of accuracy. In addition, the downstream boundary shows satisfactory absorbing performance, with no evidence of non-physical oscillations or flow recirculation.

However, in the upstream region (the right and lower boundaries in Figure~\ref{fig:hyperdomain}), there remain zones where the flow is evidently not fully developed, characterized by weak turbulence features. This behavior can be attributed to the fact that the boundary conditions do not inject a physically consistent turbulent kinetic energy spectrum, but instead prescribe only the inflow velocity. During advection, particularly after interaction with buildings, the flow undergoes instabilities and evolves into broadband velocity fluctuations, thereby generating turbulent kinetic energy and intermittency spontaneously within the domain. Prior to full development, the wind field is clearly inaccurate, leading to the appearance of quasi-stationary streaks. As shown in the sectional wind speed, the spatial extent of this development region is strongly influenced by building density. Denser urban configurations facilitate more rapid flow development and promote spatial convergence. For computational domains of 2 km and 4 km, the inlet buffer region on the eastern side is approximately 700 m, whereas, owing to the sparser building distribution, the 8~km domain requires more than 1~km for adequate flow development.

The same boundary effects are quantitatively examined for higher-order quantities and are manifested more prominently. Here, we present the x-direction distribution of length-normalized vorticity after integration in the y-direction, defined by
\begin{equation}
\langle \|\omega\|\rangle_y
= \frac{1}{L_y}\int_{y_1}^{y_2}
\sqrt{\varepsilon_{ijk}\varepsilon_{ilm}\,\partial_j u_k\,\partial_l u_m}
\,~\mathrm{d}y,
\end{equation}
where $u_k$ are the velocity components and $\varepsilon$ is the Levi--Civita symbol under summation convention.
The integration range in the y-direction is defined by the smallest domain configuration. Compared with the 8-km computational domain, the smaller domain leads to an underestimation of the inlet vorticity on the right-hand side, owing to the fact that the Dirichlet boundary condition does not introduce turbulent kinetic energy. As the flow develops downstream, the vorticity distribution obtained in the small-domain simulation gradually converges to that of the large-domain case. This observation further demonstrates that LES is capable of overcoming erroneous features present in the mesoscale output, which is consistent with the conclusions reported in \cite{bc-Mazzaro}.
On the other hand, boundary effects are also observed at the outlet, where an accumulation of turbulent kinetic energy occurs. Although the balanced boundary condition provides a certain degree of absorption, it remains insufficient to fully remove the excess turbulent kinetic energy through advection. Consequently, vorticity exhibits an artificial buildup near the outlet, with an affected region of approximately 300 m, reaching its maximum value at the downstream boundary.
From a numerical methods perspective, velocity Dirichlet outlet conditions based on equilibrium reconstruction constrain the non-equilibrium component of the distribution functions \cite{Zou1997OnPressureVelocityBC}, which can spuriously reflect turbulent fluctuations and cause an artificial accumulation of turbulent kinetic energy near the outlet at high Reynolds numbers \cite{Hecht2010OnSiteVelocityBC,Guo2002ExtrapolationBC,Junk2008OutflowBC}. Nevertheless, this effect is considered acceptable when the region of interest lies sufficiently far from the outlet.

\begin{figure}[b]
    \centering

    \begin{minipage}[b]{0.49\linewidth}
        \centering
        \begin{minipage}[b]{0.48\linewidth}
            \centering
            \includegraphics[width=\linewidth]{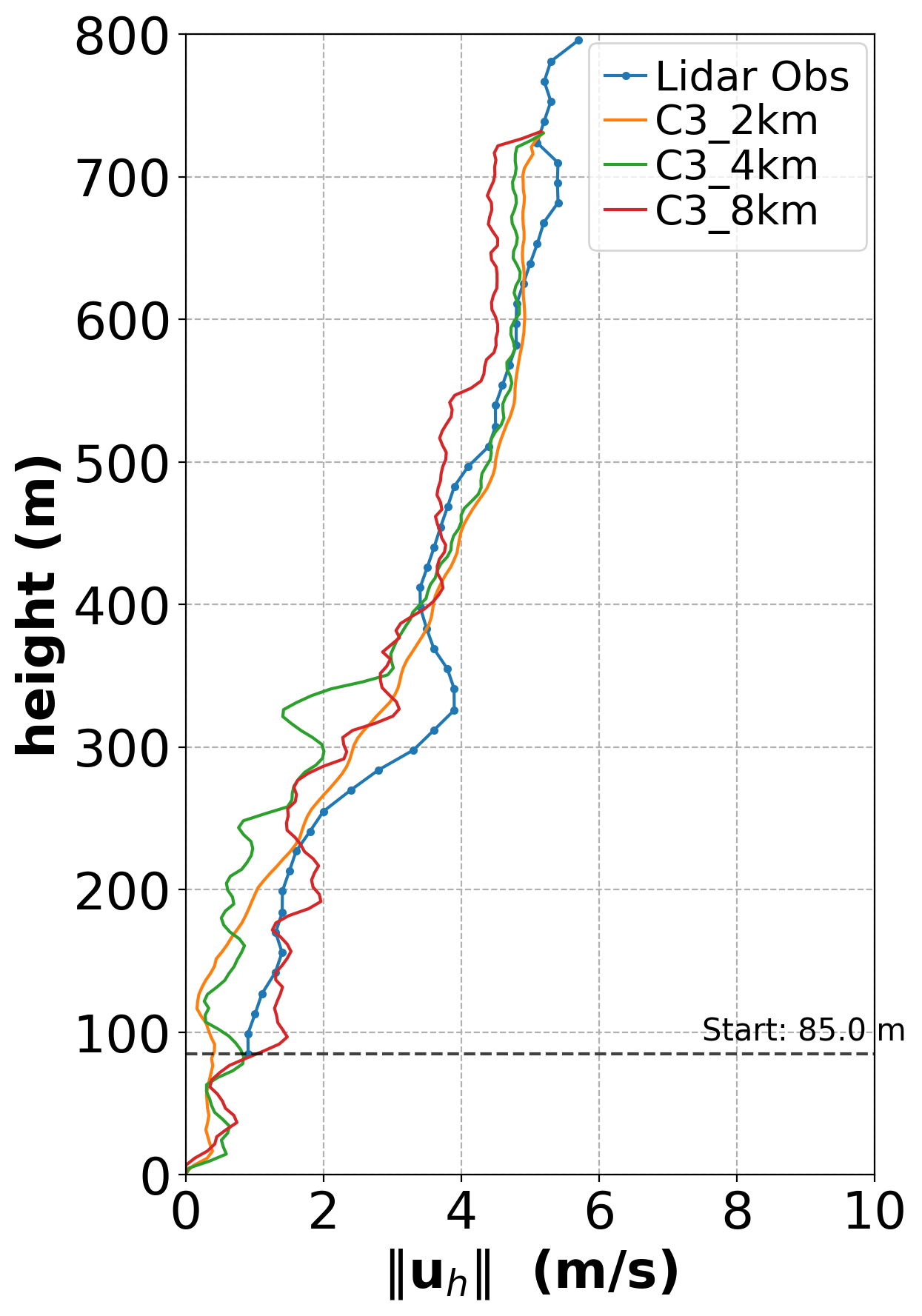}
    
        \par\smallskip
        \textup{(a)}
        \end{minipage}
        \hfill
        \begin{minipage}[b]{0.48\linewidth}
            \centering
            \includegraphics[width=\linewidth]{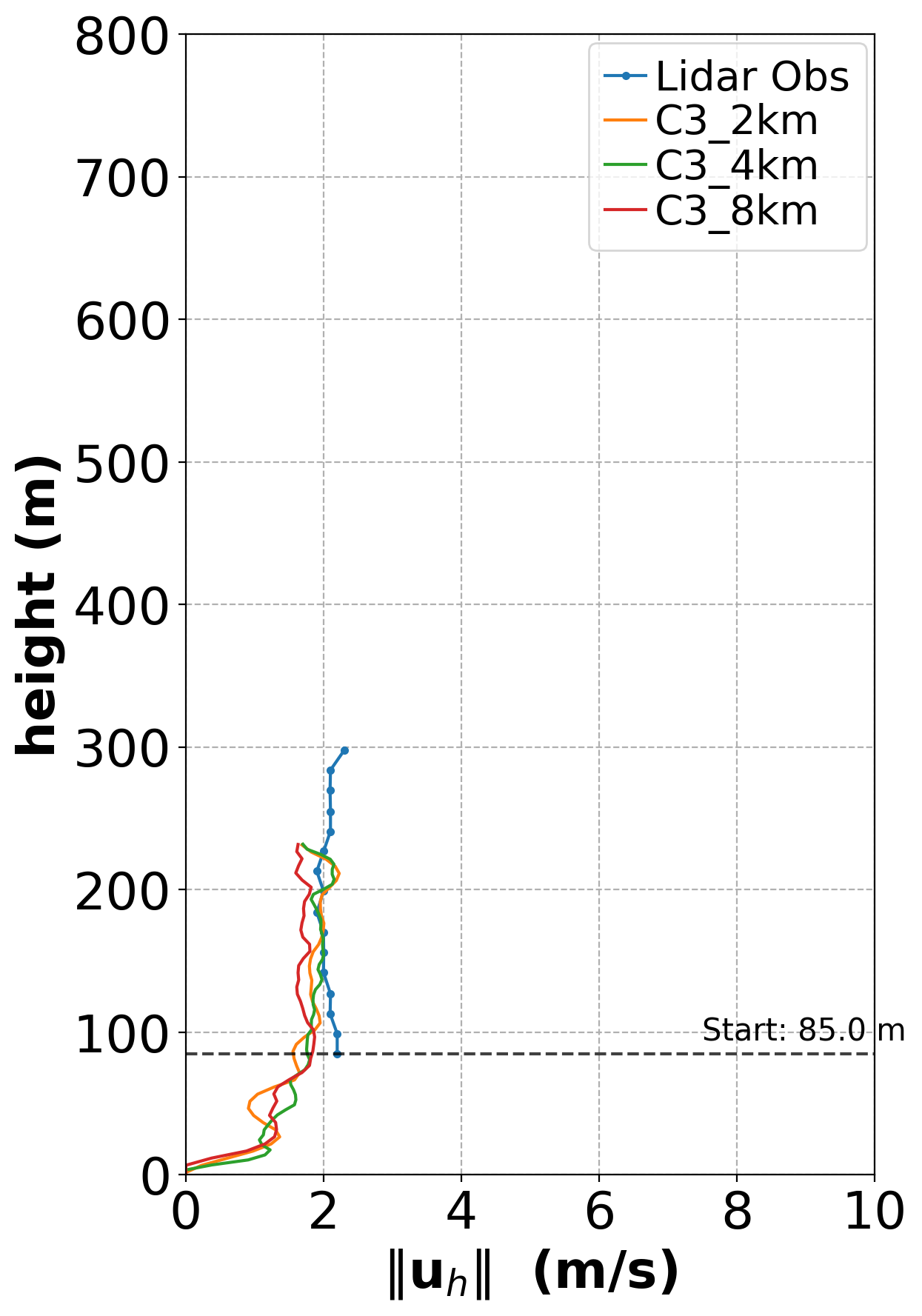}

        \par\smallskip
        \textup{(b)}
        \end{minipage}

    \end{minipage}
    \hfill
    \begin{minipage}[b]{0.49\linewidth}
        \centering
        \begin{minipage}[b]{0.48\linewidth}
            \centering
            \includegraphics[width=\linewidth]{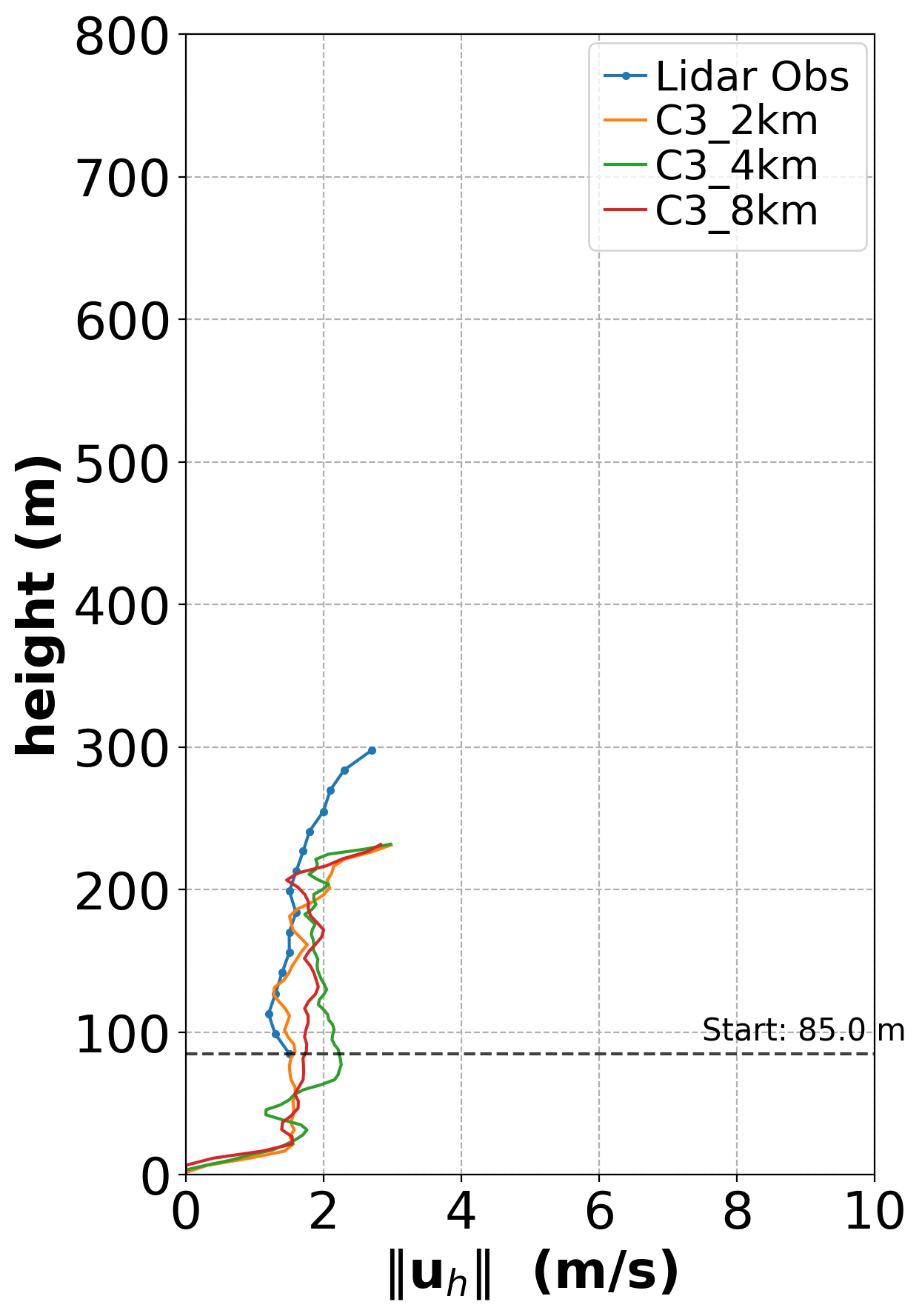}
            
        \par\smallskip
        \textup{(c)}
        \end{minipage}
        \hfill
        \begin{minipage}[b]{0.48\linewidth}
            \centering
            \includegraphics[width=\linewidth]{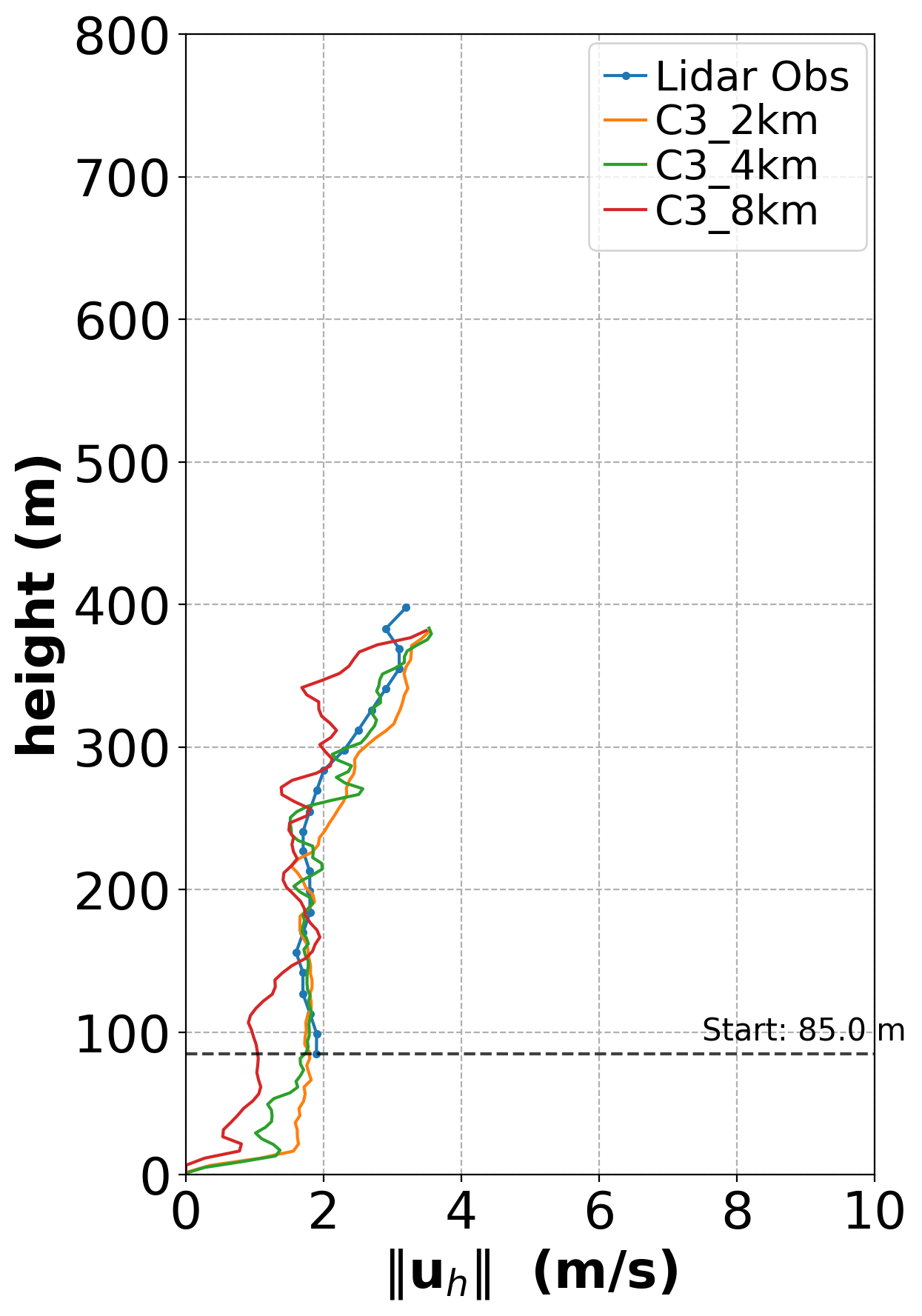}
            
        \par\smallskip
        \textup{(d)}
        \end{minipage}

    \end{minipage}
    \caption{Wind profile at GAW105 position with various precursor domain strategies, comparing with observations (Case 3). (a) 17:10, (b) 18:10, (c) 18:30, and (d) 18:40.}
    \label{fig:hyperprofile}
\end{figure}

{We further compared the horizontal wind profiles obtained at four representative time instants for different precursor domain sizes} (Figure~\ref{fig:hyperprofile}). The results indicate that, under large-domain configurations where the wind field is fully developed, the flow exhibits more pronounced small-scale spatial structures under strong-wind conditions. As shown in Figure~\ref{fig:hyperprofile}(a), both the 4 km and 8 km computational domains display significant fluctuation characteristics, whereas these features cannot be adequately captured in the 2 km domain.
On the other hand, in the 8 km precursor domain configuration, the lateral boundaries are located relatively far from the focus region, which reduces the effectiveness of data assimilation. More importantly, the Kriging-based boundary constructor degrades into long-distance data extrapolation, resulting in increased uncertainty at the side boundaries. Consequently, its profile paradoxically exhibits poorer agreement than the other two sizes.

\subsection{Wind fields characteristics}

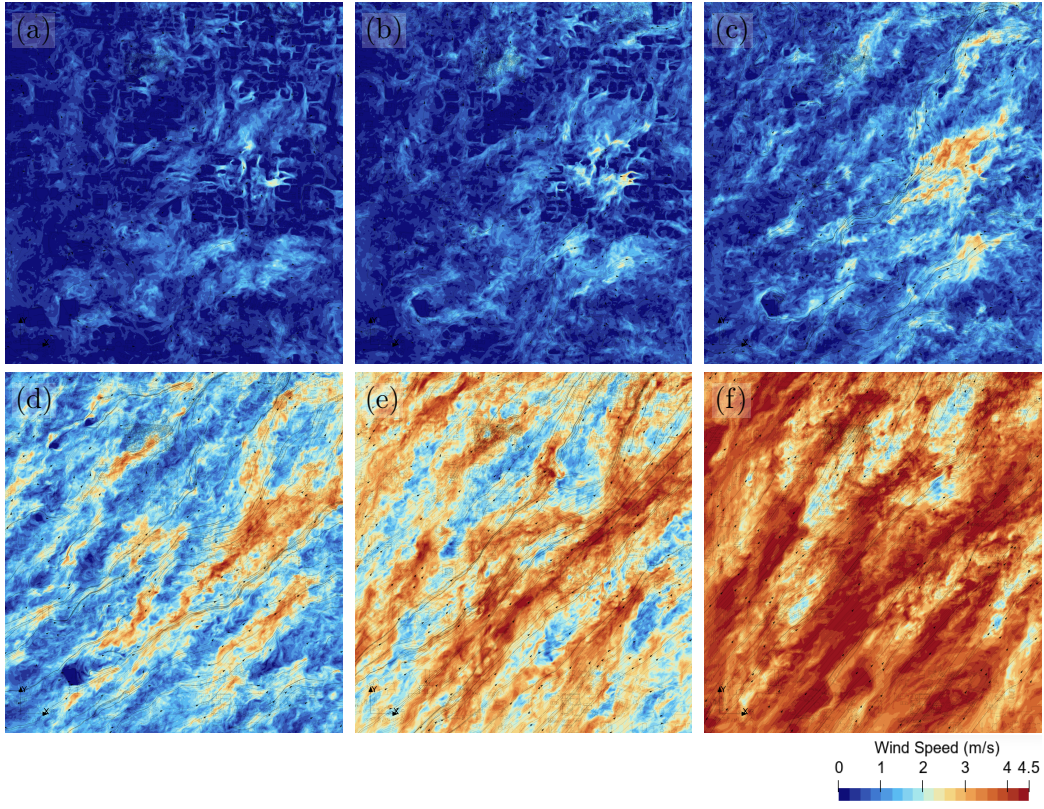
\begin{figure}[t]
    \centering
    \begin{minipage}[b]{0.325\linewidth}
        \centering
        \begin{tikzpicture}
            \node[anchor=south west, inner sep=0] (img) at (0,0)
                {\includegraphics[width=\linewidth]{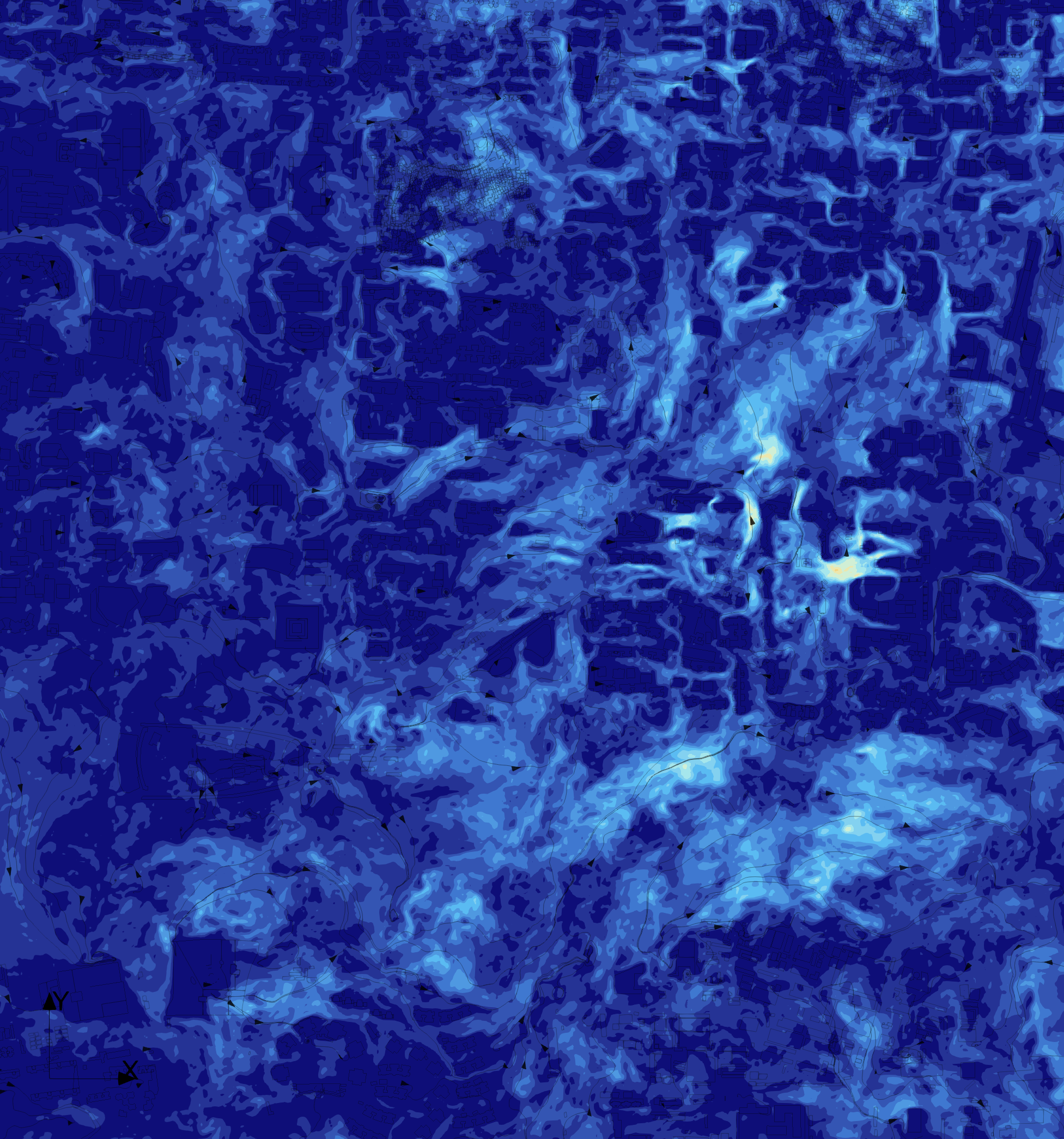}};
            \begin{scope}[x={(img.south east)}, y={(img.north west)}]
                \node[anchor=north west, inner sep=1pt, fill=white, fill opacity=0.2, text opacity=1, font=\small]
                    at (0.02,0.97) {(a)};
            \end{scope}
        \end{tikzpicture}
    \end{minipage}
    \hfill
    \begin{minipage}[b]{0.325\linewidth}
        \centering
        \begin{tikzpicture}
            \node[anchor=south west, inner sep=0] (img) at (0,0)
                {\includegraphics[width=\linewidth]{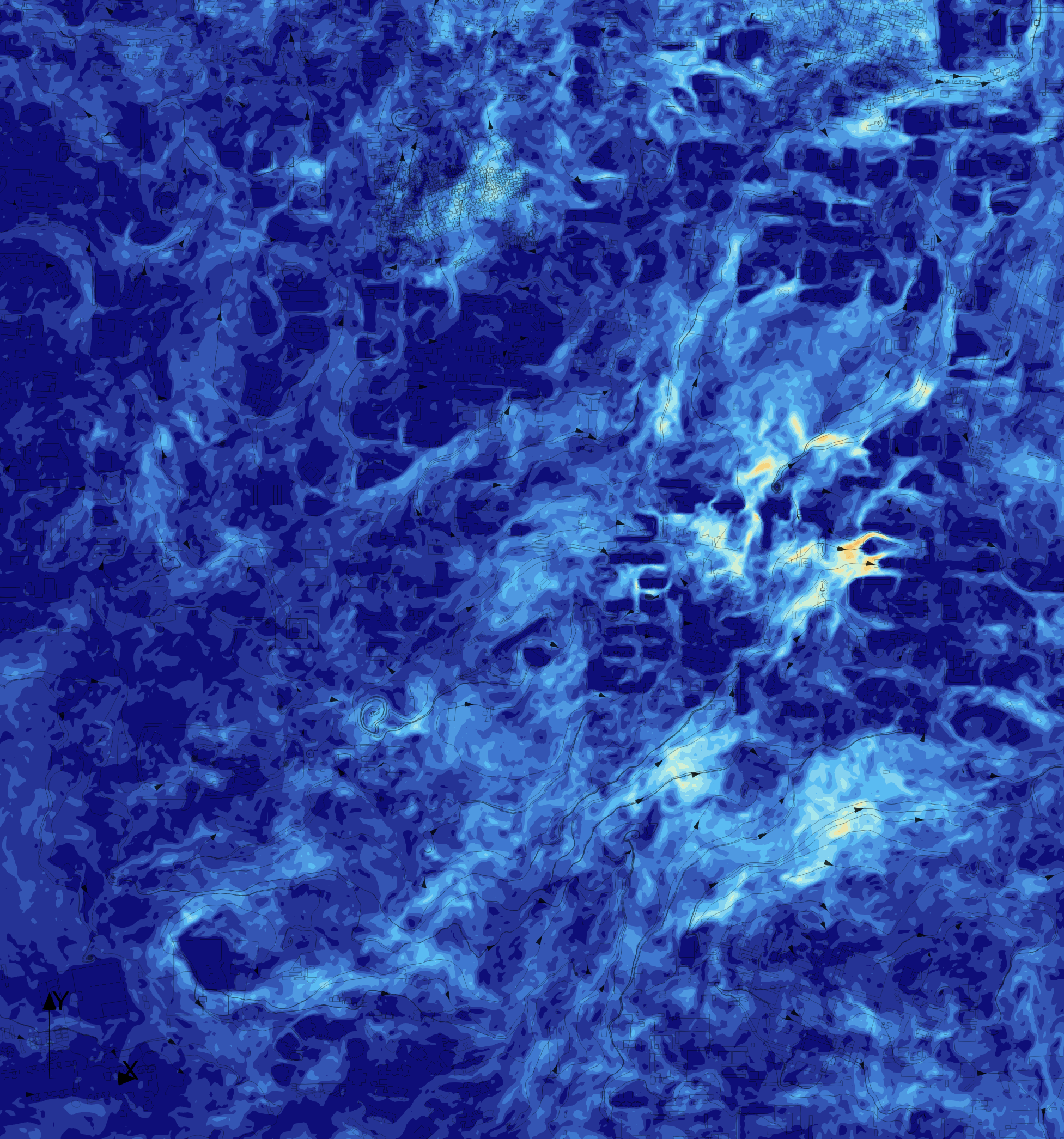}};
            \begin{scope}[x={(img.south east)}, y={(img.north west)}]
                \node[anchor=north west, inner sep=1pt, fill=white, fill opacity=0.2, text opacity=1, font=\small]
                    at (0.02,0.97) {(b)};
            \end{scope}
        \end{tikzpicture}
    \end{minipage}
    \hfill
    \begin{minipage}[b]{0.325\linewidth}
        \centering
        \begin{tikzpicture}
            \node[anchor=south west, inner sep=0] (img) at (0,0)
                {\includegraphics[width=\linewidth]{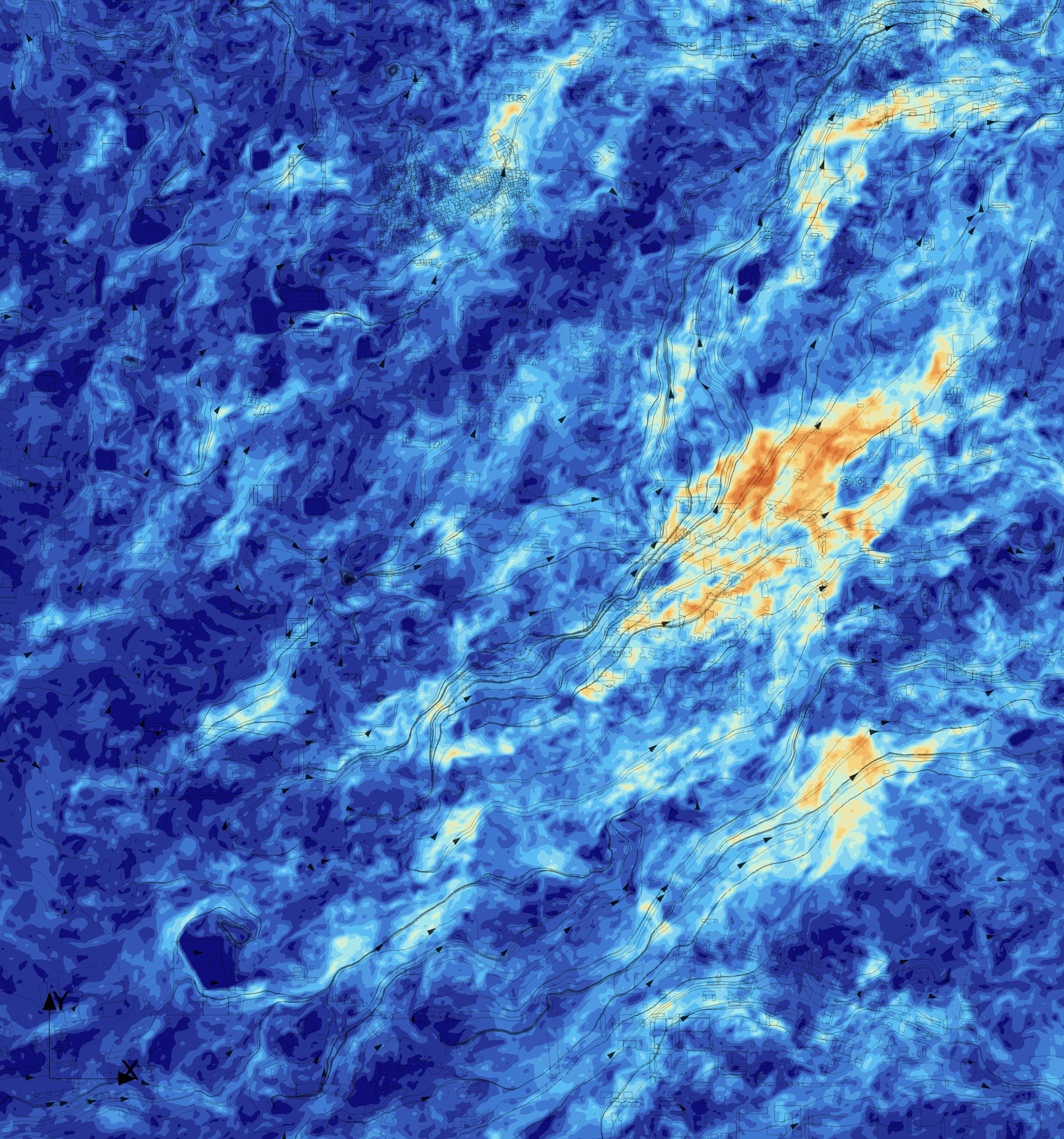}};
            \begin{scope}[x={(img.south east)}, y={(img.north west)}]
                \node[anchor=north west, inner sep=1pt, fill=white, fill opacity=0.2, text opacity=1, font=\small]
                    at (0.02,0.97) {(c)};
            \end{scope}
        \end{tikzpicture}
    \end{minipage}

\vspace{0.2em}

    \begin{minipage}[b]{0.325\linewidth}
        \centering
        \begin{tikzpicture}
            \node[anchor=south west, inner sep=0] (img) at (0,0)
                {\includegraphics[width=\linewidth]{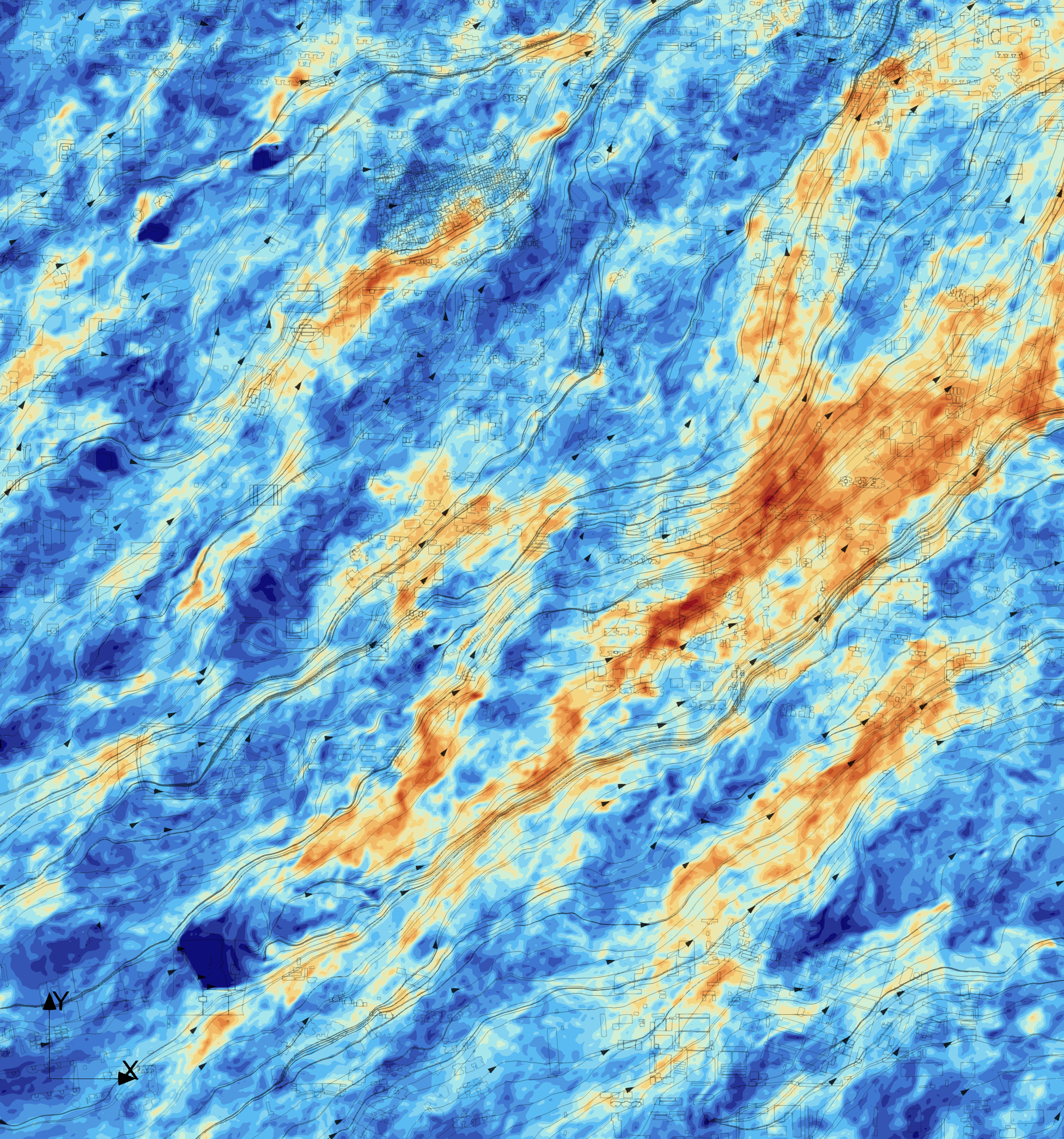}};
            \begin{scope}[x={(img.south east)}, y={(img.north west)}]
                \node[anchor=north west, inner sep=1pt, fill=white, fill opacity=0.2, text opacity=1, font=\small]
                    at (0.02,0.97) {(d)};
            \end{scope}
        \end{tikzpicture}
    \end{minipage}
    \hfill
    \begin{minipage}[b]{0.325\linewidth}
        \centering
        \begin{tikzpicture}
            \node[anchor=south west, inner sep=0] (img) at (0,0)
                {\includegraphics[width=\linewidth]{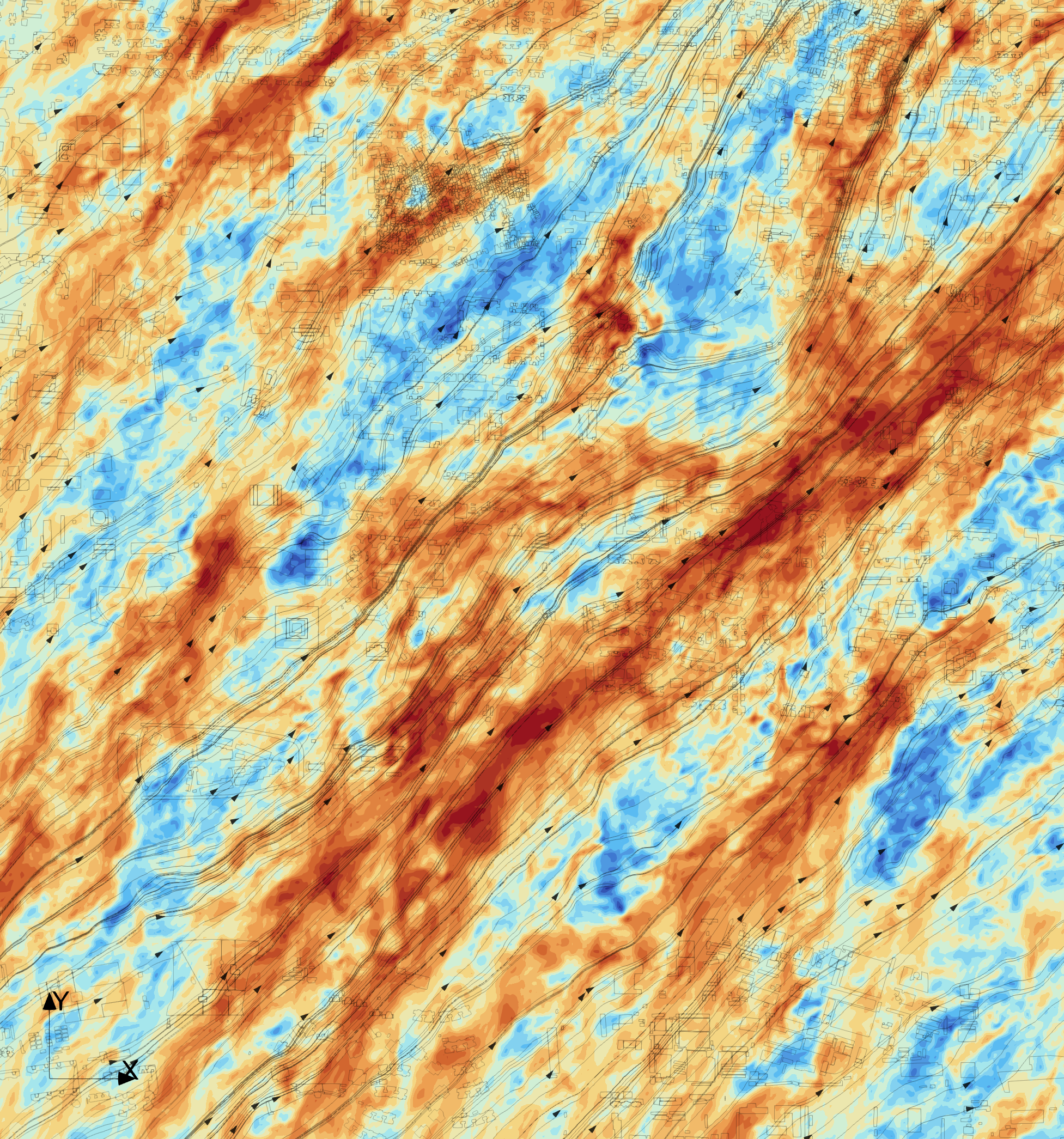}};
            \begin{scope}[x={(img.south east)}, y={(img.north west)}]
                \node[anchor=north west, inner sep=1pt, fill=white, fill opacity=0.2, text opacity=1, font=\small]
                    at (0.02,0.97) {(e)};
            \end{scope}
        \end{tikzpicture}
    \end{minipage}
    \hfill
    \begin{minipage}[b]{0.325\linewidth}
        \centering
        \begin{tikzpicture}
            \node[anchor=south west, inner sep=0] (img) at (0,0)
                {\includegraphics[width=\linewidth]{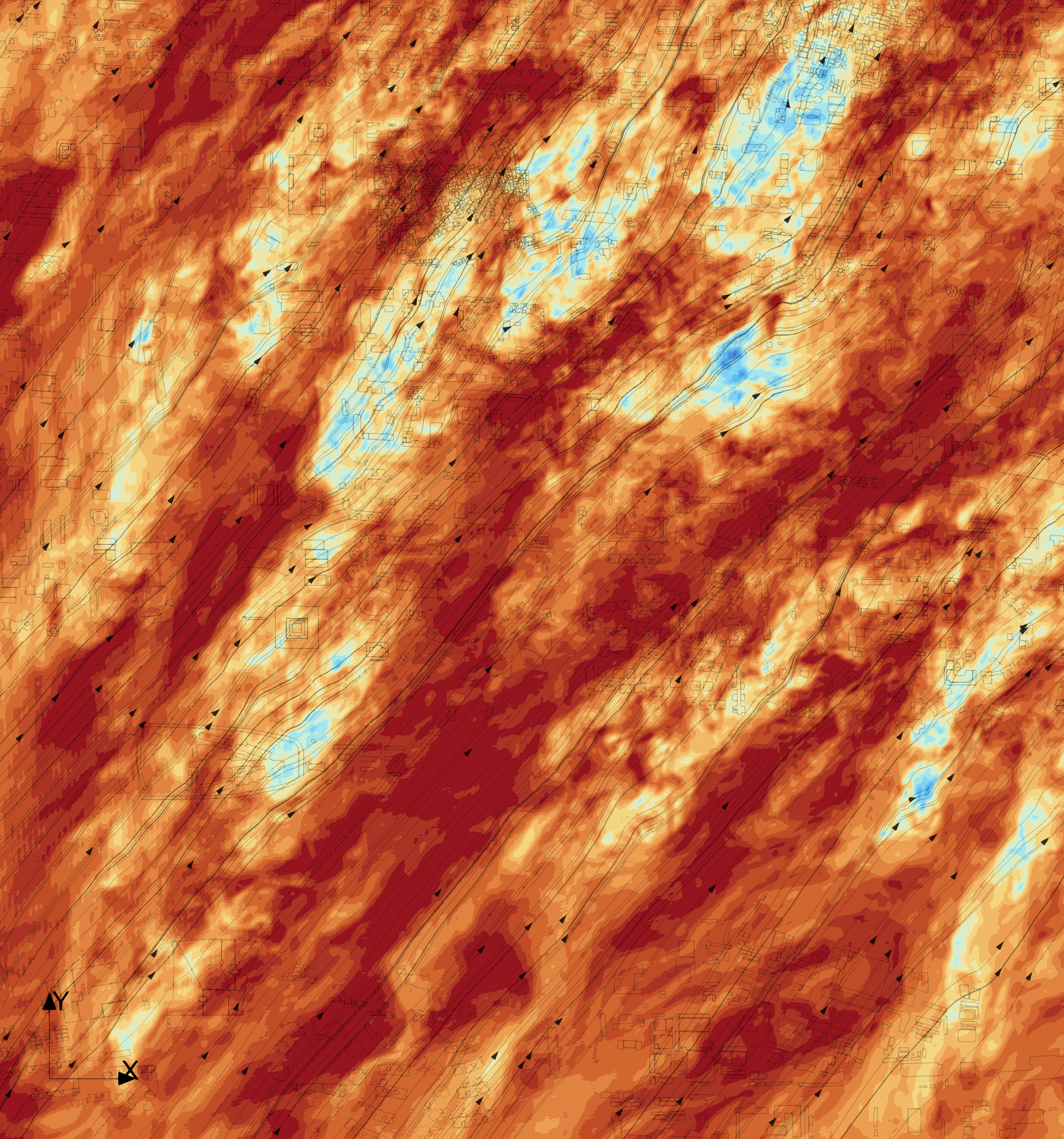}};
            \begin{scope}[x={(img.south east)}, y={(img.north west)}]
                \node[anchor=north west, inner sep=1pt, fill=white, fill opacity=0.2, text opacity=1, font=\small]
                    at (0.02,0.97) {(f)};
            \end{scope}
        \end{tikzpicture}
    \end{minipage}

    \hfill
    \begin{minipage}[b]{0.2\linewidth}
        \centering
        \includegraphics[width=\linewidth]{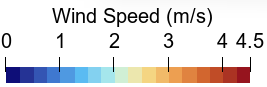}
    \end{minipage}
    
\caption{{Instantaneous wind speed on horizontal sections represented by contours and streamlines.} (a) $z=20~m$, (b) $z=50~m$, (c) $z=100~m$, (d) $z=200~m$, (e) $z=300~m$, (f) $z=400~m$.}
\label{fig:sectionsz}
\end{figure}

We present the numerically simulated wind response of urban buildings and terrain under moderate wind conditions observed at 18:30.
Figure~\ref{fig:sectionsz} illustrates the wind velocity field structures at various horizontal cross-sectional heights, represented by streamlines. At lower elevations between 20 and 100 meters, the wind field exhibits a highly complex structure, strongly modulated by the underlying terrain and built environment. The streamlines are frequently distorted, with localized convergence, bifurcation, and recirculation patterns, indicating pronounced horizontal directional variability. The intricate geometries across different regions manifest substantial disparities in surface roughness. Within this vertical range, wind speeds remain relatively low overall, suggesting significant frictional dissipation within the near-surface layer due to the influence of buildings, and implying that momentum transfer remains underdeveloped. In certain areas aligned with the topography, channelized wind corridors and valley effects are discernible. Streamlines are observed to become elongated and densely packed within some narrow zones, indicating the presence of terrain-induced acceleration zones.
Specifically, as shown in Figure~\ref{fig:sectionsz}(a), at an ultra-low altitude (z = 20 m), numerous small-scale vortex structures and coherent transport features are observed, driven by the wakes and surrounding flows of low-rise buildings.

At the intermediate height of approximately 200 meters (Figure~\ref{fig:sectionsz}(d)), the wind field structure undergoes a marked transition. The streamlines become more coherent overall, with the predominant wind direction emerging more distinctly. The curvature scale of the flow increases significantly, while small-scale turbulent structures diminish. This indicates that the flow at this altitude is gradually decoupling from the direct influence of surface roughness elements characteristic of the lower atmospheric boundary layer, and is transitioning into the more thoroughly mixed mid-boundary layer. The mean wind speed increases substantially; however, residual weak disturbances corresponding to underlying terrain features remain discernible, reflecting the upward propagation and gradual attenuation of terrain-induced effects.

At elevations exceeding 300 meters, the wind field exhibits pronounced vertical coherence and anisotropy. Streamlines are nearly parallel and form a quasi-two-dimensional, stratified structure. The dominant wind direction is highly uniform across the entire domain, indicative of characteristics typical of a potential flow region. Wind speed increases monotonically with height, consistent with the canonical wind profile of the atmospheric boundary layer. The influence of terrain and underlying surface roughness on streamline morphology is largely smoothed out at this altitude, with only minimal residual disturbances or yaw evident in isolated areas. The flow is predominantly governed by the large-scale background wind.

From the perspective of mechanisms, the transition from the friction layer to the logarithmic layer and further into the higher mixed layer is clearly delineated in our study. In the lower levels, momentum exchange is primarily governed by turbulence and surface drag, resulting in low wind speeds and highly scattered wind directions. As altitude increases, turbulent mixing facilitates the downward transport of higher momentum from aloft, while the influence of surface drag diminishes, leading to increased wind speeds and greater directional uniformity. Correspondingly, streamlines evolve from fragmented and meandering patterns to more rectilinear and organized structures. This signifies a transition from a regime dominated by locally CFD-resolved geometric flow patterns (buildings and terrain) to one increasingly governed by data assimilation of the background flow (mesoscale boundary data).
The dynamical characteristics of the boundary layer turbulent diffusion in heterogeneous surface boundary layers remain subjects requiring further investigation.

\begin{figure}[t]
\centering

\begin{minipage}[b]{0.49\linewidth}
    \centering
    \begin{tikzpicture}
        \node[anchor=south west, inner sep=0] (img) at (0,0)
            {\includegraphics[width=\linewidth]{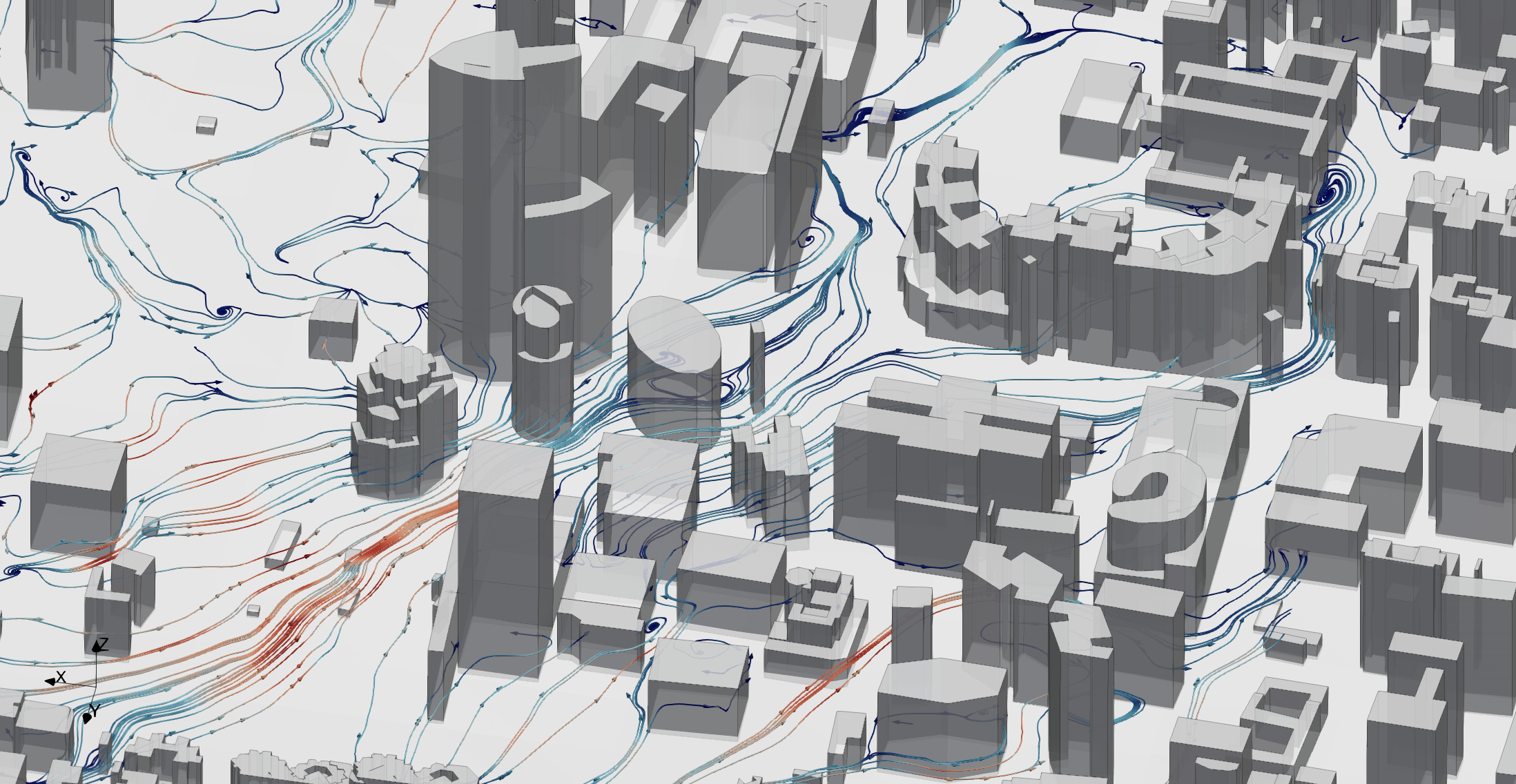}};
        \begin{scope}[x={(img.south east)}, y={(img.north west)}]
            \node[anchor=north west, inner sep=1pt, fill=white, fill opacity=0.2, text opacity=1, font=\small]
                at (0.01,0.98) {(a)};
        \end{scope}
    \end{tikzpicture}
\end{minipage}
\begin{minipage}[b]{0.49\linewidth}
    \centering
    \begin{tikzpicture}
        \node[anchor=south west, inner sep=0] (img) at (0,0)
            {\includegraphics[width=\linewidth]{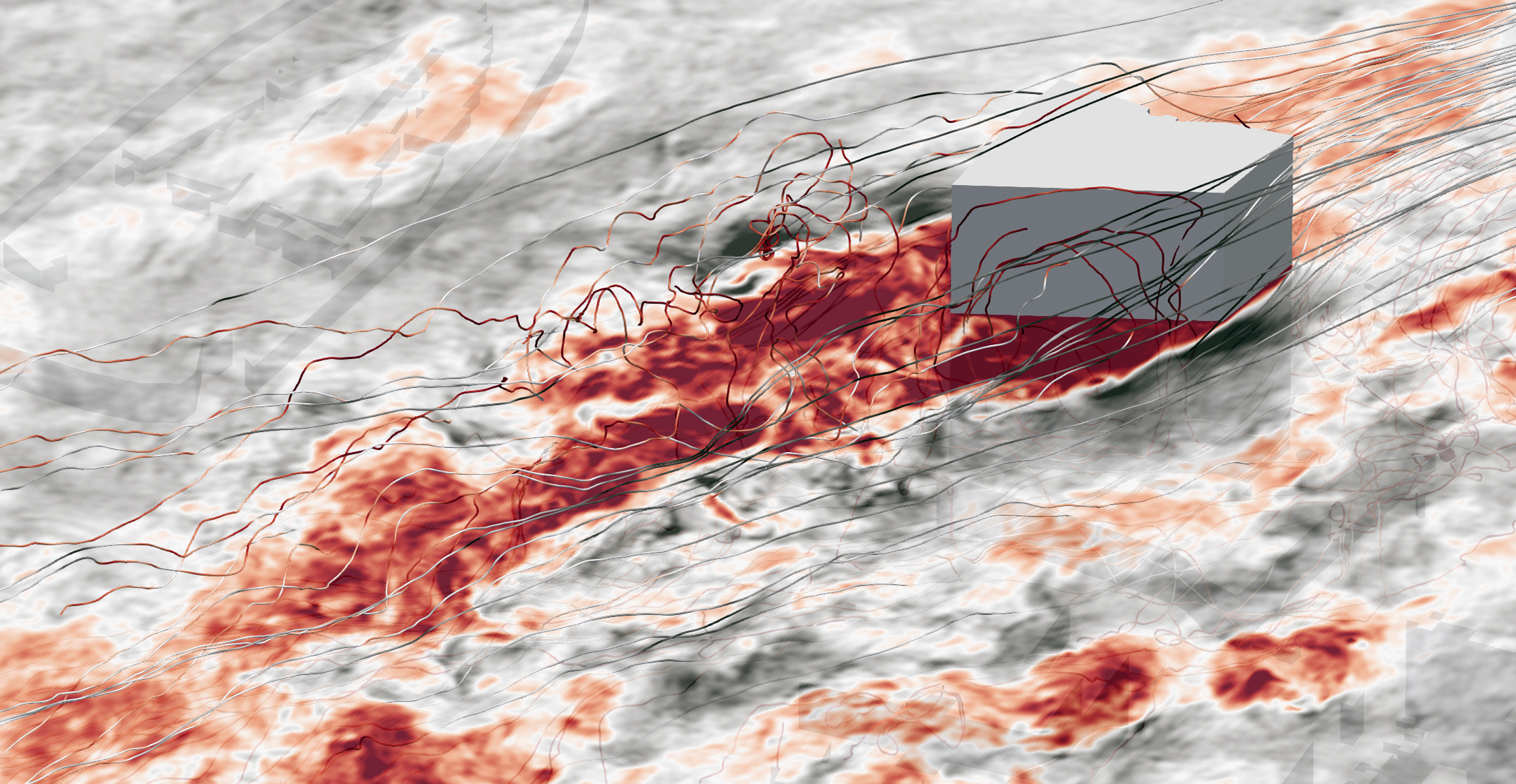}};
        \begin{scope}[x={(img.south east)}, y={(img.north west)}]
            \node[anchor=north west, inner sep=1pt, fill=white, fill opacity=0.2, text opacity=1, font=\small]
                at (0.01,0.98) {(b)};
        \end{scope}
    \end{tikzpicture}
\end{minipage}

\begin{minipage}[b]{0.49\linewidth}
    \centering
    \begin{tikzpicture}
        \node[anchor=south west, inner sep=0] (img) at (0,0)
            {\includegraphics[width=\linewidth]{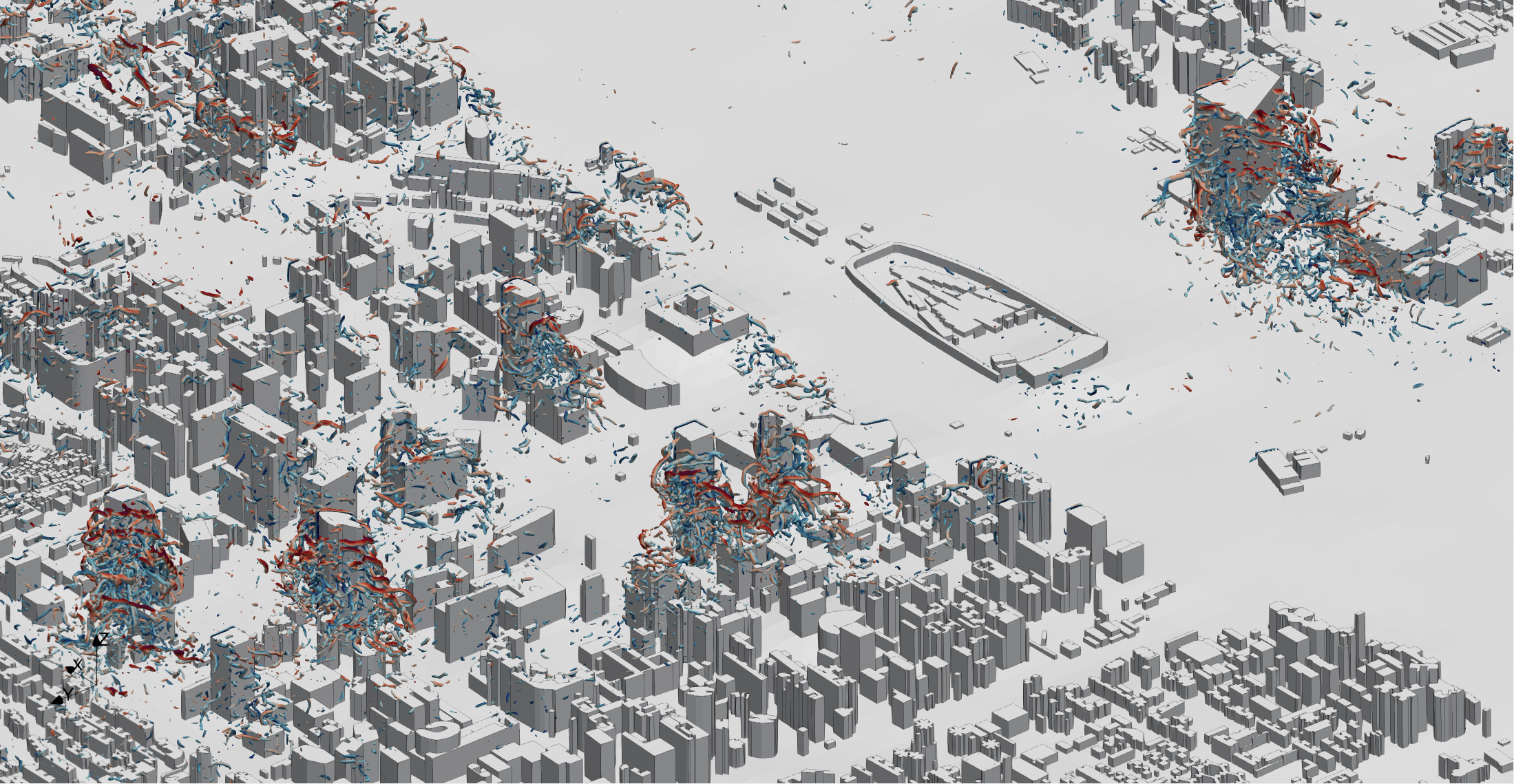}};
        \begin{scope}[x={(img.south east)}, y={(img.north west)}]
            \node[anchor=north west, inner sep=1pt, fill=white, fill opacity=0.2, text opacity=1, font=\small]
                at (0.01,0.98) {(c)};
        \end{scope}
    \end{tikzpicture}
\end{minipage}
\begin{minipage}[b]{0.49\linewidth}
    \centering
    \begin{tikzpicture}
        \node[anchor=south west, inner sep=0] (img) at (0,0)
            {\includegraphics[width=\linewidth]{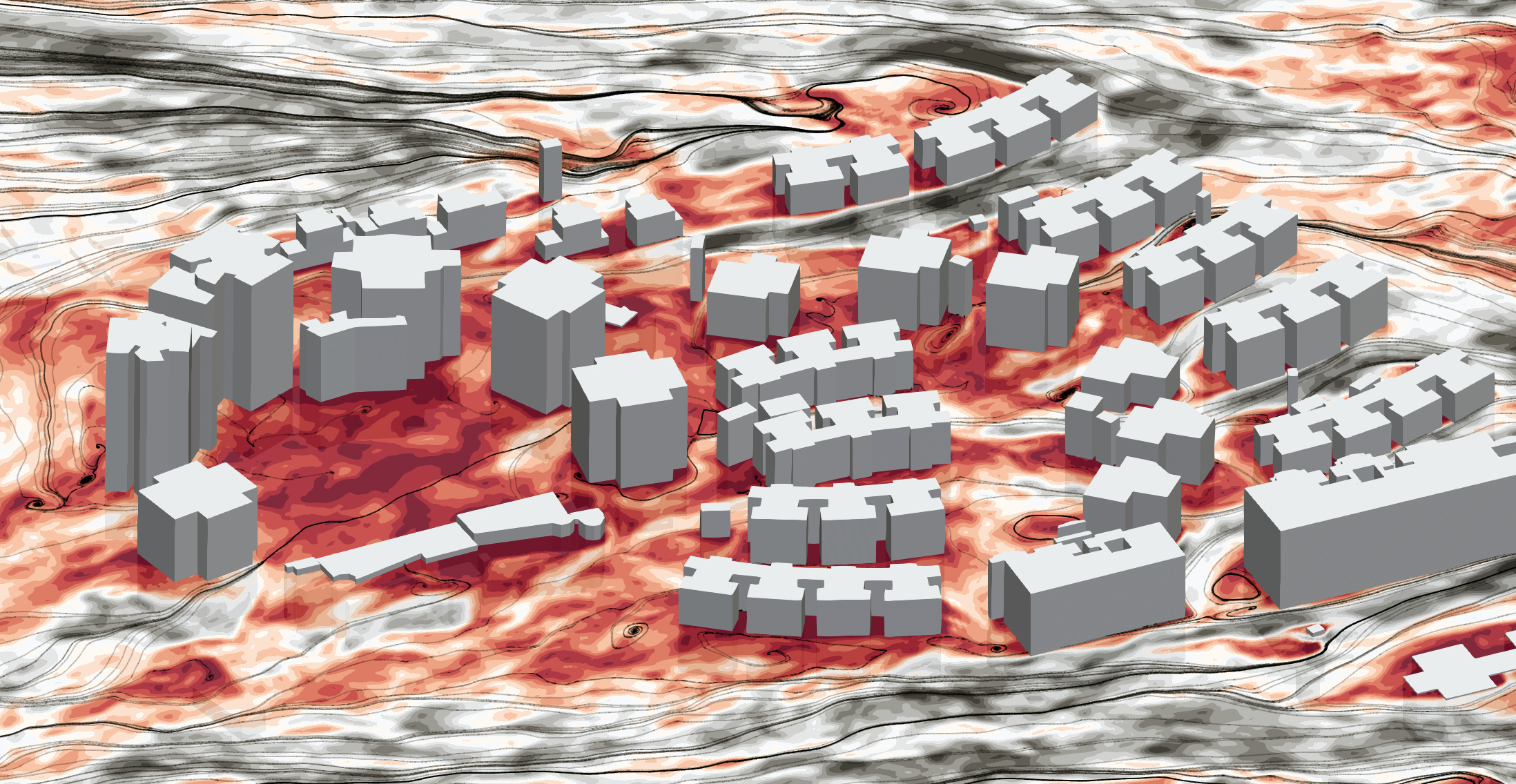}};
        \begin{scope}[x={(img.south east)}, y={(img.north west)}]
            \node[anchor=north west, inner sep=1pt, fill=white, fill opacity=0.2, text opacity=1, font=\small]
                at (0.01,0.98) {(d)};
        \end{scope}
    \end{tikzpicture}
\end{minipage}
\caption{{Low-altitude urban wind event (instantaneous). (a) Street canyon wind effect, (b) shedding wake region of a single building, (c) wake of multiple tall buildings represented by $Q=0.1s^{-2}$, and (d) large wind past a residential community.}}
\label{fig:sectionsy}
\end{figure}
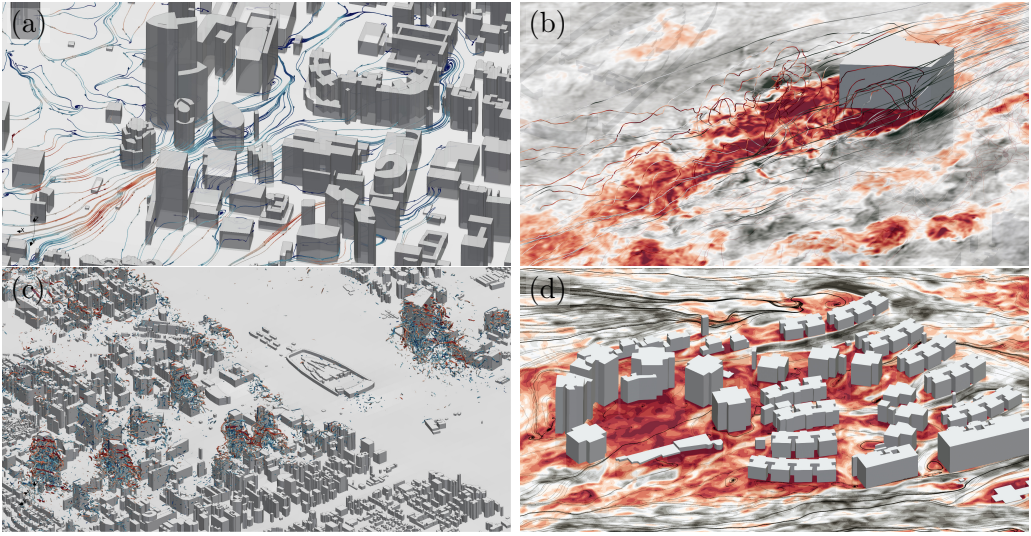

Owing to the high spatial resolution of the simulations, numerous intricate building-induced flow micrometeorological structures and urban wind phenomena have been simultaneously captured, including winds in street canyon (Figure~\ref{fig:sectionsy}(a)), complex unsteady wakes of buildings by streamlines and Q-criterion (Figure~\ref{fig:sectionsy}(b)-(c)), and localized winds in a residential community.
Although this study does not primarily focus on the complex small-scale flow structures themselves, their clear resolution further substantiates the spatial-resolving capability of this approach.

\subsection{{Horizontal wind profile and quality control}}

\begin{figure}[t]
    \centering
    \begin{minipage}[b]{0.5\linewidth}
        \centering
        \begin{minipage}[b]{0.592\linewidth}
            \centering
            \includegraphics[width=\linewidth]{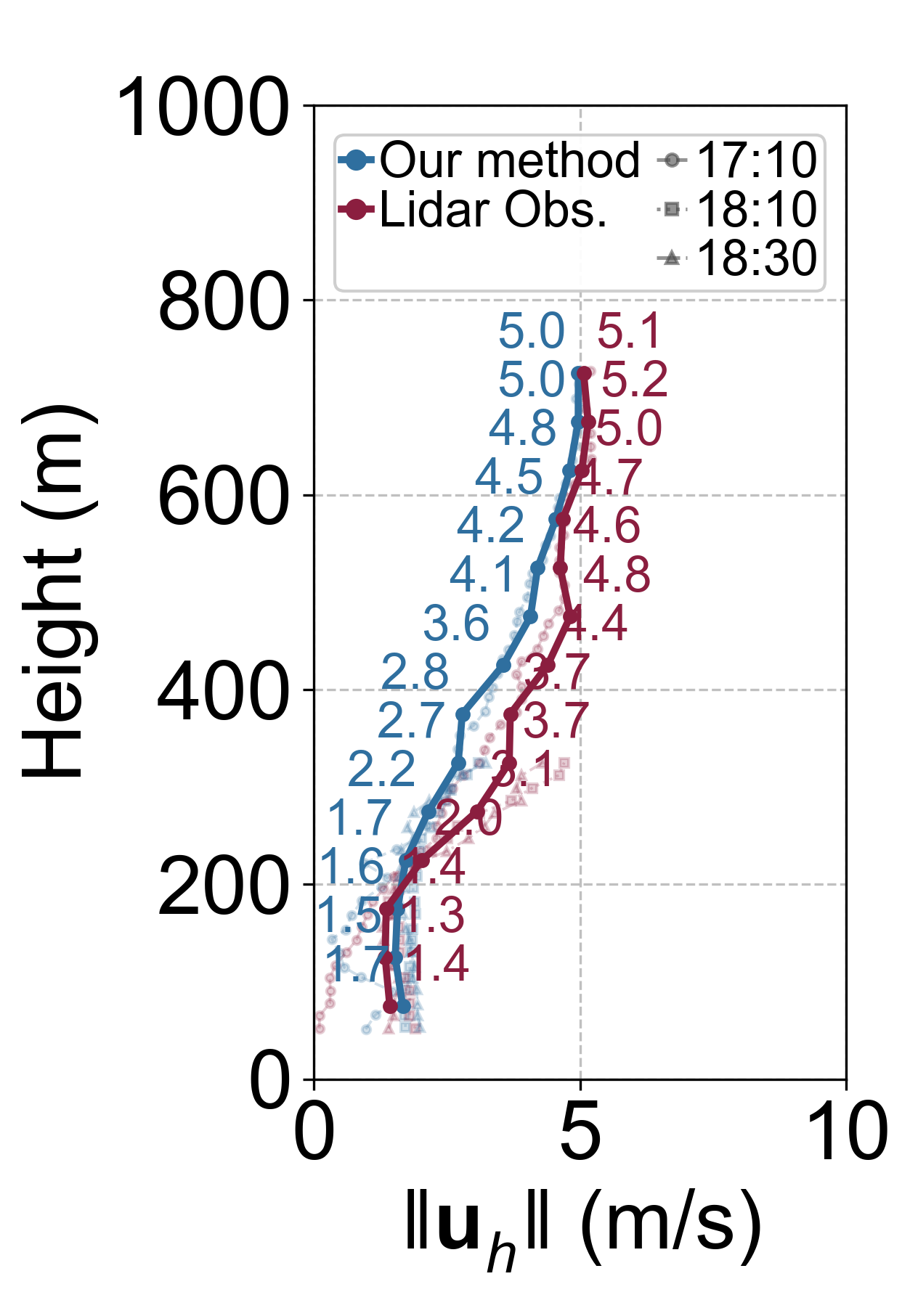}

            \par\smallskip
            \textup{(a)}
        \end{minipage}%
        \begin{minipage}[b]{0.408\linewidth}
            \centering
            \includegraphics[width=\linewidth,trim=0 0 93.5bp 0,clip]{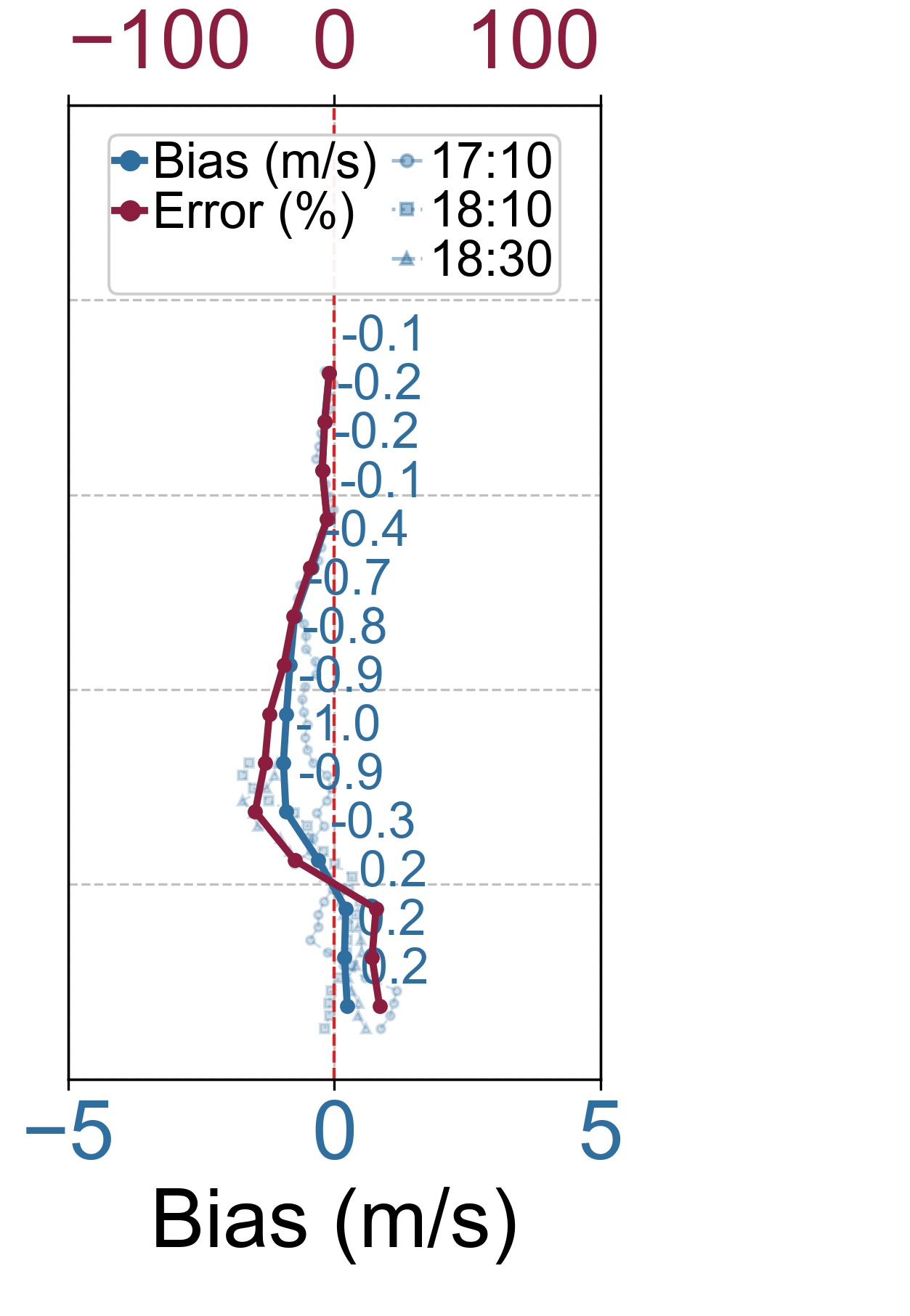}

            \par\smallskip
            \textup{(b)}
        \end{minipage}
    \end{minipage}

\caption{{Wind-speed profiles and error quantification for Case 1, validated against GAW111. (a) The 50-m bin-averaged profiles and the non-binned profiles at 17:10, 18:10, and 18:30; and (b) the corresponding bin-averaged bias and relative error, together with the non-binned bias at the three instants.}}
\label{fig:case1}
\end{figure}

For the validation of horizontal wind profiles, the time instants at 17:10, 18:10, and 18:30 are selected. The lidar measurement ceiling in this study is dynamically adjusted in real time by the quality-control procedure rather than prescribed as a fixed height, which is determined by the lidar manufacturer using the echo signal intensity, retrieval availability, the aggregated availability of instants within the time window as well as a series of uncertainty control. As a result, the valid height coverage can vary during the observation period, according to signal intensity, retrieval availability, and uncertainty constraints. This design is motivated by the operational requirement to support low-altitude flight operations by providing the highest possible reliable wind-field information, thereby preserving temporally continuous and trustworthy boundary input for near-real-time application. Accordingly, across the six lidar stations, the highest valid observation height available for analysis is 800~m at 17:10, 400~m at 18:10, and 400~m at 18:30. {Figure~\ref{fig:case1}(a) shows the quasi-steady instantaneous vertical profiles at the listed validation times together with the corresponding time-aggregated, 50-m bin-averaged curves, while Figure~\ref{fig:case1}(b) presents the associated bias and relative-error profiles.} The local turning points appearing in the non-binned profiles should therefore not be interpreted as plotting artifacts. They mainly reflect vertically localized flow distortions associated with building wakes, shear layers, and other small-scale vortical structures resolved by the simulation and partly retained in the lidar retrievals. {The bin-averaged comparison suppresses these fine-scale fluctuations to provide a clearer visualization of profile-scale agreement; the quantitative statistics reported below use the original pointwise pairs.}
Because the selected validation instants were screened to avoid abrupt wind-direction transitions, the nearby lidar stations are considered to be under the same background inflow regime at each instant with strong potential-flow tendency as indicated by Figure~\ref{fig:sectionsz}. Therefore, the inter-station differences discussed in wind profiles are mainly by local urban and terrain modulation rather than changes in the large-scale wind direction.

First, five sets of lidar observations are employed to construct a Kriging interpolation field (see Table~\ref{tab:2}), which is subsequently used as the boundary input for the model. The validation is conducted exclusively using data from the single lidar system GAW111. To prevent self-validation and the consequent artificial inflation of accuracy, no overlap exists between the lidar observations used for boundary construction and those used for validation in this case and in all subsequent computations.
As indicated by the Figure~\ref{fig:case1}(a), the profile validation at 17:10 demonstrates a relatively high level of accuracy. {Compared with the validating lidar observational data, the deviations remain below 0.5 m/s across most height levels.} It is additionally noteworthy that, although the input data comprise only boundary information and include only the wind speed at the upper boundary at this location, the combined effects of boundary data assimilation and CFD simulations successfully reconstruct the characteristic features of the wind profile. This finding indicates that, through sufficiently accurate CFD computations, the low-level wind field is reliably resolved and exhibits an inherent corrective capability.
{For the two time instants with lower computational heights, whose profiles are also shown in Figure~\ref{fig:case1}(a), the predicted wind speeds generally follow the vertical variations in the lidar profiles.} At 18:10, the solver successfully captures small-scale features of the low-level flow, thereby downscaling the assimilated input data. However, for both cases, substantial discrepancies are observed when compared with the lidar measurements, particularly above 200~m, where a strong vertical gradient is present. These discrepancies may originate from uncertainties inherent in the validating lidar observations, which in turn increase the overall uncertainty of the coupled computation–validation procedure.

{For clearer profile visualization while suppressing the influence of spatial fluctuations, successive 50~m altitude intervals are used for the bin-averaged curves. The same bin width is used for all remaining cases.}
{As observed in Figure~\ref{fig:case1}(a)--(b), good agreement is achieved in the ultra-low-altitude region below 200 m as well as near the upper boundary above 500 m.} In contrast, within the intermediate altitude range, the simulated results are slightly lower than the corresponding observations, inheriting the errors within this height range from the time instances characterized by the lower computational domain.
However, wind profile at only one location is used for validation, so we lack sufficient confidence to provide quantitative quality control metrics. This indicates that more lidar data should be reserved for independent validation rather than used to construct assimilation boundaries.

\begin{figure}[t]
    \centering

    \begin{minipage}[b]{0.5\linewidth}
        \centering
        \begin{minipage}[b]{0.592\linewidth}
            \centering
            \includegraphics[width=\linewidth]{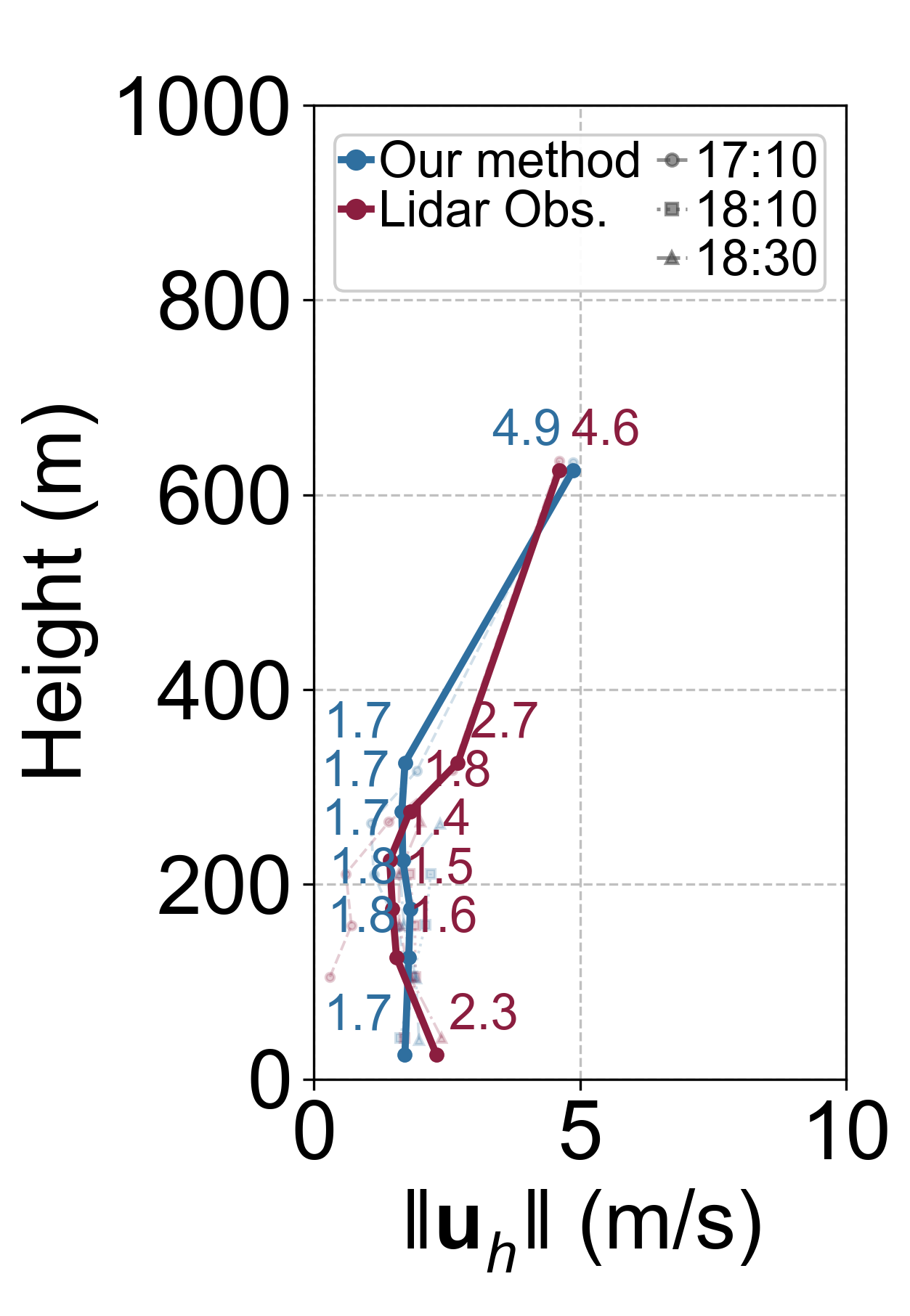}
        \end{minipage}%
        \begin{minipage}[b]{0.408\linewidth}
            \centering
            \includegraphics[width=\linewidth,trim=0 0 93.5bp 0,clip]{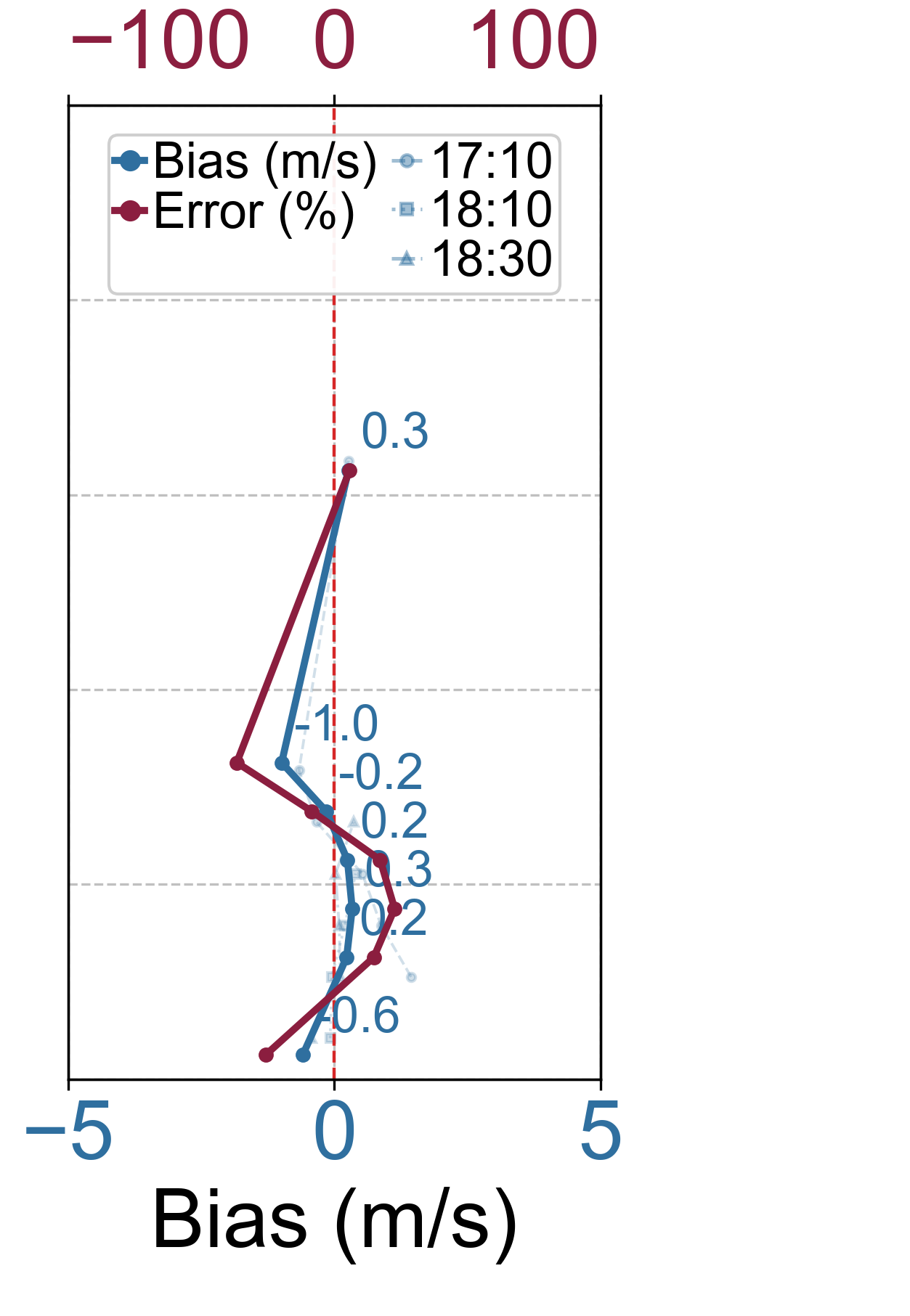}
        \end{minipage}

        \par\smallskip
        \textup{(a)}
    \end{minipage}%
    \begin{minipage}[b]{0.5\linewidth}
        \centering
        \begin{minipage}[b]{0.592\linewidth}
            \centering
            \includegraphics[width=\linewidth]{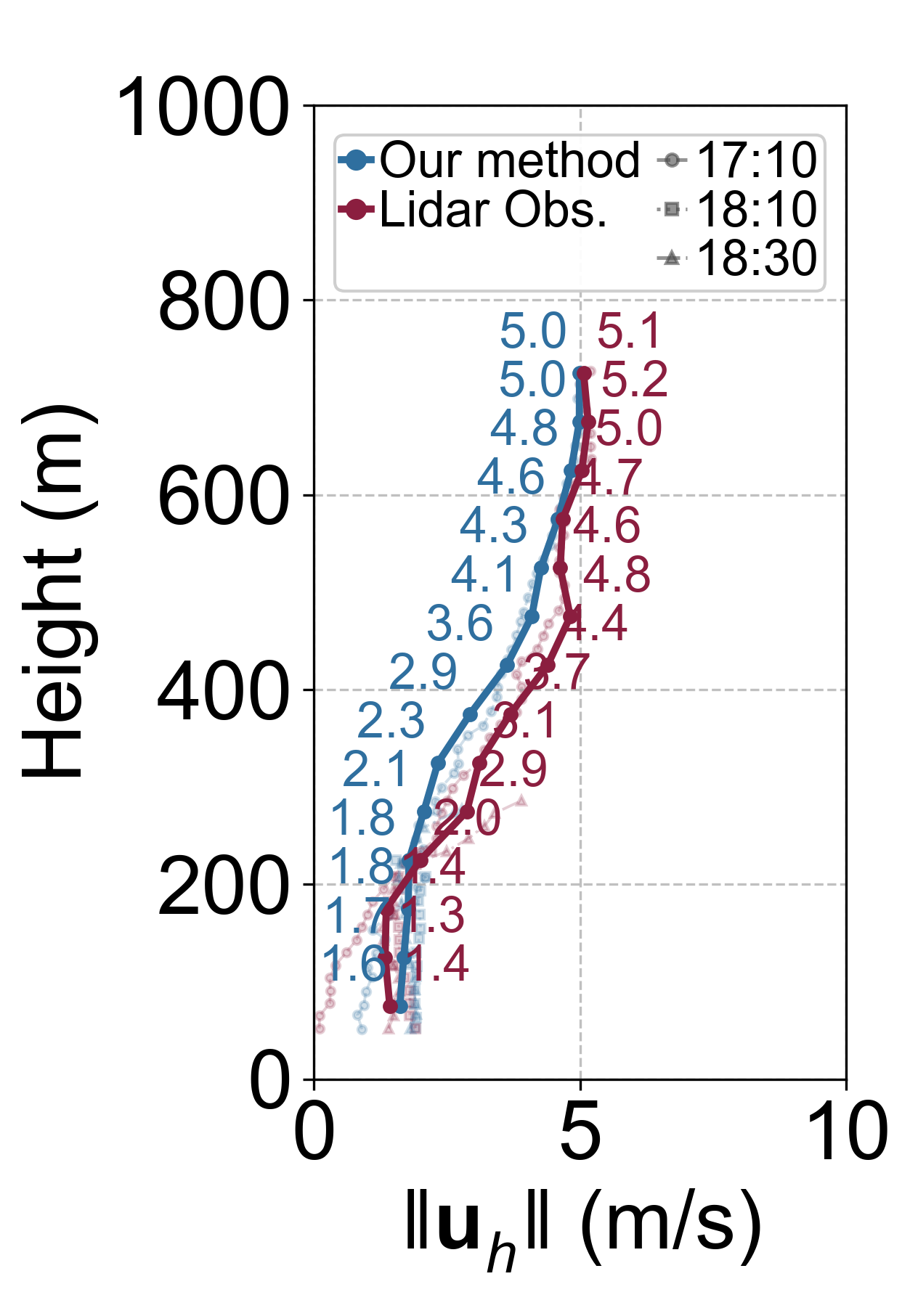}
        \end{minipage}%
        \begin{minipage}[b]{0.408\linewidth}
            \centering
            \includegraphics[width=\linewidth,trim=0 0 93.5bp 0,clip]{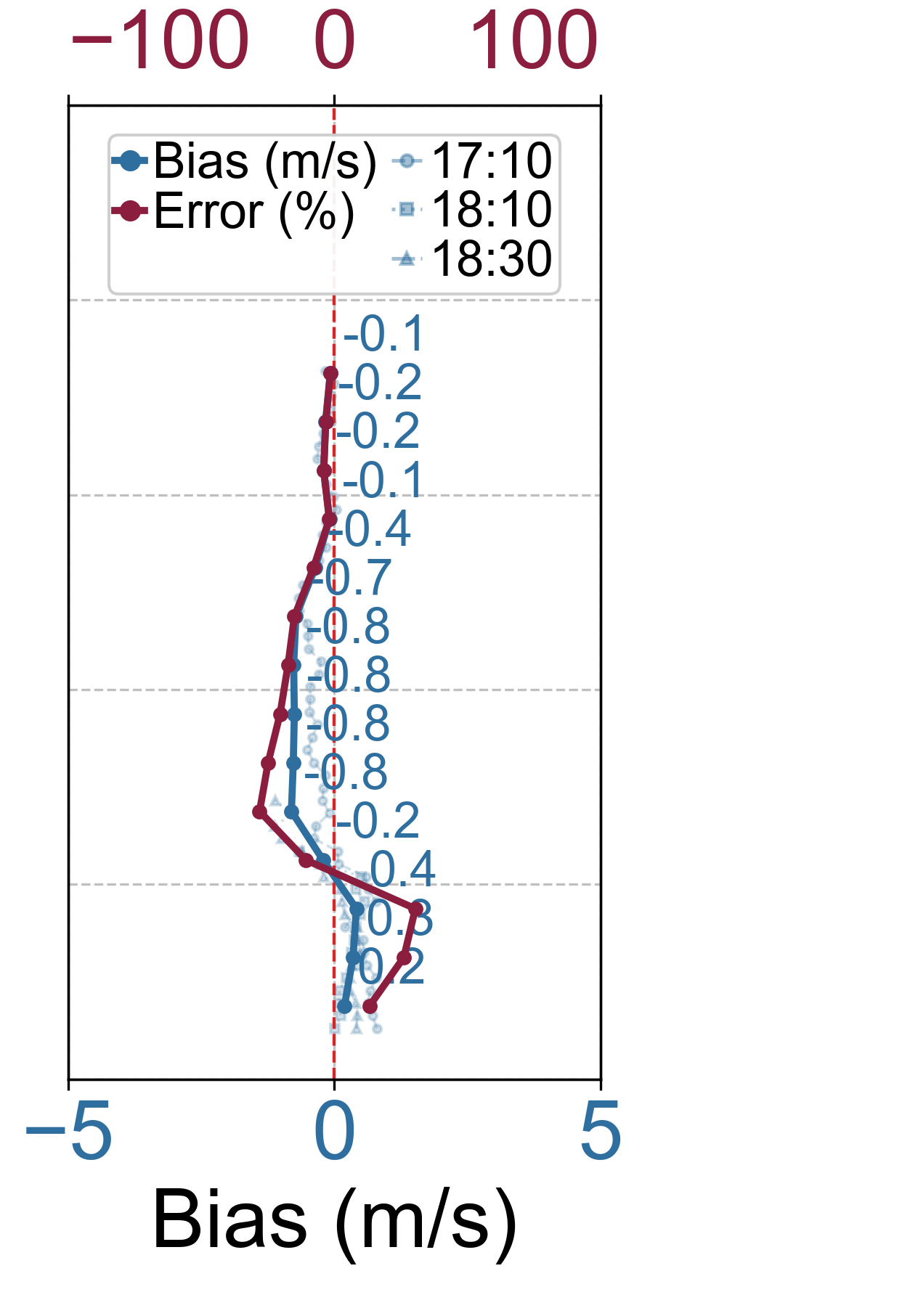}
        \end{minipage}

        \par\smallskip
        \textup{(b)}
    \end{minipage}
\caption{{Wind-speed profiles and error quantification for Case 2, validated against (a) GAW110 and (b) GAW111. For each station, the left plot presents the 50-m bin-averaged and non-binned profiles, and the right plot presents the corresponding error quantification.}}
\label{fig:case2}
\end{figure}

\begin{figure}[t]
    \centering

    \begin{minipage}[b]{0.5\linewidth}
        \centering
        \begin{minipage}[b]{0.592\linewidth}
            \centering
            \includegraphics[width=\linewidth]{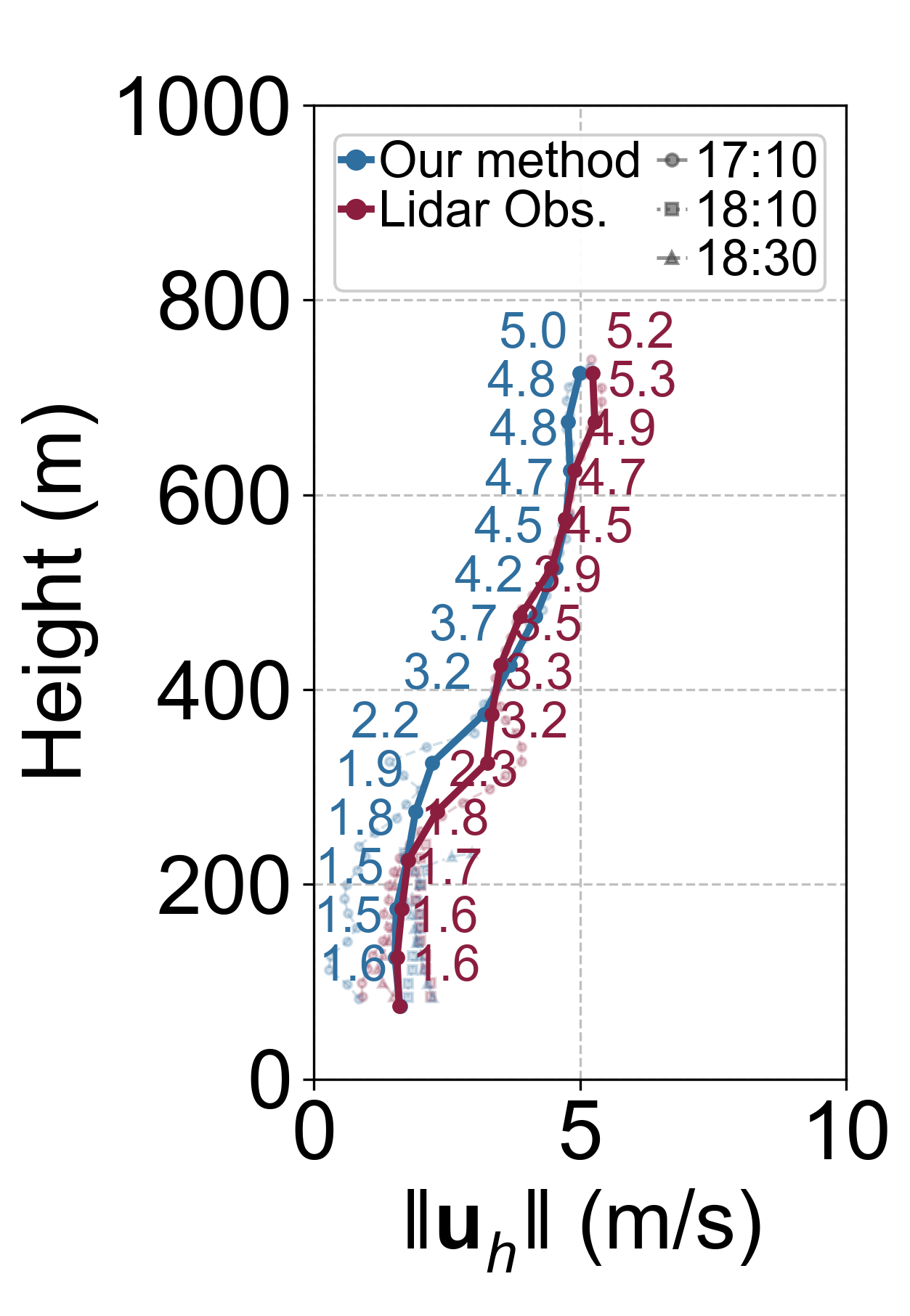}
        \end{minipage}%
        \begin{minipage}[b]{0.408\linewidth}
            \centering
            \includegraphics[width=\linewidth,trim=0 0 93.5bp 0,clip]{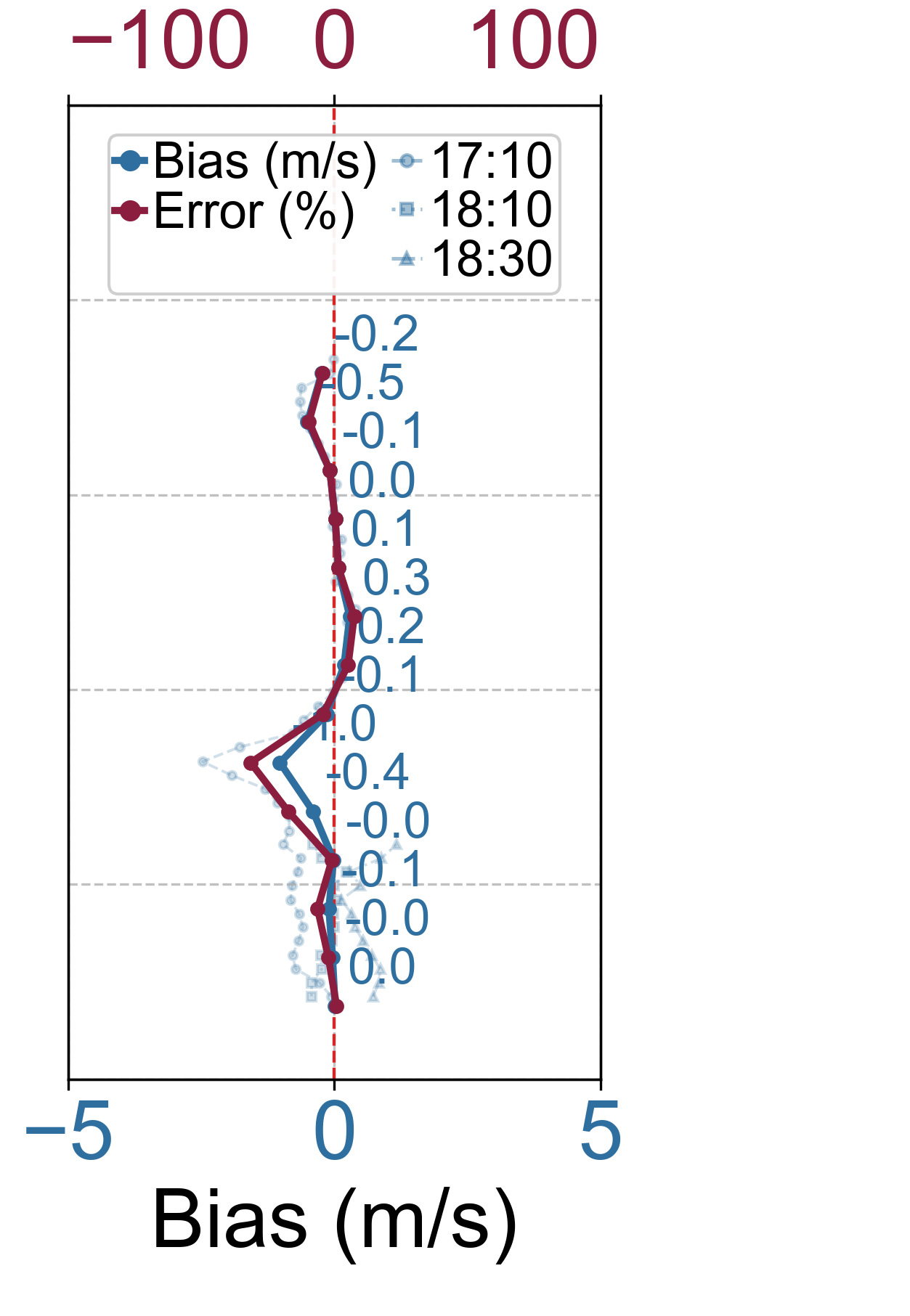}
        \end{minipage}

        \par\smallskip
        ~~~\textup{(a)}
    \end{minipage}%
    \begin{minipage}[b]{0.5\linewidth}
        \centering
        \begin{minipage}[b]{0.592\linewidth}
            \centering
            \includegraphics[width=\linewidth]{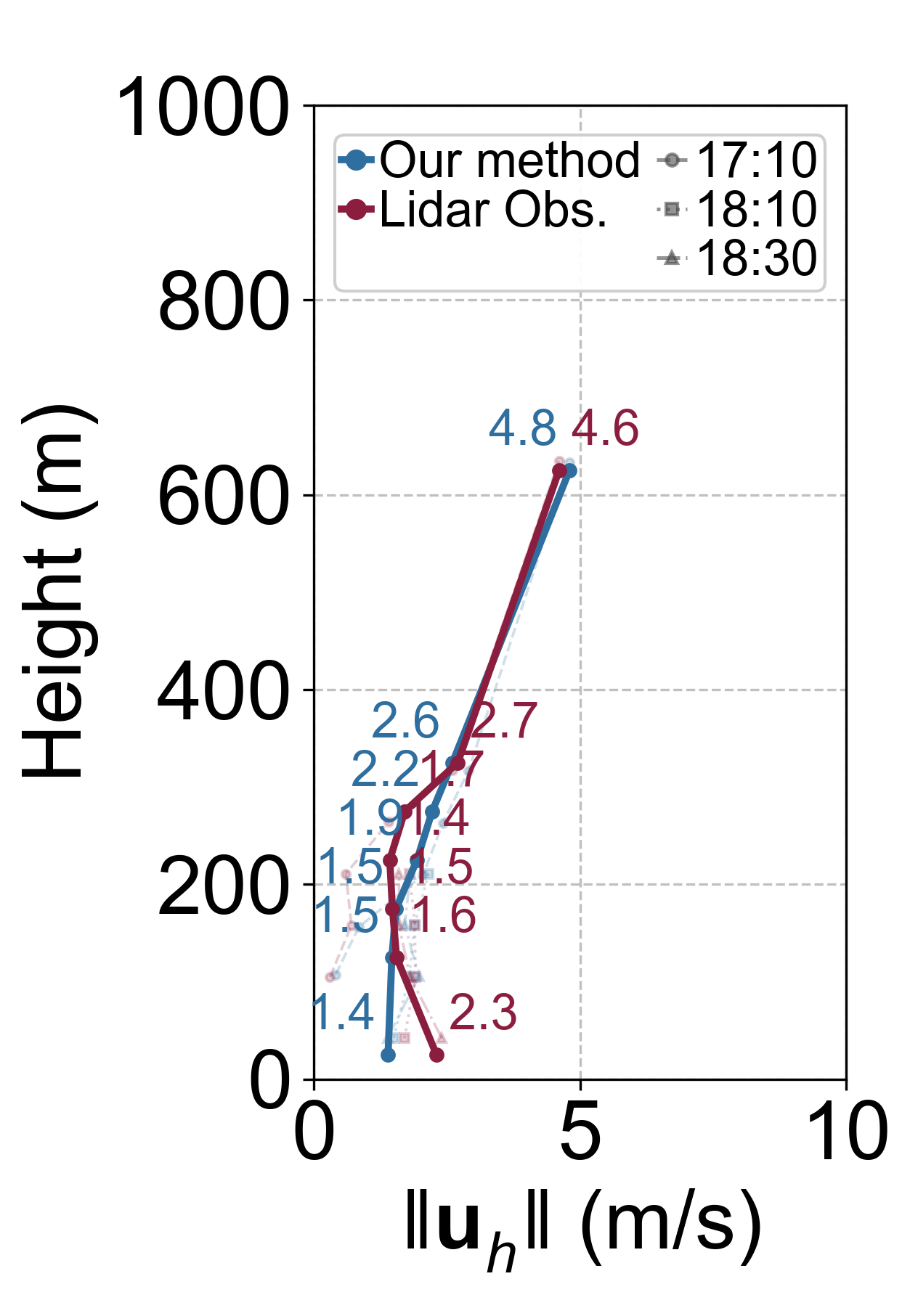}
        \end{minipage}%
        \begin{minipage}[b]{0.408\linewidth}
            \centering
            \includegraphics[width=\linewidth,trim=0 0 93.5bp 0,clip]{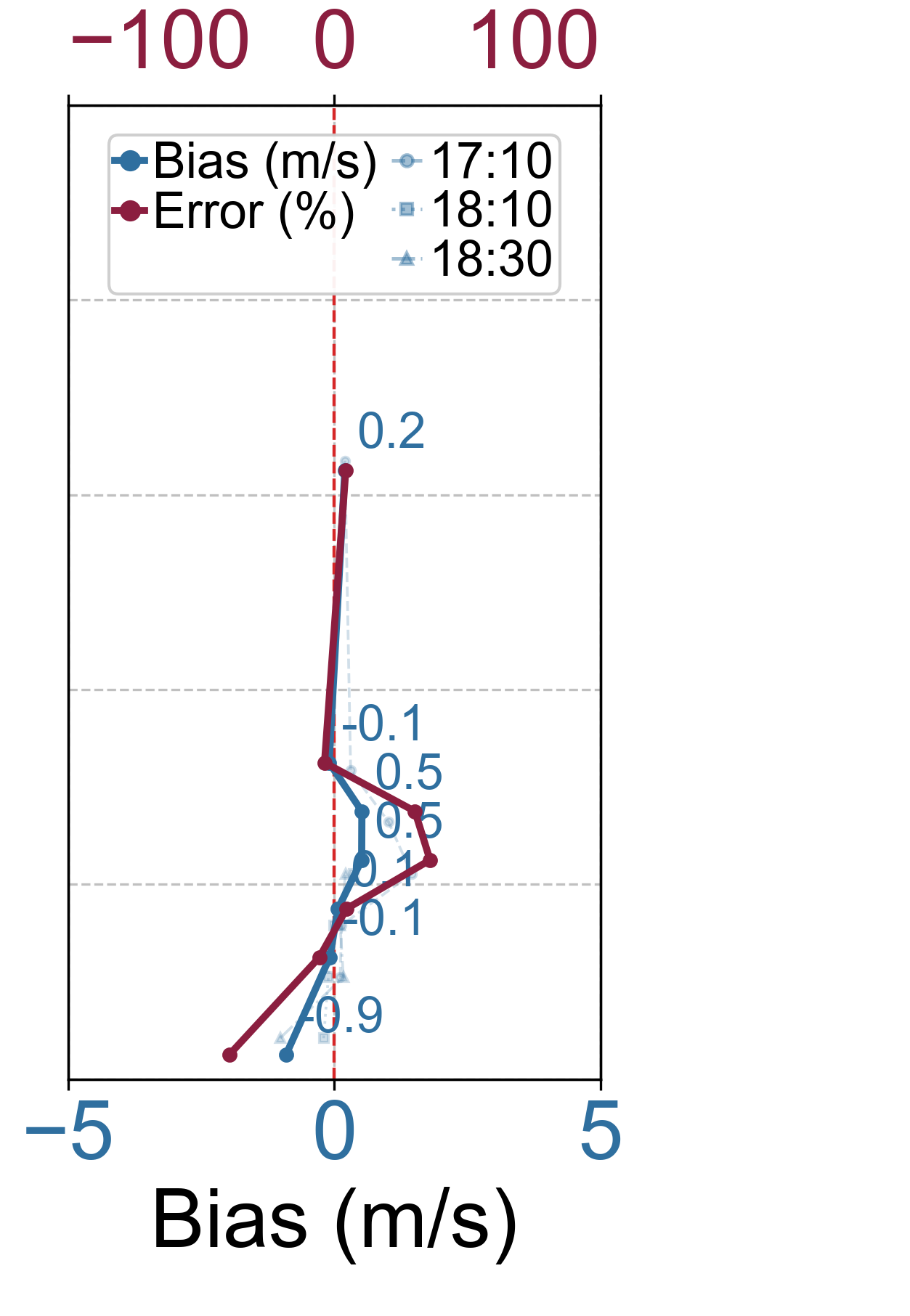}
        \end{minipage}

        \par\smallskip
        ~~~\textup{(b)}
    \end{minipage}

    \par\medskip

    \begin{minipage}[b]{0.5\linewidth}
        \centering
        \begin{minipage}[b]{0.592\linewidth}
            \centering
            \includegraphics[width=\linewidth]{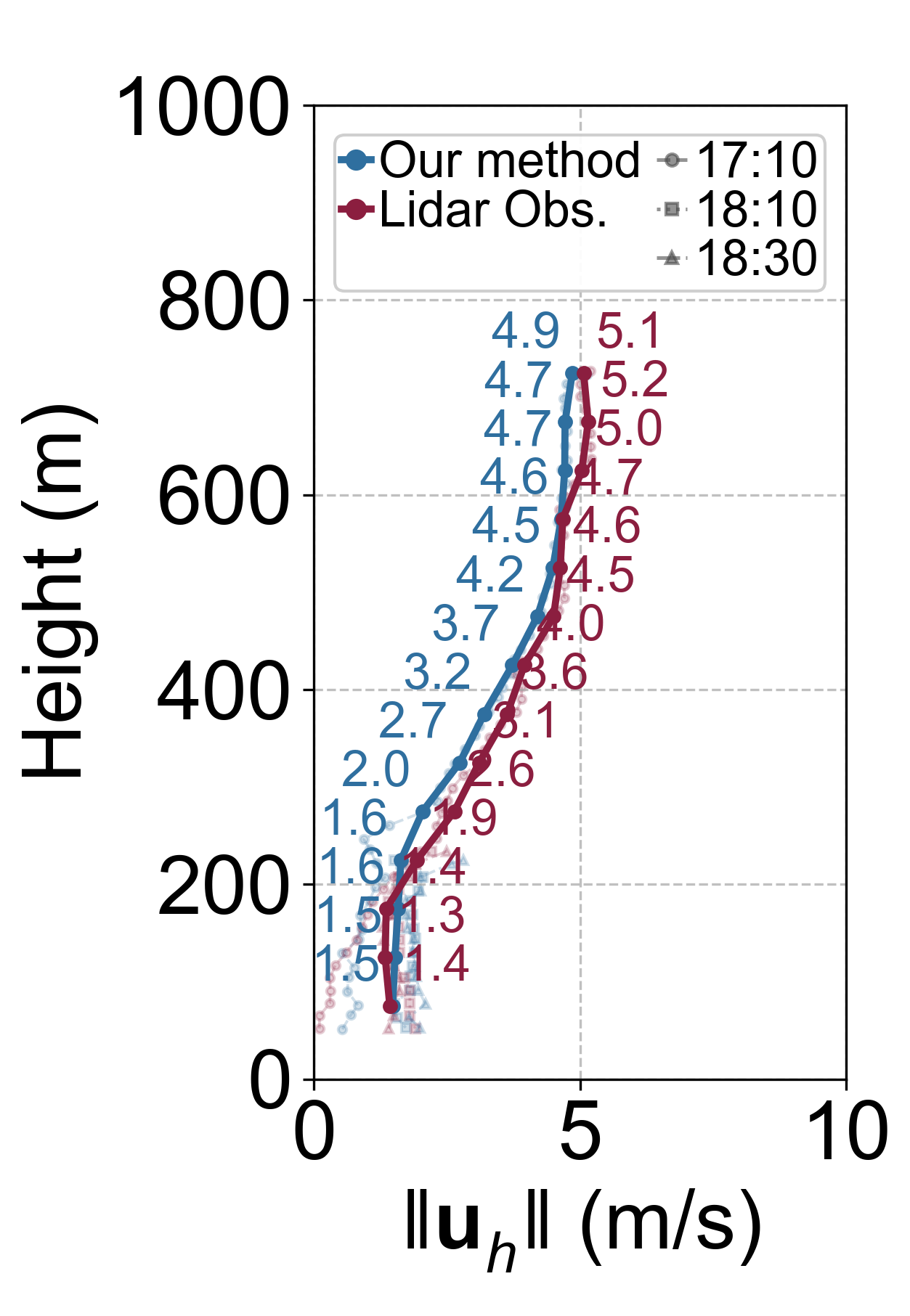}
        \end{minipage}%
        \begin{minipage}[b]{0.408\linewidth}
            \centering
            \includegraphics[width=\linewidth,trim=0 0 93.5bp 0,clip]{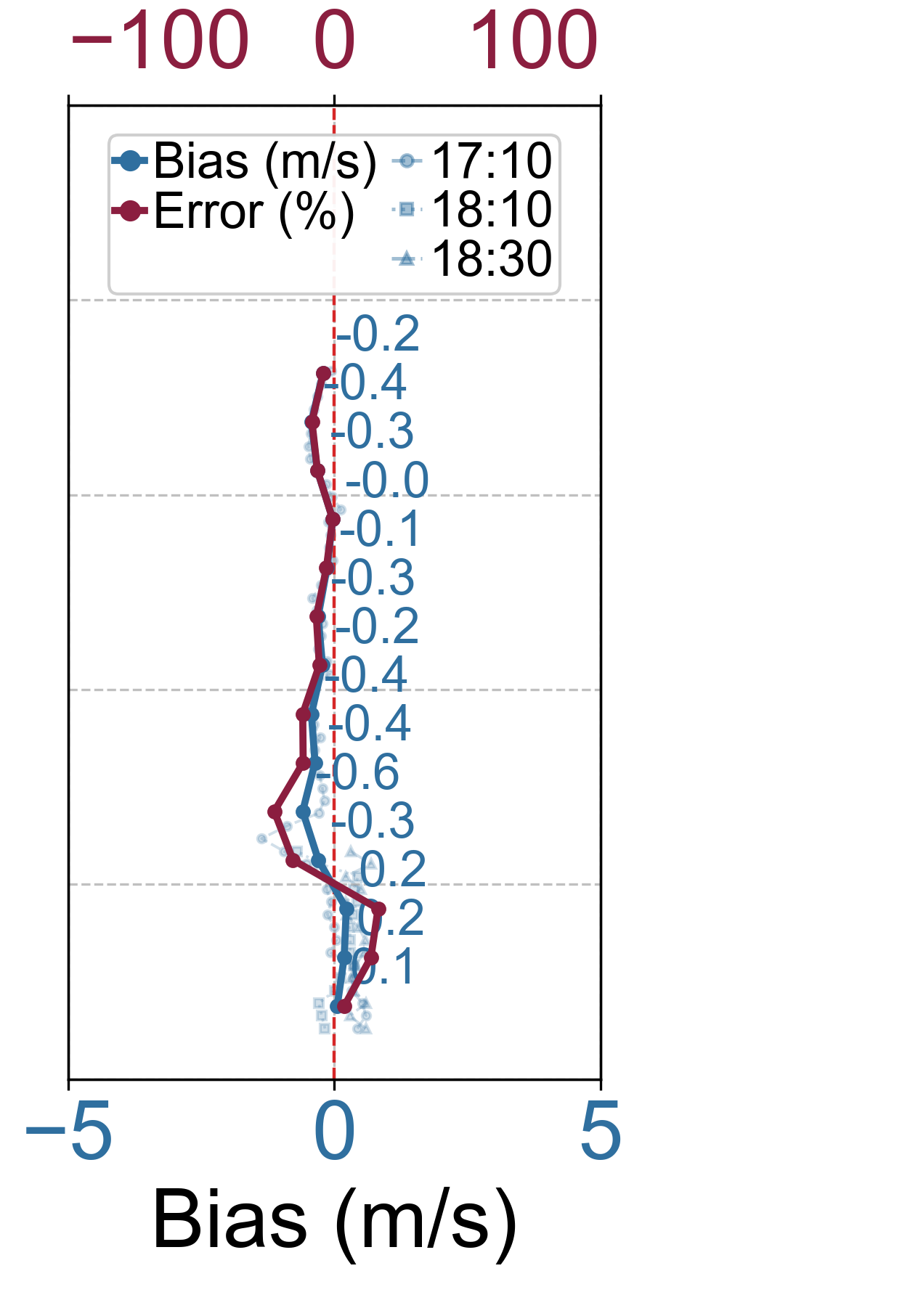}
        \end{minipage}

        \par\smallskip
        ~~~\textup{(c)}
    \end{minipage}

\caption{{Wind-speed profiles and error quantification for Case 3, validated against (a) GAW105, (b) GAW110, and (c) GAW111. For each station, the left plot presents the 50-m bin-averaged and non-binned profiles, and the right plot presents the corresponding error quantification.}}
\label{fig:case3}
\end{figure}

The bias characteristics of Case 2 (Figure~\ref{fig:case2}) are similar to those of Case 1, while incorporating a new set of validation locations. The maximum deviation is likewise observed at an altitude of approximately 400 m, and it is introduced near the upper boundary during periods when the computational domain height is relatively low. This behavior is attributed to a slight degradation in measurement accuracy of individual lidar systems near the top of the observational line of sight.

As for Case 3 shown in Figure~\ref{fig:case3}, lidar data are further allocated to observation processing from data assimilation, and the overall verification metrics exhibit additional improvement. In comparison between the GAW105 wind profiler and the model simulations, the absolute errors are further reduced to within 0.3 m/s, with only the 300–400 m altitude range exhibiting a bias reaching approximately 0.5 m/s. For GAW110, except for a slightly larger discrepancy at the lowest observation level of the Doppler wind lidar, errors at all other heights remain below 0.5 m/s.
Observed wind speed distribution by GAW110 in Figure~\ref{fig:case3}(b) indicates that it records a wind speed of 2.3 m/s at the lowest level (approximately 50 m above ground), which decreases to 1.5 m/s with increasing height to 150 m. This vertical variation differs from the trends observed by other instruments. A preliminary assessment suggests that, at this particular instant, the observational instrument might have been influenced by external environmental factors or other interfering elements.
Overall, all three validation sites demonstrate a satisfactory level of agreement. This consistency indicates that, in practical low-level wind field support events, low-level wind field computation and validation can be conducted with a sufficient level of confidence.

\begin{figure}[t]
    \centering

    \begin{minipage}[b]{0.5\linewidth}
        \centering
        \begin{minipage}[b]{0.592\linewidth}
            \centering
            \includegraphics[width=\linewidth]{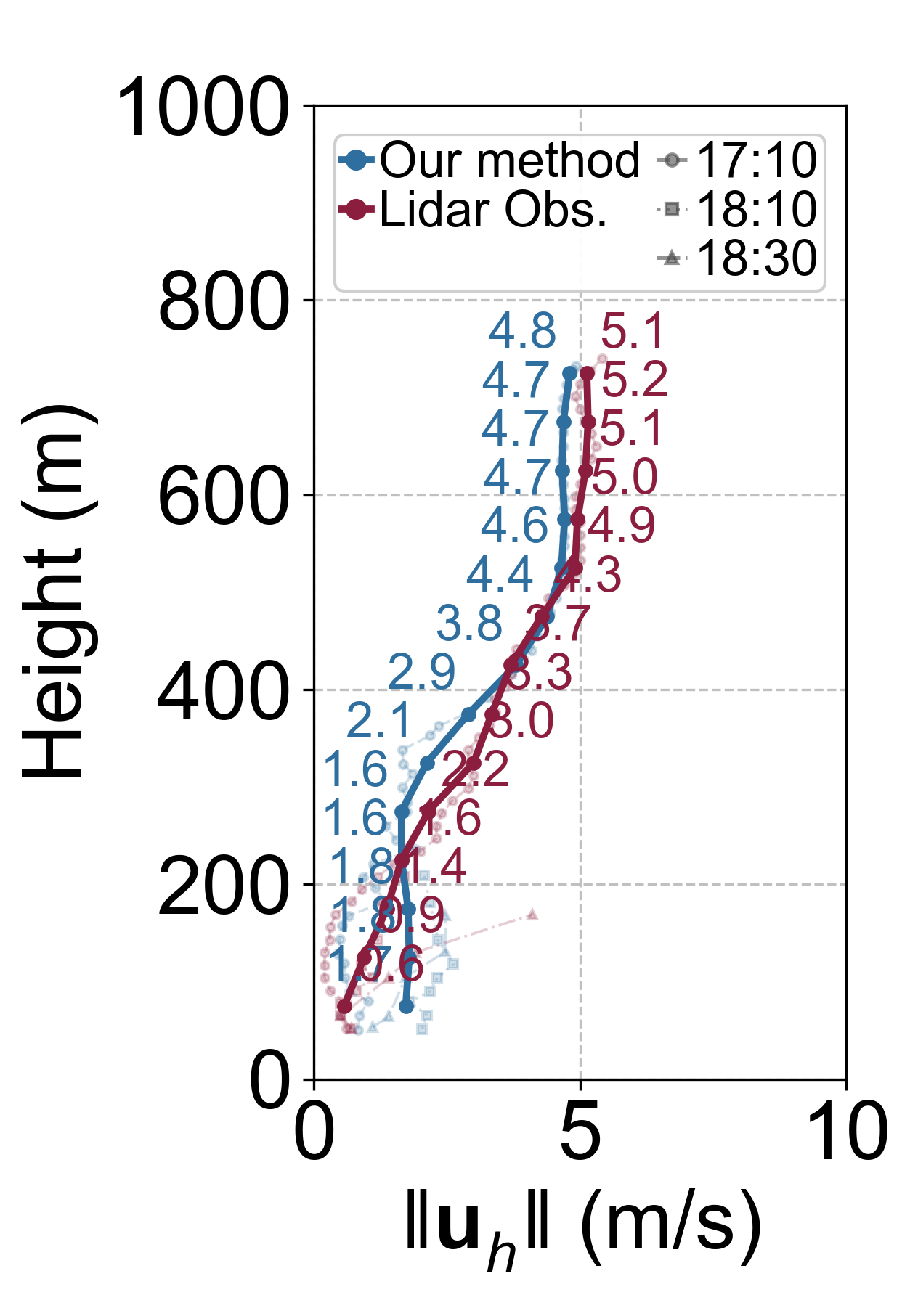}
        \end{minipage}%
        \begin{minipage}[b]{0.408\linewidth}
            \centering
            \includegraphics[width=\linewidth,trim=0 0 93.5bp 0,clip]{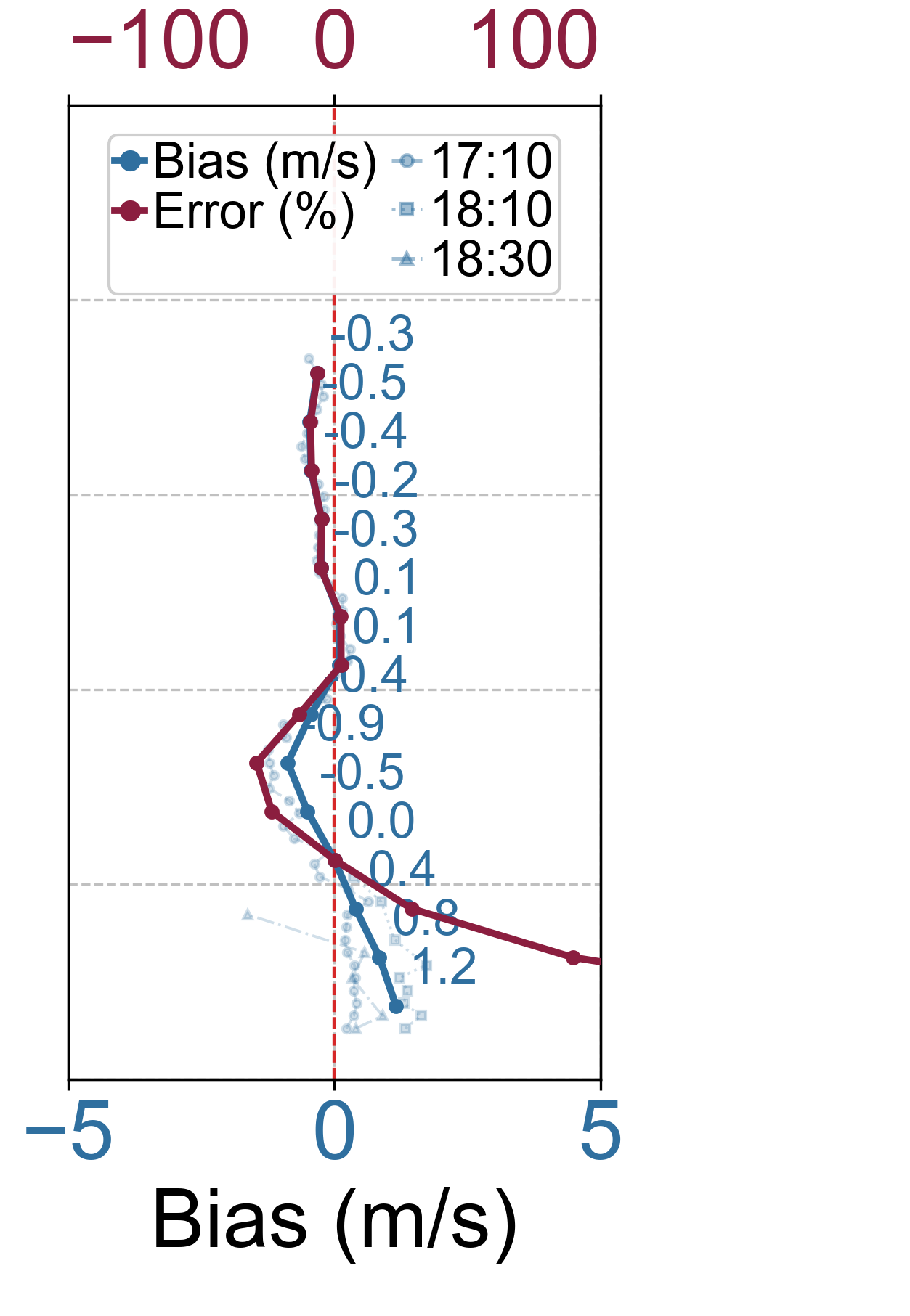}
        \end{minipage}

        \par\smallskip
        ~~~\textup{(a)}
    \end{minipage}%
    \begin{minipage}[b]{0.5\linewidth}
        \centering
        \begin{minipage}[b]{0.592\linewidth}
            \centering
            \includegraphics[width=\linewidth]{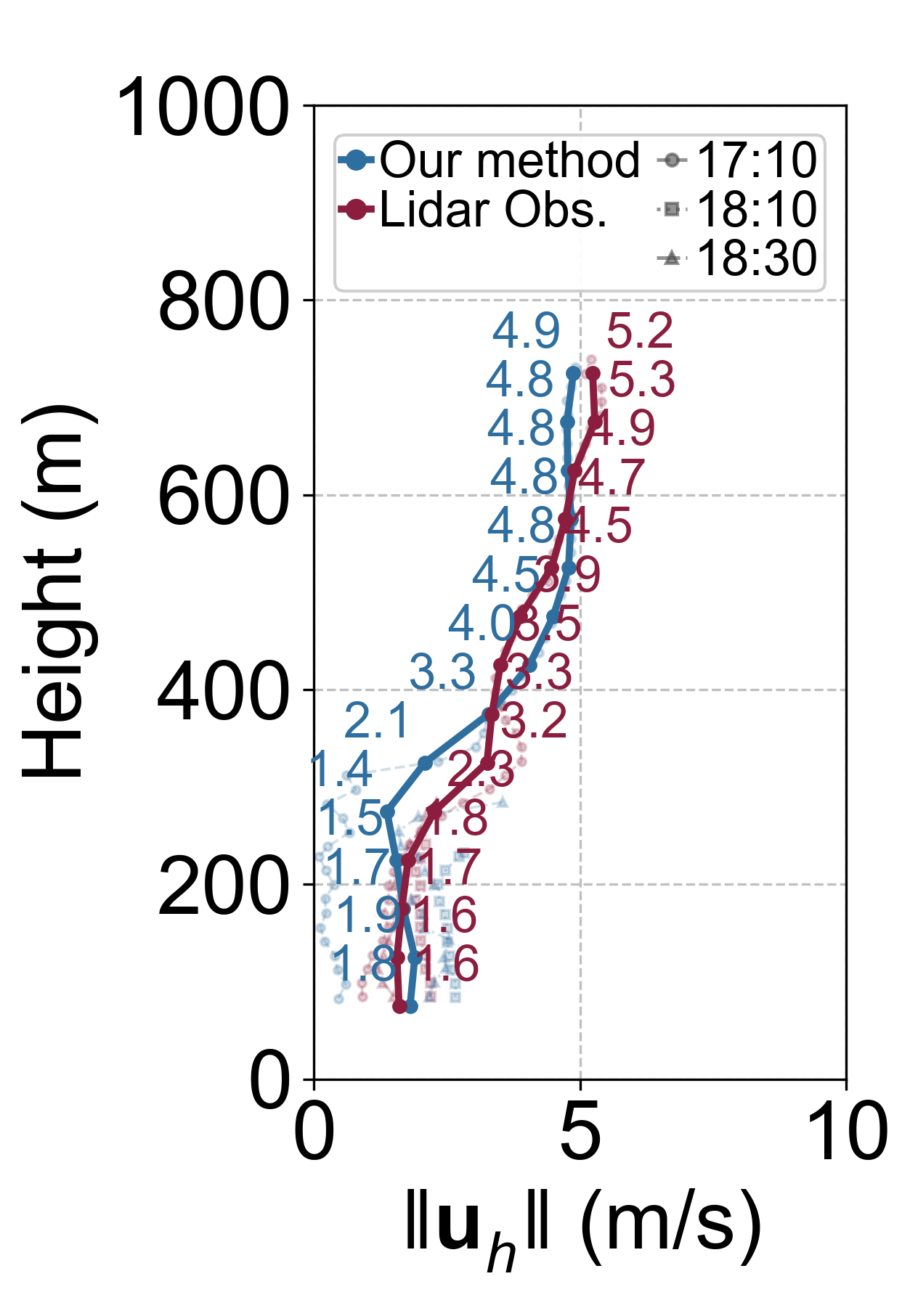}
        \end{minipage}%
        \begin{minipage}[b]{0.408\linewidth}
            \centering
            \includegraphics[width=\linewidth,trim=0 0 93.5bp 0,clip]{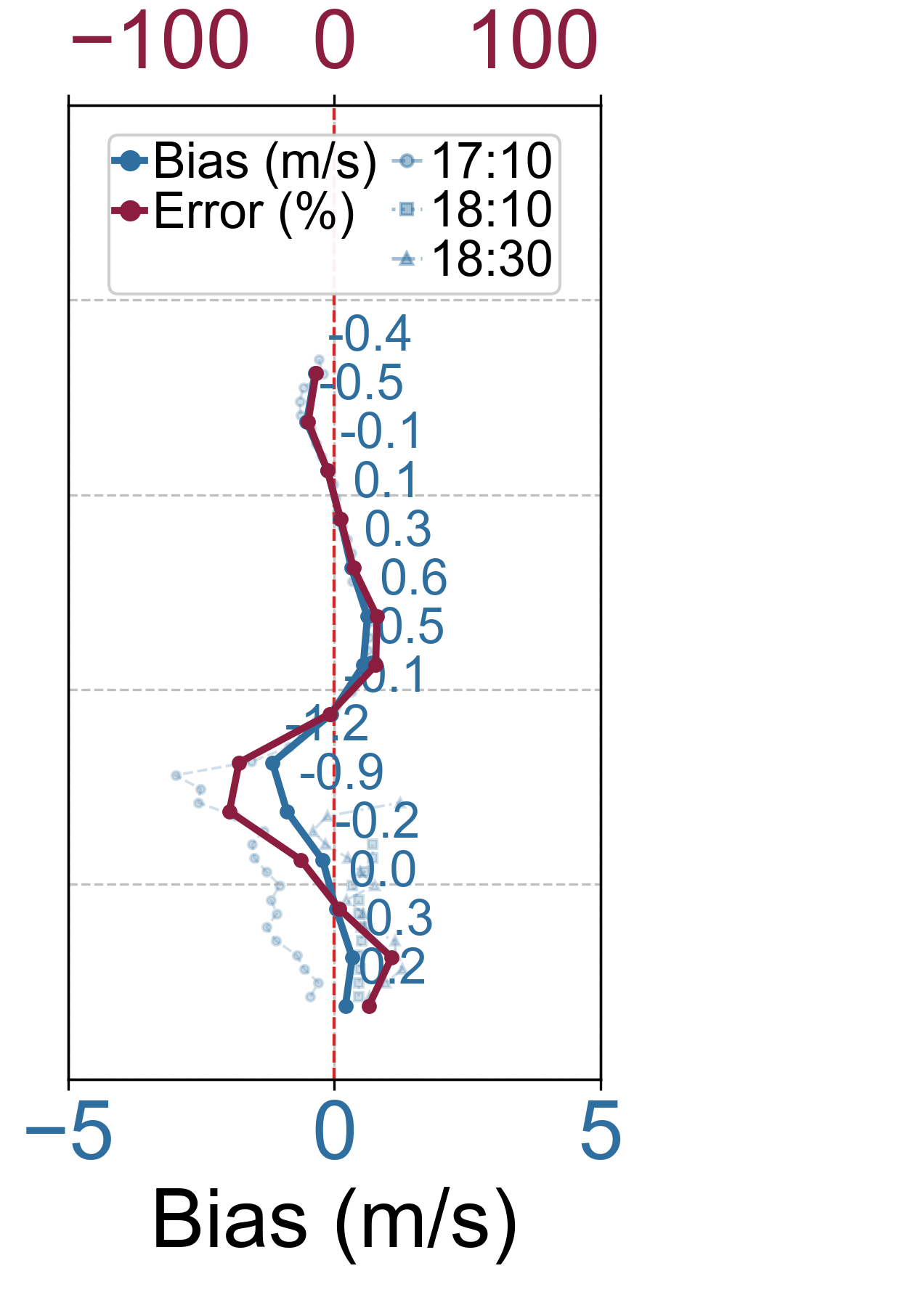}
        \end{minipage}

        \par\smallskip
        ~~~\textup{(b)}
    \end{minipage}

    \par\medskip

    \begin{minipage}[b]{0.5\linewidth}
        \centering
        \begin{minipage}[b]{0.592\linewidth}
            \centering
            \includegraphics[width=\linewidth]{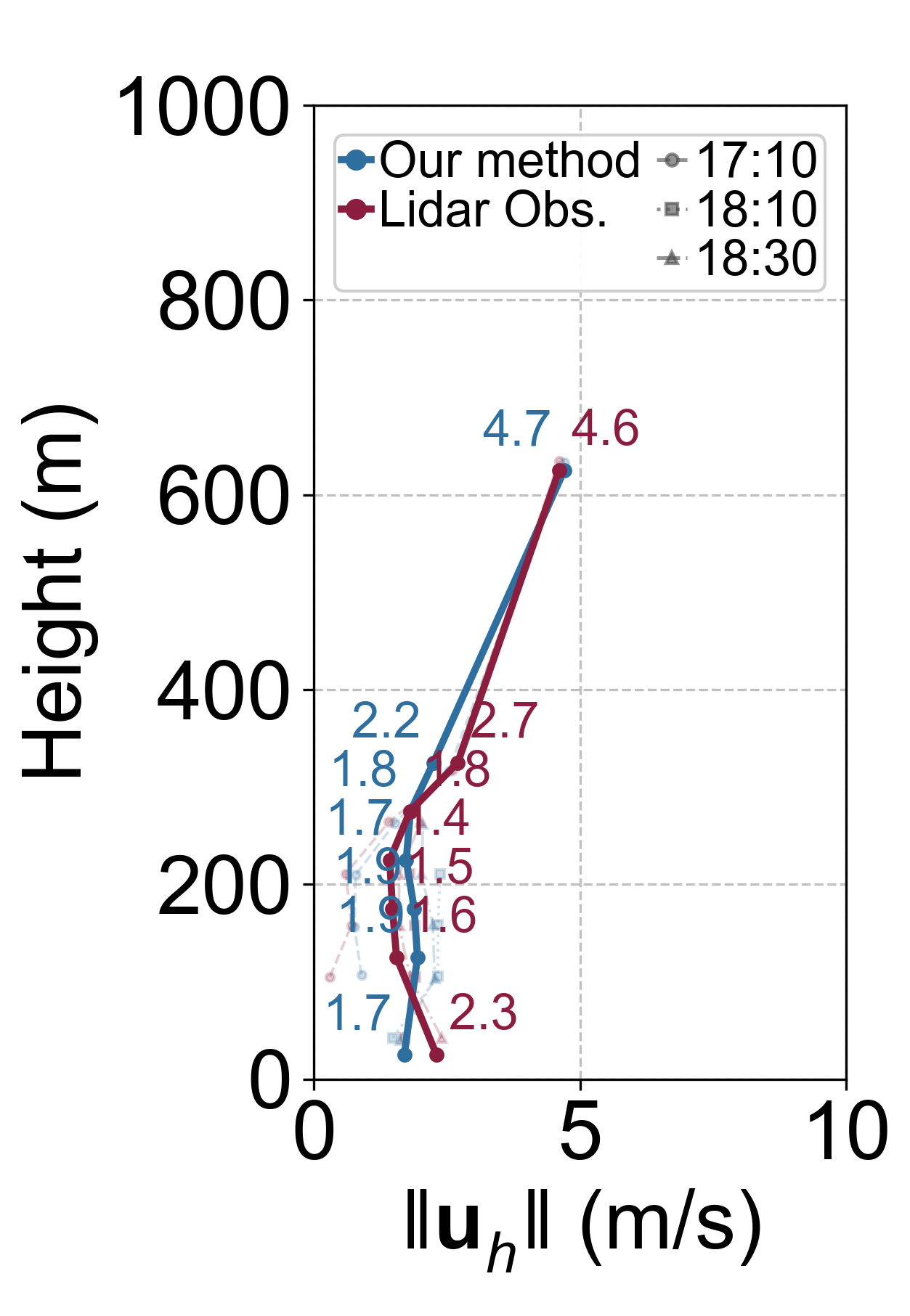}
        \end{minipage}%
        \begin{minipage}[b]{0.408\linewidth}
            \centering
            \includegraphics[width=\linewidth,trim=0 0 93.5bp 0,clip]{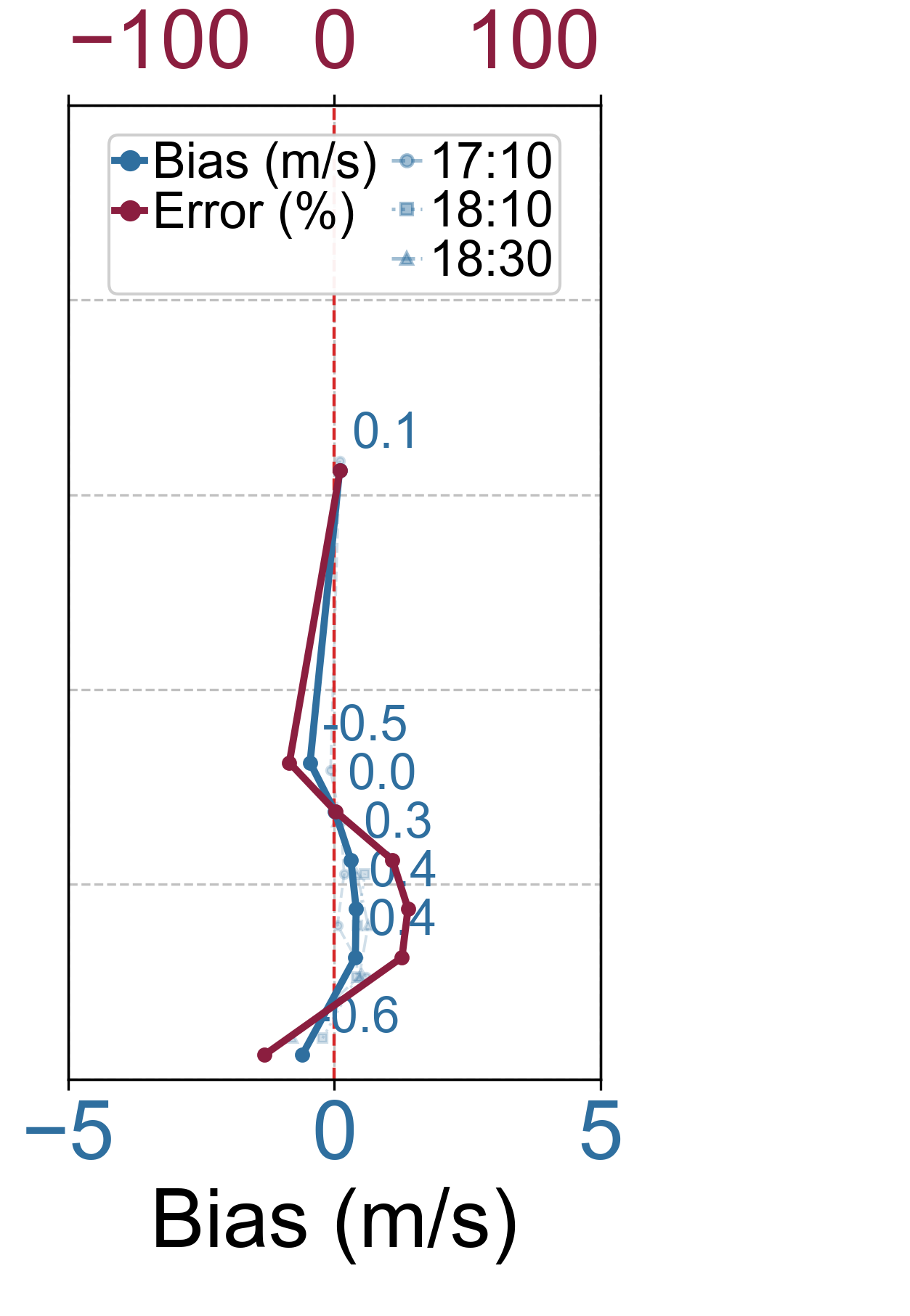}
        \end{minipage}

        \par\smallskip
        ~~~\textup{(c)}
    \end{minipage}%
    \begin{minipage}[b]{0.5\linewidth}
        \centering
        \begin{minipage}[b]{0.592\linewidth}
            \centering
            \includegraphics[width=\linewidth]{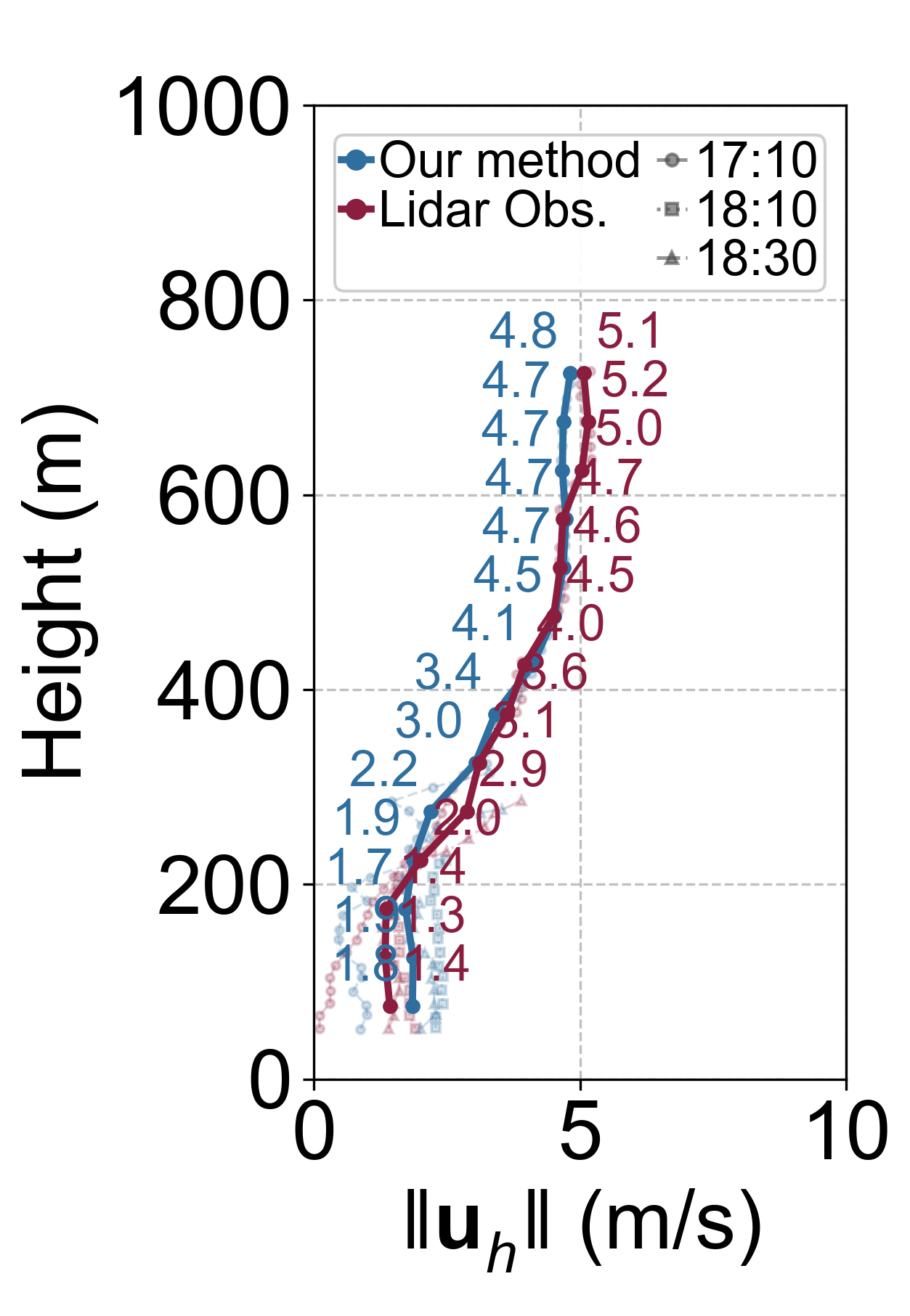}
        \end{minipage}%
        \begin{minipage}[b]{0.408\linewidth}
            \centering
            \includegraphics[width=\linewidth,trim=0 0 93.5bp 0,clip]{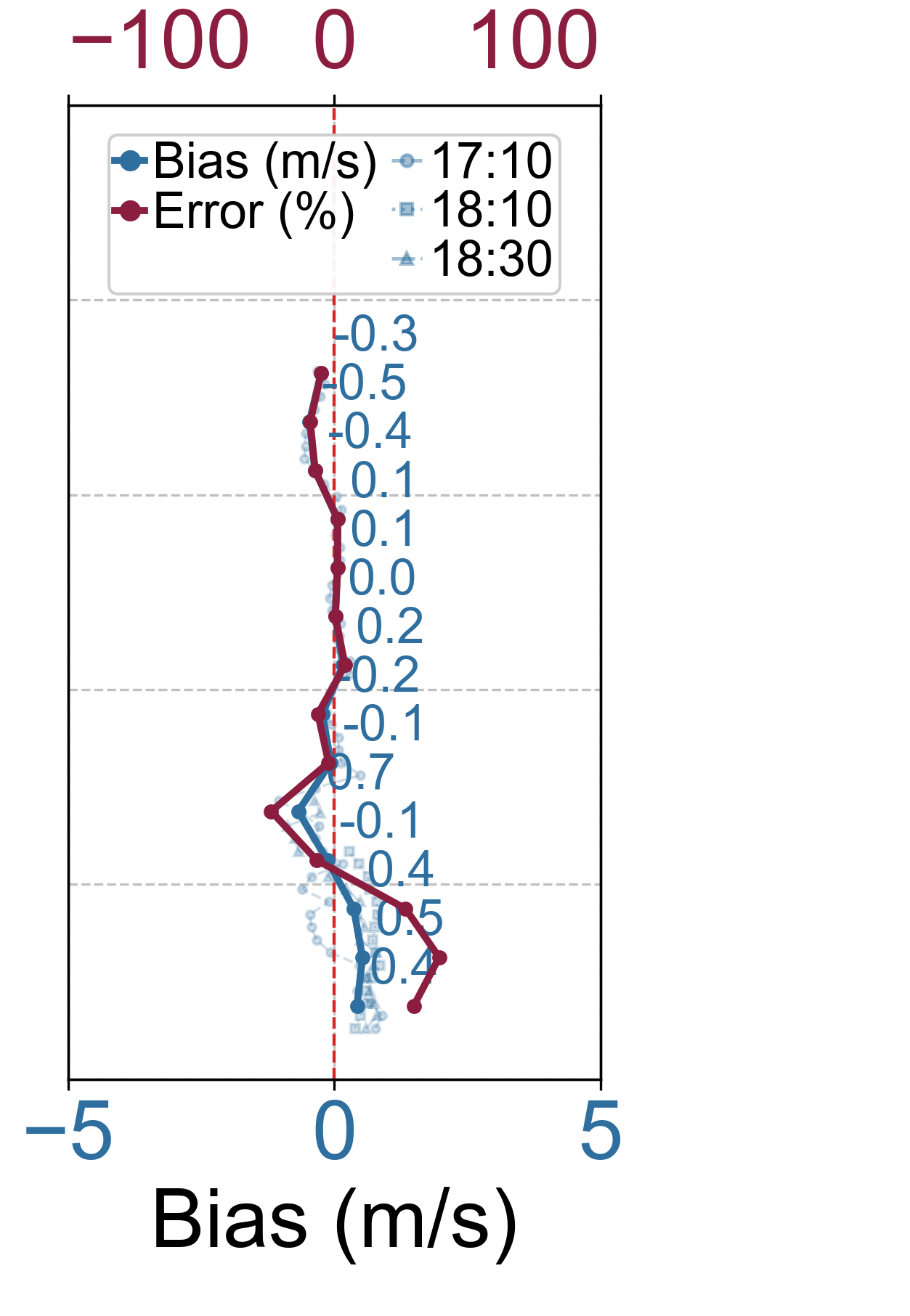}
        \end{minipage}

        \par\smallskip
        ~~~\textup{(d)}
    \end{minipage}

\caption{{Wind-speed profiles and error quantification for Case 4, validated against (a) GAW104, (b) GAW105, (c) GAW110, and (d) GAW111. For each station, the left plot presents the 50-m bin-averaged and non-binned profiles, and the right plot presents the corresponding error quantification.}}
\label{fig:case4}
\end{figure}

Case 4 further reduces the number of inputs to 2 lidars, while simultaneously expanding the spatial extent of the evaluation domain. Validation is conducted over four locations, including GAW104, GAW105, GAW110, and GAW111. The results shown in Figure~\ref{fig:case4} demonstrate that the simulations in all four regions are able to stably reproduce the vertical variation trends of the wind field.
However, compared to Case 3, both the mean and maximum values of the overall error have increased slightly. This indicates that when observations from the two radars are used as model inputs, there is insufficient boundary information, leading to a decline in the accuracy of the simulation results.
Although this validation method provides more comparative locations and attempts to reduce uncertainty, the accuracy of the simulation itself is compromised due to the reduction in effective data assimilation.

Among the four cases, Case 3 exhibits the highest degree of consistency in the validation results while providing enough validation locations. Therefore, an appropriate partitioning of observational data into subsets used for boundary-condition construction and for independent validation can effectively reduce the uncertainty in low-level wind-field prediction. Nevertheless, this does not necessarily imply an improvement in the absolute accuracy of the simulated results.



{To avoid inflating the quantitative assessment through profile averaging, all statistics reported below are computed from the original paired lidar--simulation wind-speed points before any vertical or temporal averaging. For a case containing }\mathcolorbox{yellow}{N}{ paired points, the pointwise error is }\mathcolorbox{yellow}{e_i=\hat{y}_i-y_i}{, where }\mathcolorbox{yellow}{y_i}{ is the observed horizontal wind speed and }\mathcolorbox{yellow}{\hat{y}_i}{ is the corresponding simulated value. The dimensional error metrics are}
\begin{equation}
\text{\mathcolorbox{yellow}{\displaystyle
\mathrm{MAE}=\frac{1}{N}\sum_{i=1}^{N}|e_i|,\qquad
\mathrm{RMSE}=\sqrt{\frac{1}{N}\sum_{i=1}^{N}e_i^2},\qquad
\mathrm{Bias}=\frac{1}{N}\sum_{i=1}^{N}e_i.}}
\end{equation}
{The two coefficients of determination are recomputed from the same pointwise pairs as}
\begin{equation}
\text{\mathcolorbox{yellow}{\displaystyle
R_1^2 = 1-\frac{\sum_{i=1}^{N}(y_i-\hat{y}_i)^2}{\sum_{i=1}^{N}(y_i-\bar{y})^2},}}
\end{equation}
\begin{equation}
\text{\mathcolorbox{yellow}{\displaystyle
R_2^2=r^2,\qquad
r=\frac{\sum_{i=1}^{N}(y_i-\bar{y})(\hat{y}_i-\bar{\hat{y}})}
{\sqrt{\sum_{i=1}^{N}(y_i-\bar{y})^2}\sqrt{\sum_{i=1}^{N}(\hat{y}_i-\bar{\hat{y}})^2}}.}}
\end{equation}
{Here, }\mathcolorbox{yellow}{\bar{y}}{ and }\mathcolorbox{yellow}{\bar{\hat{y}}}{ are the means of the observed and simulated pointwise samples, respectively. }\mathcolorbox{yellow}{R_1^2}{ measures the reduction in residual variance relative to the observed variance, whereas }\mathcolorbox{yellow}{R_2^2}{ is the squared Pearson correlation coefficient.}

{Table~\ref{tab:case_r2} reports the sample count and pointwise statistics for each case. The four cases contain 611 model--observation pairs in total; because a withheld lidar profile can be evaluated under more than one assimilation case, }\mathcolorbox{yellow}{N}{ denotes the number of paired comparisons within each case rather than the number of unique lidar observations. The MAE ranges from 0.383 to 0.534~m/s, and the RMSE ranges from 0.466 to 0.696~m/s. Table~\ref{tab:height_errors} provides the height-resolved MAE and RMSE in 100-m height bands. The error is calculated at every original pair before the values are summarized within each height band.}

\begin{table}[!htbp]
\centering
\caption{{Pointwise case-level horizontal-wind validation statistics: MAE, RMSE, Bias, and coefficients of determination.}}
\label{tab:case_r2}
\footnotesize
\setlength{\tabcolsep}{5pt}
\renewcommand{\arraystretch}{1.08}

\begin{tabular}{lrrrrrr}
\toprule
{Case} & \mathcolorbox{yellow}{N} & \shortstack{{MAE}\\{(m/s)}} & \shortstack{{RMSE}\\{(m/s)}} & \shortstack{{Bias}\\{(m/s)}} & \mathcolorbox{yellow}{R_1^2} & \mathcolorbox{yellow}{R_2^2} \\
\midrule
{Case 1} & {97}  & {0.478} & {0.630} & \mathcolorbox{yellow}{-0.232} & {0.827} & {0.859} \\
{Case 2} & {102} & {0.383} & {0.466} & {0.016}  & {0.904} & {0.924} \\
{Case 3} & {168} & {0.402} & {0.546} & \mathcolorbox{yellow}{-0.096} & {0.868} & {0.872} \\
{Case 4} & {244} & {0.534} & {0.696} & {0.008}  & {0.807} & {0.809} \\
\bottomrule
\end{tabular}
\end{table}
\FloatBarrier

{To isolate the contribution of interpolation smoothness from the dynamically resolved response, two Kriging-only baselines are evaluated against the same pointwise observations. The complete Kriging field samples the original observation-driven three-dimensional field directly at each validation location. The outer-boundary Kriging baseline retains only the four lateral faces and the top face of that field; the interior is reconstructed by layer-wise ordinary Kriging, the stored top face is retained as the upper boundary, and the station profiles are obtained by linear interpolation in height. Table~\ref{tab:kriging_cfd_comparison} compares both baselines with our method using the same 611 paired points. For the six GAW110 samples at 42~m, below the stored 50~m bottom plane, both Kriging-only predictions are assigned a horizontal wind speed of 0~m/s rather than vertically extrapolated.}

\begin{table}[!htbp]
\centering
\caption{{Pointwise comparison of the Kriging-only baselines and our method against the same 611 lidar observations: MAE, RMSE, Bias, and coefficients of determination.}}
\label{tab:kriging_cfd_comparison}
\footnotesize
\setlength{\tabcolsep}{5pt}
\renewcommand{\arraystretch}{1.08}
\begin{tabular}{lrrrrrr}
\toprule
{Method} & \shortstack{{MAE}\\{(m/s)}} & \shortstack{{RMSE}\\{(m/s)}} & \shortstack{{Bias}\\{(m/s)}} & \mathcolorbox{yellow}{R_1^2} & \mathcolorbox{yellow}{R_2^2} \\
\midrule
{Complete Kriging field} & {0.370} & {0.516} & \mathcolorbox{yellow}{-0.132} & {0.888} & {0.898} \\
{Outer-boundary Kriging} & {0.357} & {0.501} & \mathcolorbox{yellow}{-0.113} & {0.894} & {0.900} \\
{Our method}            & {0.464} & {0.612} & \mathcolorbox{yellow}{-0.058} & {0.842} & {0.843} \\
\bottomrule
\end{tabular}
\end{table}

{Both Kriging-only baselines yield lower MAE and RMSE and higher }\mathcolorbox{yellow}{R_2^2}{ than our method for the pointwise mean-horizontal-speed comparison, quantifying the substantial contribution of the smooth observation-driven interpolation to withheld-profile agreement. Conversely, the magnitude of the signed Bias is smallest for our method: 0.058~m/s, compared with 0.132 and 0.113~m/s for the complete and outer-boundary Kriging fields, respectively. Because Bias retains the error sign, this smaller net offset partly reflects cancellation between positive and negative local discrepancies, whereas MAE and RMSE retain their magnitudes.}

{The additional pointwise scatter produced by our method is associated with the explicit reconstruction of local flow instabilities, vertically layered small-scale structures, local recirculation, shear-generated turbulence, and secondary motions displaced from the dominant building-height layer. These resolved motions create bidirectional departures from a smooth mean profile and redistribute kinetic energy across height, which is then expressed in the velocity profile. A further contribution arises from the mismatch between the boundary forcing used in the calculation and the instantaneous atmospheric boundary. The present engineering configuration holds the time-collocated mean velocity field fixed during each integration and superposes synthetic turbulent fluctuations. In the real environment, the boundary velocity varies continuously and has neither a constant value nor a fixed turbulence intensity. Consequently, the phase and amplitude of the small-scale layered structures, local reverse flow, and shear-generated turbulence cannot be synchronized exactly with the instantaneous lidar profile. Meaningful reduction of this mismatch would require second-scale, high-frequency, time-dependent three-dimensional boundary forcing together with contemporaneous turbulence information. The fixed-mean-boundary treatment with synthetic turbulence remains a commonly used engineering assumption. The comparison therefore quantifies the mean-profile contribution inherited from interpolation, while the building-resolved unsteady response and the independent validations in Section~\ref{sec:validation-in-principle} establish the complementary physical and numerical contribution of the solver.}

\begin{table}[!htbp]
\centering
\caption{{Height-resolved pointwise horizontal-wind errors across all case-specific paired comparisons: MAE and RMSE. Each height interval is left-inclusive and right-exclusive.}}
\label{tab:height_errors}
\footnotesize
\setlength{\tabcolsep}{5pt}
\renewcommand{\arraystretch}{1.08}
\begin{tabular}{lrrr}
\toprule
{Height interval (m)} & \mathcolorbox{yellow}{N} & \shortstack{{MAE}\\{(m/s)}} & \shortstack{{RMSE}\\{(m/s)}} \\
\midrule
\mathcolorbox{yellow}{0\le z<100}   & {76}  & {0.531} & {0.632} \\
\mathcolorbox{yellow}{100\le z<200} & {171} & {0.459} & {0.561} \\
\mathcolorbox{yellow}{200\le z<300} & {126} & {0.618} & {0.790} \\
\mathcolorbox{yellow}{300\le z<400} & {56}  & {0.662} & {0.906} \\
\mathcolorbox{yellow}{400\le z<500} & {54}  & {0.338} & {0.396} \\
\mathcolorbox{yellow}{500\le z<600} & {54}  & {0.174} & {0.227} \\
\mathcolorbox{yellow}{600\le z<700} & {52}  & {0.326} & {0.370} \\
\mathcolorbox{yellow}{700\le z<800} & {22}  & {0.228} & {0.283} \\
\bottomrule
\end{tabular}
\end{table}
\FloatBarrier

{The component-wise Ordinary-Kriging variance is also evaluated at each withheld lidar location on the retained 50-m target levels for the three validation times. Because the variance changes with location, height, and analysis time, Table~\ref{tab:kriging_variance} reports its median and 95th percentile for each case. Here, }\mathcolorbox{yellow}{N_K}{ counts combinations of withheld lidar location, target level, and analysis time and is distinct from the number of original lidar--simulation pairs used for the pointwise error statistics.}

\begin{table}[!htbp]
\centering
\caption{{Component-wise Ordinary-Kriging variance at withheld lidar locations. Values are medians, with 95th percentiles in parentheses.}}
\label{tab:kriging_variance}
\footnotesize
\setlength{\tabcolsep}{7pt}
\renewcommand{\arraystretch}{1.08}
\begin{tabular}{lrrr}
\toprule
{Case} & \mathcolorbox{yellow}{N_K} & \shortstack{\mathcolorbox{yellow}{\sigma_{K,u}^2}\\{(m$^2$/s$^2$)}} & \shortstack{\mathcolorbox{yellow}{\sigma_{K,v}^2}\\{(m$^2$/s$^2$)}} \\
\midrule
{Case 1} & {29}  & {0.066 (0.182)} & {0.091 (0.661)} \\
{Case 2} & {52}  & {0.078 (0.189)} & {0.088 (0.743)} \\
{Case 3} & {75}  & {0.092 (0.330)} & {0.100 (1.498)} \\
{Case 4} & {104} & {0.030 (0.394)} & {0.049 (0.932)} \\
\bottomrule
\end{tabular}
\end{table}
\FloatBarrier

Moreover, even by using only a limited amount of sparse observational data as input, this set of experiments successfully infers both the vertical structure and the quantitative characteristics of the wind field in other regions. These results effectively demonstrate that the proposed simulation framework possesses strong capability for wind-field analysis and reconstruction under complex underlying surfaces with dense building coverage.

\subsection{\texorpdfstring{{Physical-time convergence and quasi-steady treatment}}{Physical-time convergence and quasi-steady treatment}\label{sec:quasi-steady}}

{For each analysis time, the fixed reconstructed boundary state drives an unsteady LBM--LES calculation. The purpose is not to reproduce the chronological evolution of the atmospheric wind field during the integration or to provide a fully time-evolving meteorological forecast. Instead, the calculation obtains the building-resolved turbulent response that develops under the instantaneous reconstructed boundary state, thereby providing a quasi-steady urban wind-field solution for low-altitude operations.}

\begin{figure}[!htbp]
    \centering
    \includegraphics[width=\linewidth]{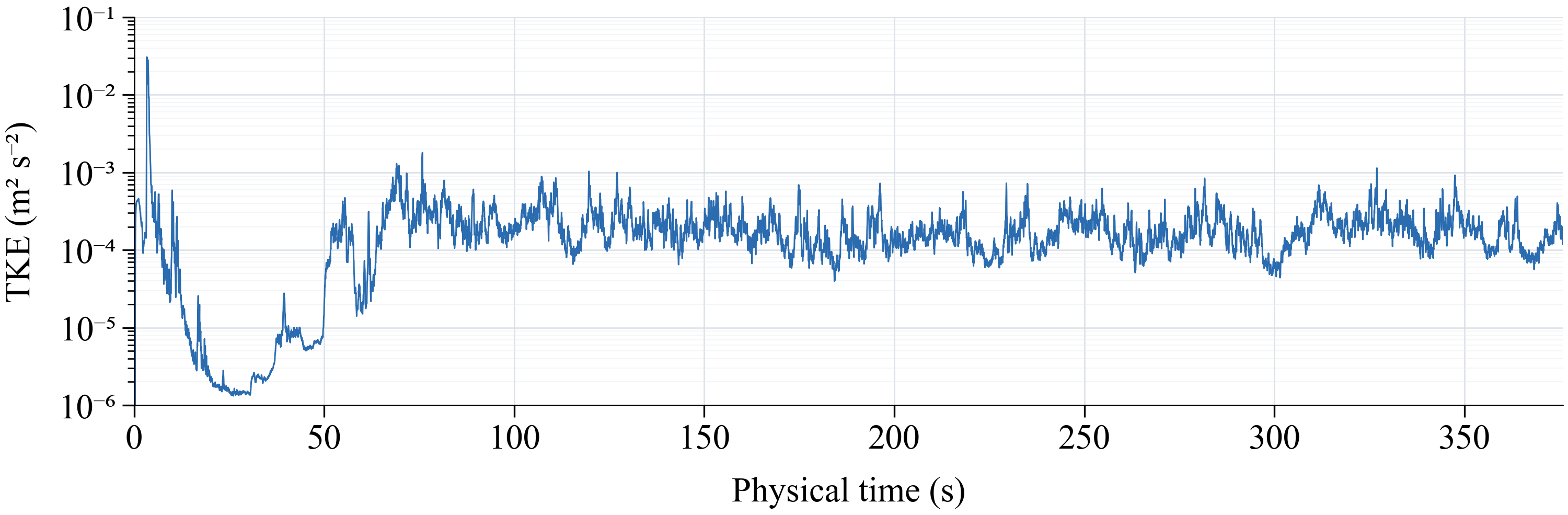}
    \caption{{Temporal evolution of the 1-s sliding-window mean TKE along the measurement line during the LBM--LES integration.}}
    \label{fig:temporal-convergence}
\end{figure}

{Figure~\ref{fig:temporal-convergence} shows the temporal evolution of the 1-s sliding-window mean TKE along the measurement line. During the first approximately 30~s, the TKE response is governed by the formation and dissipation of the LBM start-up/acoustic (onset) vortex generated at initialization. From approximately 30 to 70~s, interaction with the resolved near-ground buildings and their friction reconstructs the flow structure and drives its development from the initial laminar state toward turbulent structures. After approximately 100~s, the TKE fluctuates about a persistent turbulent plateau. This transition demonstrates convergence in physical time and entry into a quasi-steady turbulent state; the remaining fluctuations are the unsteady turbulent content of that state. The 40,000-step integration, corresponding to 400~s of physical time, is therefore substantially longer than the time required to pass the initial flow-development stage and provides an ample post-development interval for obtaining the reported quasi-steady results.}

\section{Conclusions\label{sec:4_concl}}

In this study, we develop and validate a lattice Boltzmann method with large-eddy simulation (LBM-LES) framework that leverages sparse station observations to reconstruct urban wind fields with near-real-time updates and continuous spatial coverage for low-altitude applications.
The primary conclusions are as follows:

\begin{enumerate}[1.] 

\item Based on LBM-LES, our \textsc{LatticeUrbanWind} efficiently handles low-altitude wind field simulations, considering three-dimensional complex flow patterns, including building-induced flow, street canyon, and terrain influences.

\item The synergistic integration of lidar and LBM-LES technology provides an effective solution for full spatial and temporal coverage of urban micrometeorological winds.

\item The numerical results by the developed LBM-LES pipeline show good agreement with observations, demonstrating its accuracy and reliability. Within minutes, it completes meter-level resolution GPU-accelerated wind field simulations spanning several kilometers.

\item {The grid strategy exerts a pronounced influence on the simulation outcomes. In the present urban setting, 5 m is the coarsest tested resolution that still preserves the essential building geometry and near-ground profile structure, whereas coarser grids underestimate ground shear stress and yield an underdeveloped wind-velocity profile.}

\item The size of the precursor domain determines whether the flow is fully developed at the location of the focus region. A sufficiently large precursor domain provides an adequate buffer zone, enabling the reconstruction of turbulent characteristics. Moreover, the buffer length required for flow development is strongly influenced by the building density.

\end{enumerate}


{The present field assessment is limited to a single precipitation-free urban study window, an isothermal configuration, and the available horizontal-wind lidar observations. The available observations do not include fluctuation velocities, turbulence intensity, gust statistics, or direct vertical-velocity measurements. Therefore, these quantities cannot be independently evaluated against observations in the present study. Although the reproduction of the observed mean horizontal-wind profiles is dynamically consistent with the simulated flow development and turbulent structures that shape them, the available observations do not provide sufficient evidence to establish the accuracy of these quantities. Further field validation should incorporate vertically resolved temperature observations across multiple atmospheric-stability regimes, together with fluctuation-velocity, gust, turbulence-intensity, and vertical-velocity observations; extend to sites with more pronounced topographic relief; and assess performance during strong-wind and precipitation events. These tests are needed to delineate the conditions over which the framework can be applied across a wider range of meteorological and topographic settings.}

In summary, this paper presents a comprehensive verification and analysis of near-real-time simulation of low-altitude micrometeorological winds with CPU-GPU hybrid LBM-LES approach. 
Based on our existing findings, our subsequent research will further focus on the influence of multi-physics field coupling at the LBM drive boundary, {including explicit treatment of vertical-velocity information,} thermal (buoyancy) effects \cite{EDT-Thermal,EDT-Breview} and high-order turbulent fields. Additionally, we are to establish standardized workflows, datasets and benchmarks for numerical simulations in urban environments, comparative to the open-access AIJ Wind Tunnel Database and Guideline \cite{tominaga2008aijguidelines,AIJLES}, in order to reduce the barrier to entry for new users and peers, and promote repeatability in continuously operating low-altitude support systems. 
{What's more, although the platform has been applied in real-world scenarios such as The World Games 2025 in Chengdu and for the national low-altitude economy zone in China, its performance under complex meteorological conditions still requires further verification and iteration to meet the growing, ever-changing and challenging support needs of the low-altitude operations.}

%

\section*{Declarations}
{
\small
\begin{itemize}


    \item \textbf{Funding:} \\
    This work was supported by Jing-Jin-Ji Regional Integrated Environmental Improvement - National Science and Technology Major Project (2025ZD1201\\902), Guangzhou Low-Altitude 628 Flight Meteorological Support Volunteer Observation Experiments Scheme, and the Open Foundation of Key Laboratory of Meteorological Software, China Meteorological Administration (Grant No. GHMA2024Y05).

    \item \textbf{Acknowledgements:} \\
    The authors thank Yi Zhang and Taofeng Gu of the Guangzhou Meteorological Integrated Security Center for technical support, observational data resources, and discussions.
    
    \item \textbf{Conflicts of interest:} \\
    The authors have no conflicts of interest to declare that are relevant to the content of this article.
    
    \item \textbf{Data availability:} \\
    Data supporting the findings of this study are available from the corresponding author on a reasonable request. Our \textsc{LatticeUrbanWind} platform is made open-source permanently under a customized non-commercial license and maintained at https://github.com/hweifluids/LatticeUrbanWind. The release version used in this study is LUW-v2.8. The necessary files to reproduce a complete end-to-end computational process are included to enhance usability and reproducibility.
\end{itemize}
}

\providecommand{\noopsort}[1]{}\providecommand{\singleletter}[1]{#1}%

\clearpage
\newcounter{supplementarymaterial}
\renewcommand{\thesupplementarymaterial}{S\arabic{supplementarymaterial}}
\providecommand{\theHsupplementarymaterial}{\arabic{supplementarymaterial}}
\newcommand{\supplementarymaterialsection}[3]{%
  \refstepcounter{supplementarymaterial}%
  \section*{Supplementary Material~#1. #2}%
  \label{#3}%
}

For the following supplementary materials, please refer to the separate document from the publisher.

\supplementarymaterialsection{S1}{Kriging boundary construction from lidar}{supp:Kriging}

\supplementarymaterialsection{S2}{Construction of terrain layer}{supp:TEInt}

\supplementarymaterialsection{S3}{Velocity interpolation}{supp:VInt}

\supplementarymaterialsection{S4}{Lattice Boltzmann method in FluidX3D}{supp:LBM}

\supplementarymaterialsection{S5}{\texorpdfstring{{Immersed Boundary Method}}{Immersed Boundary Method}}{supp:surface-momentum-ibm}

\supplementarymaterialsection{S6}{Coriolis forcing term}{supp:sourceterm}

\supplementarymaterialsection{S7}{Turbulence Generation Approach}{supp:STG}

\supplementarymaterialsection{S8}{2D Wavenumber Spectra}{supp:wn}


\vfill

\end{document}